\documentclass[aps,prd,floatfix,superscriptaddress,preprintnumbers]{revtex4}
\usepackage{amssymb}
\usepackage{amsmath}
\usepackage{amsfonts}
\usepackage{epsfig}             
\usepackage{graphicx}
\usepackage{tabularx}
\usepackage{bm}
\usepackage{color}
\newcommand{\seq}{\begin{subequations}}
\newcommand{\sen}{\end{subequations}}
\newcommand{\eq}{\begin{eqnarray}}
\newcommand{\en}{\end{eqnarray}}
\newcommand{\la}{\langle}
\newcommand{\ra}{\rangle}
\newcommand{\bfn}{{\bf 0}_{\perp}}

\newcommand{\bfp}{{\bf p}_{\perp}}

\newcommand{\bfk}{{\bf k}_{\perp}}
\newcommand{\bfb}{{\bf b}_{\perp}} 

\begin{document}

\title{Gluon parton densities in soft-wall AdS/QCD} 

\author{Valery E. Lyubovitskij} 
\affiliation{Institut f\"ur Theoretische Physik,
Universit\"at T\"ubingen, 
Kepler Center for Astro and Particle Physics,  
Auf der Morgenstelle 14, D-72076 T\"ubingen, Germany}
\affiliation{Departamento de F\'\i sica y Centro Cient\'\i fico
Tecnol\'ogico de Valpara\'\i so-CCTVal, Universidad T\'ecnica
Federico Santa Mar\'\i a, Casilla 110-V, Valpara\'\i so, Chile} 
\affiliation{Department of Physics, Tomsk State University,
634050 Tomsk, Russia}
\affiliation{Tomsk Polytechnic University, 634050 Tomsk, Russia} 
\author{Ivan Schmidt}
\affiliation{Departamento de F\'\i sica y Centro Cient\'\i fico
Tecnol\'ogico de Valpara\'\i so-CCTVal, Universidad T\'ecnica
Federico Santa Mar\'\i a, Casilla 110-V, Valpara\'\i so, Chile}

\date{\today}

\begin{abstract}

We study the gluon parton densities [parton distribution functions (PDFs), 
transverse momentum distributions (TMDs), generalized parton distributions (GPDs)]  
and form factors in soft-wall AdS/QCD.  
We show that the power behavior of gluon parton distributions 
and form factors at large values of the light-cone variable and 
large values of square momentum is  
consistent with quark counting rules. We also show that the 
transverse momentum distributions derived in our approach 
obey the model-independent Mulders-Rodrigues inequalities without 
referring to specific model parameters. 
All gluon parton distributions are defined in terms of the unpolarized 
and polarized gluon PDFs and profile functions. The latter are related 
to gluon PDFs via differential equations. 

\end{abstract}

\maketitle

\section{Introduction}

The soft-wall AdS/QCD model~\cite{Karch:2006pv}-\cite{Andreev:2006vy}, 
based on breaking of conformal symmetry due a quadratic dilaton field, 
has achieved important progress in the description and understanding of 
hadron structure (mass spectrum, parton distributions, form factors, thermal 
properties, etc.)~\cite{Brodsky:2014yha}. One of the main advantages of 
the soft-wall AdS/QCD is the analytical implementation of quark counting 
rules~\cite{Brodsky:1973kr}, in the description of hadronic form factors at large 
$Q^2$ (power scaling)~\cite{Brodsky:2014yha}-\cite{Lyubovitskij:2020gjz}.  
Together with form factors, parton distributions of quarks and gluons in hadrons 
play important role in the QCD description of hadron structure and spin physics (see, e.g., 
Refs.~\cite{Lin:2017snn,Lin:2020rut,Angeles-Martinez:2015sea,Diehl:2013xca,Boer:2011fh}
for reviews). Based on QCD factorization, one can separate effects
of strong interactions at small and long distances, characterizing respectively 
perturbative and nonperturbative dynamics of quarks and gluons. In particular, 
the nonperturbative part is parametrized by parton distribution functions, which are
universal functions for each hadron and independent of the specific process.
Since these universal parton distributions cannot be directly calculated in QCD,
they are either extracted from data (world data analysis) or 
calculated using lattice QCD, or in QCD motivated approaches 
(light-front QCD, AdS/QCD, quark and potential models, etc.), which  
have been applied to extract or predict the PDFs, TMDs, and GPDs 
(for a recent overview see, e.g., Ref.~\cite{Lin:2020rut}). 

As happens with form factors, the partonic distributions 
obey model-independent scaling rules at large $x$. The starting point for these rules was 
the derivation of the Drell-Yan-West (DYW) relation~\cite{Drell:1969km} 
between the large-$Q^2$ behavior of nucleon electromagnetic form factors and 
the large-$x$ behavior of the structure functions (see also 
Ref.~\cite{Bloom:1970xb} for the extension to inelastic scattering) and 
quark counting rules~\cite{Brodsky:1973kr}. Based on the results of 
Refs.~\cite{Drell:1969km,Bloom:1970xb,Brodsky:1973kr} 
the behavior of the quark PDF in nucleon 
$q_v(x) \sim (1-x)^p$ at $x \to 1$ was related with the scaling of 
the proton Dirac form factor $F_1^p(Q^2) \sim 1/(Q^2)^{(p+1)/2}$ at large $Q^2$, where 
the parameter $p$ is related to the number of constituents in the proton 
(or twist $\tau$) as $p = 2 \tau - 3$~\cite{Drell:1969km,Blankenbecler:1974tm}. 
The model-independent predictions of perturbative QCD (pQCD) for 
the GPDs of pion ${\cal H}_q^\pi(x,Q^2)$ and nucleon ${\cal H}_q^N(x,Q^2)$, 
${\cal E}_q^N(x,Q^2)$, are given in Ref.~\cite{Yuan:2003fs}, at large $x$ and finite $Q^2$, as: 
\eq
\label{Hpi_N}
{\cal H}_q^\pi(x,Q^2) \sim (1-x)^2\,, \quad 
{\cal H}_q^N(x,Q^2)   \sim (1-x)^3\,, \quad {\cal E}_q^N(x,Q^2) 
\sim (1-x)^5 \,.
\en 
The prediction of pQCD for the pion PDF $q_\pi(x) \sim (1-x)^2$ 
at large $x$, which follows from the prediction for GPDs~\cite{Yuan:2003fs}),
was supported by the updated analysis~\cite{Aicher:2010cb} 
of the E615 data~\cite{Conway:1989fs} on the cross section 
of the Drell-Yan (DY) process $\pi^- N \to \mu^+ +\mu^- X$,  
including next-to-leading logarithmic threshold resummation effects: 
$q_\pi(x) \sim (1-x)^{2.03}$ at the initial scale 
$\mu_0 = 0.63$ GeV~\cite{Aicher:2010cb}.    

In the case of gluons the large and small $x$ behavior of parton densities was 
studied in~\cite{Brodsky:1989db,Brodsky:1994kg}. 
In particular, the QCD constraints on unpolarized $G(x)$ and polarized $\Delta G(x)$ 
gluon PDFs in nucleons have been derived in~\cite{Brodsky:1989db}: 
\eq\label{SetI} 
       G(x) &=& \frac{N}{x} \, \biggl[ 5 (1-x)^4 - 4 (1-x)^5 + (1-x)^6 \biggr] 
= \frac{N}{x} \, (1-x)^4 \, \biggl[ 2 + 2x + x^2 \biggr] \,, \nonumber\\
\Delta G(x) &=& \frac{N}{x} \, \biggl[ 5 (1-x)^4 - 4 (1-x)^5 - (1-x)^6 \biggr] 
= N \, (1-x)^4 \, \biggl[ 6 - x\biggr] 
\,,  
\en 
where $N$ is the normalization constant. 
 
On the other hand, the densities $G(x)$ and $\Delta G(x)$ can be written as combinations 
of the helicity-aligned $G^+(x) = G_{g\uparrow/N\uparrow}(x)$ and 
helicity-antialigned $G^-(x) = G_{g\downarrow/N\uparrow}(x)$ gluon distributions
\eq
G(x) = G^+(x) + G^-(x)\,, \quad
\Delta G(x) = G^+(x) - G^-(x) \,,
\en
where 
\eq 
G^+(x) &=& \frac{N}{x} \, (1-x)^4 \, \biggl[ 1 + 4x \biggr] \,, \nonumber\\
G^-(x) &=& \frac{N}{x} \, (1-x)^6  \,. 
\en 
Thus, at large $x$ the power scaling of gluon PDFs and their ratios read~\cite{Brodsky:1989db}
\eq 
G^+(x) \sim (1-x)^4\,, \quad G^-(x) \sim (1-x)^6\,, \quad 
\frac{G^-(x)}{G^+(x)} \sim (1-x)^2 \,,
\en 
which is consistent with QCD constraints~\cite{Bjorken:1969mm,Gribov:1972ri}, 
dictated matching the signs of the quark and gluon helicities and 
the even power scaling of gluon PDFs. 

At small $x$ the gluon asymmetry ratio $\Delta G/G$ behaves as 
\eq\label{smallx} 
\frac{\Delta G(x)}{G(x)} \to N_q x \,, 
\en 
where $N_q$ is the number of valence quarks in a specific hadron (e.g., 
$N_q = 3$ in case of nucleon). The scaling rule (\ref{smallx}) is consistent 
with Reggeon exchange arguments~\cite{Brodsky:1988ip}. 

The moments of the gluon PDFs, the momentum fraction 
$\la x_g \ra$ and the helicity $\Delta G$ carried by intrinsic gluons 
in nucleon, obtained in Ref.~\cite{Brodsky:1988ip}, are
\eq\label{gluon_momentsI} 
\la x_g \ra = \int\limits_0^1 dx x G(x) = \frac{10}{21} N\,, 
\quad 
\Delta G = \int\limits_0^1 dx \Delta G(x) = \frac{7}{6} N\,. 
\en 
The ratio of these two moments $\Delta G/\la x_g \ra = \frac{49}{20}$ 
is independent of the $N$. 

In Ref.~\cite{Brodsky:1994kg}  a slightly different small 
$x$ constraint on the ratio $\Delta G/G$, due to color coherence of gluon 
couplings, has been implemented, leading to the following form of gluon PDFs 
\eq\label{SetII} 
       G(x) &=& \frac{N}{x} \, (1-x)^4 \biggl[ 1 + (1-x)^2 \biggr] \,, 
\nonumber\\
\Delta G(x) &=& \frac{N}{x} \, (1-x)^4 \biggl[ 1 - (1-x)^2 \biggr] \,.
\en 
One can see that the two sets of gluon PDFs, presented in Eqs.~(\ref{SetI}) 
and~(\ref{SetII}), differ by the presence of an extra linear 
term $+4x$ in the set of Ref.~\cite{Brodsky:1989db}. As a consequence, the two models 
considered in Refs.~\cite{Brodsky:1989db,Brodsky:1994kg} 
produce different results for the moments of gluon PDFs. 
In particular, the model considered in Ref.~\cite{Brodsky:1994kg} gives: 
\eq\label{gluon_momentsII} 
1 = \la x_g \ra = \frac{12}{35} N\,, 
\quad 
\Delta G = \int\limits_0^1 dx \Delta G(x) = \frac{11}{30} N\,, 
\quad \frac{\Delta G}{\la x_g \ra} = \frac{77}{72} \,.
\en 

As we know, the PDFs are related to the transverse momentum dependent (TMD) 
parton distributions, upon integration of the latter over the struck parton transverse 
momentum $\bfk$. TMDs provide a three-dimensional picture 
of hadrons and their knowledge is important for the description of 
QCD processes, using TMD factorization at small values of the transverse 
momentum of particles produced in hadronic collisions. 
At present the TMDs are under intensive study both experimentally 
and theoretically (for recent progress see, e.g., 
Refs.~\cite{Lin:2020rut,Angeles-Martinez:2015sea}). 
Leading twist quark TMDs have been proposed 
in a series of papers~in Refs.~\cite{Ralston:1979ys,Brodsky:2002cx}. Gluon TMDs 
have been introduced in Ref.~\cite{Mulders:2000sh} and later considered 
in Ref.~\cite{Meissner:2007rx}-\cite{Kaur:2020pvc}. 

The calculation of partonic densities in hadrons, using soft-wall AdS/QCD, 
can be done indirectly using an integral representation for the hadronic 
form factors or normalization conditions for the 
hadronic wave functions~\cite{Vega:2009zb}-\cite{Lyubovitskij:2020otz}.  
One should stress  that the study of parton densities in the soft-wall approach 
is closely related to other important problems such as the construction of hadronic 
effective wave functions~\cite{Brodsky:2014yha,Brodsky:2007hb,Vega:2009zb,%
Branz:2010ub,Brodsky:2011xx,Lyubovitskij:2020otz,Gutsche:2014zua,deTeramond:2018ecg,%
Lyubovitskij:2013ski,Gutsche:2013zia,Gutsche:2014yea,%
Gutsche:2016gcd,Brodsky:2020ajy,Vega:2020ctz}. 
First results in soft-wall AdS/QCD on parton densities -- quark GPDs -- 
have been obtained in Ref.~\cite{Vega:2009zb}. 
The idea for the extraction of GPDs in Ref.~\cite{Vega:2009zb} was based on the 
use of the integral representation of the hadronic form factor with twist 
$\tau$~\cite{Brodsky:2007hb,Vega:2010ns,Gutsche:2011vb}.  
It can also be written in closed form in terms of the beta function $B(\alpha,\beta)$
\eq 
F_\tau(Q^2) = \int\limits_0^1 dy \, (\tau - 1) \, (1-y)^{\tau-2} \, y^a 
= (\tau-1) \, B(\tau-1,a+1) 
\en 
Upon identification of the $y$ variable with the light-cone momentum fraction $x$, 
this gives an expression for both the PDFs $q_\tau(x)$ and the GPDs 
${\cal H}_\tau(x,Q^2)$~\cite{Brodsky:2007hb,Vega:2010ns,Gutsche:2013zia}: 
\eq 
q_\tau(x) = (\tau - 1) \, (1-x)^{\tau-2}\,, \quad {\cal H}_\tau(x,Q^2) = q_\tau(x) \, x^a \,. 
\en 
Nevertheless, this $x$ dependence of the PDFs and GPDs contradicts 
model-independent results: the DY inclusive counting rules 
for $q_\tau(x)$ at $x \to 1$~\cite{Drell:1969km,Blankenbecler:1974tm,Yuan:2003fs} 
and the prediction of pQCD for GPDs  --- pion ${\cal H}_q^\pi(x,Q^2)$ 
and nucleon ${\cal H}_q^N(x,Q^2)$, ${\cal E}_q^N(x,Q^2)$ at large $x$ and 
finite $Q^2$~\cite{Yuan:2003fs}. 

To solve the problem of large-$x$ scaling of parton densities in soft-wall AdS/QCD, 
it was found in Ref.~\cite{Lyubovitskij:2013ski} that the interpretation 
of the variable $y$ 
in the integral representation~(\ref{VInt}) as light-cone variable is not truly correct 
and that one should propose a {\it generalized $x$ dependent light-cone variable $y(x)$}. 
With this assumption the power behavior of hadronic PDFs and GPDs at large $x$ can be made consistent 
with the model-independent 
results of Refs.~\cite{Drell:1969km,Blankenbecler:1974tm,Yuan:2003fs}, 
provided that an appropriate choice of the $x$ dependence of the function $y(x)$ is imposed. 
In particular, the simplest choice for the function $y(x)$ was found to be: 
\eq\label{ytau} 
y_N(x) = \exp\Big[-\log(1/x) (1-x)^{2/(N-1)}\Big]
\en 
leading to the correct large-$x$ scaling of PDFs and GPDs in mesons  
\eq 
q_\tau^M(x) \sim {\cal H}_\tau^M(x,Q^2) \sim (1-x)^{2 \tau - 2}
\en 
at $N=2 \tau - 2$ and in baryons 
\eq 
q_\tau^B(x) \sim {\cal H}_\tau^B(x,Q^2) \sim (1-x)^{2 \tau - 3}
\en 
at $N=2 \tau - 3$. The function $y_\tau(x)$ obeys the following boundary 
conditions $y_\tau(0) = 0$ and $y_\tau(1) = 1$.  
Extension of ideas proposed in Ref.~\cite{Lyubovitskij:2013ski}
were further developed in Ref.~\cite{Lyubovitskij:2020otz}. 
In particular, it was explicitly demonstrated how to correctly define 
hadronic parton distributions (PDFs, TMDs, and GPDs) in the
soft-wall AdS/QCD approach, in order for them to be consistent 
with quark counting rules and Drell-Yan-West duality.
All parton distributions are defined in terms of profile functions 
universal for each specific hadron. 

Recently a similar idea was considered in the framework of light-front 
holographic QCD (LFHQCD)~\cite{deTeramond:2018ecg,Brodsky:2020ajy,Chang:2020kjj}.  
In particular, a function [named as $w(x)$] was introduced in the integral 
representation of the form factor~\cite{deTeramond:2018ecg,Brodsky:2020ajy}: 
\eq 
F_\tau(Q^2) = \frac{1}{N_\tau} \, \int\limits_0^1 dx \, w^\prime(x) \, 
[w(x)]^{Q^2/4\lambda-1/2} \, [1-w(x)]^{\tau-2}
\en 
In fact, both mathematical extensions considered in Refs.~\cite{Lyubovitskij:2013ski}   
and~\cite{deTeramond:2018ecg,Brodsky:2020ajy} are equivalent. The only difference is that 
in Refs.~\cite{deTeramond:2018ecg,Brodsky:2020ajy} an extra power $-1/2$ was included 
in the $[w(x)]^{Q^2/4\lambda-1/2}$, while in the soft-wall 
model of~\cite{Brodsky:2007hb,Vega:2010ns,Gutsche:2011vb} the factor 
is $[w(x)]^{Q^2/4\lambda}$. In other words, the soft-wall 
model~\cite{Brodsky:2007hb,Vega:2010ns,Gutsche:2011vb} 
and LFHQCD~\cite{Brodsky:2007hb,Vega:2010ns,Gutsche:2011vb} deal with slightly 
different analytical expressions for the hadronic form factors: 
$F_\tau(Q^2) \sim B(\tau - 1,1+Q^2/4\lambda)$ in the soft-wall AdS/QCD 
and $F_\tau(Q^2) \sim B(\tau - 1,1/2+Q^2/4\lambda)$ in LFHQCD.  

The main objective of this paper is to extend the ideas proposed 
in Refs.~\cite{Lyubovitskij:2013ski,Lyubovitskij:2020otz}, applying them
to gluon parton densities. 
In particular, we will derive results for gluon 
parton distributions (PDFs, TMDs, and GPDs) in hadrons 
with arbitrary quark content and spin. 
The obtained power behavior of parton distributions at large values 
of light-cone variable $x$ are then also consistent 
with quark counting rules and DYW duality. 
All parton distributions are defined in terms of profile functions depending on 
the light-cone coordinate and are fixed from PDFs and electromagnetic form factors. 

The paper is organized as follows.
In Sec.~II we present an overview of our approach and 
consider the derivation of PDFs, TMDs, and GPDs using quarks as an example. 
In Sec.~III we consider the derivation of gluon parton densities 
(PDFs, TMDs, and GPDs). In Sec.~IV we present numerical applications 
of our analytical results for gluon parton densities using 
different parametrizations for $G(x)$ and $\Delta G(x)$ PDFs.  
Finally, Sec.~V contains our summary. 

\section{Basic notions of soft-wall AdS/QCD approach} 

Here we briefly overview the soft-wall AdS/QCD approach 
(for more details see, e.g., Refs.~\cite{Vega:2010ns,Gutsche:2011vb}). 
The framework for studies of the dynamics of boson 
$\Phi_{M_1 \cdots M_J}(x,z)$ and fermion $\Psi_{M_1 \cdots M_{J-1/2}}(x,z)$ 
fields with spin $J$ (dual to mesons and baryons, respectively) 
in a five-dimensional AdS space, is specified by the metric 
\eq
ds^2 =
g_{MN} dx^M dx^N = \eta_{ab} \, e^{2A(z)} \, dx^a dx^b = e^{2A(z)}
\, (\eta_{\mu\nu} dx^\mu dx^\nu - dz^2)\,, \hspace*{1cm}
\eta_{\mu\nu} = {\rm diag}(1, -1, \ldots, -1) \,,
\en
where $M$ and
$N = 0, 1, \cdots , d$ are the space-time (base manifold) indices,
$a=(\mu,z)$ and $b=(\nu,z)$ are the local Lorentz (tangent) indices,
$g_{MN}$ and  $\eta_{ab}$ are curved and flat metric tensors, which
are related by the vielbein $\epsilon_M^a(z)= e^{A(z)} \,                     
\delta_M^a$ as $g_{MN} =\epsilon_M^a \epsilon_N^b \eta_{ab}$. Here
$z$ is the holographic coordinate, $R$ is the AdS radius, and $g =             
|{\rm det} g_{MN}| = e^{10 A(z)}$. We restrict
ourselves to a conformal-invariant metric with $A(z) = \log(R/z)$. 

The boson action is written as: 
\eq
S_B &=& \frac{(-)^J}{2}
\int d^d x dz \sqrt{g} \, e^{-\varphi(z)}
\biggl[ g^{MN} g^{M_1N_1} \cdots g^{M_JN_J} \,
\partial_M\Phi_{M_1 \cdots M_J}(x,z) \,
\partial_N\Phi_{N_1 \cdots N_J}(x,z) \nonumber\\
&-& (\mu_J^2 + V_J(z)) \, g^{M_1N_1} \cdots g^{M_JN_J}
\Phi_{M_1 \cdots M_J}(x,z) \, \Phi_{N_1 \cdots N_J}(x,z) \biggr]
\en 
where the bosonic spin-$J$ field $\Phi_{M_1 \cdots M_J}(x,z)$ is 
described 
by a symmetric, traceless tensor, satisfying the conditions
\eq
\partial^{M_1}  \Phi_{M_1M_2 \cdots M_J} = 0\,, \quad\quad
g^{M_1M_2}  \Phi_{M_1M_2 \cdots M_J} = 0\,, 
\en
Here $V_J(z) = e^{-2A(z)} U_J(z)$, where $U_J(z)$ 
is the effective dilaton potential
\eq
U_J(z) = \frac{1}{2} \varphi^{\prime\prime}(z)
\,+\, (d-1-2J) \, \varphi^{\prime}(z) A^{\prime}(z) 
\en 
and 
\eq
\mu_J^2 R^2 = (\Delta - J) (\Delta + J - 4) 
\en 
is the bulk mass. 
The quadratic dilaton field $\varphi(z)$ is specified as 
$\varphi(z) = \kappa^2 z^2$, where $\kappa$ 
is the dimensional parameter. The dimension of the boson 
AdS fields $\Delta$ is identified with twist $\tau$ as 
$\Delta = \tau = N + L$, where $N$ is the number of partons 
and $L$ is the orbital angular momentum. 

Restricting to the axial gauge $\Phi_{\cdots z \cdots}(x,z)=0$ 
and performing the Kaluza-Klein expansion 
\eq\label{KK_coord_PhiJ}
\Phi^{\mu_1 \cdots \mu_J}(x,z)
= \sum\limits_n \ \Phi^{\mu_1 \cdots \mu_J}_n(x) \ \Phi_{n}(z)
\en
one can derive the equation of motion (EOM) for the profile 
function $\phi_{n\tau}(z) =  e^{3A(z)/2} \, \Phi_n(z)$:  
\eq\label{Eq1J}
\Big[ - \frac{d^2}{dz^2} + \frac{4 (\tau-2)^2 - 1}{4z^2} + U_J(z)
\Big] \phi_{n\tau}(z) = M^2_{n\tau\!J} \phi_{n\tau}(z)
\en 
with analytical solutions for the bulk profile 
\eq
\phi_{n\tau}(z) &=& \sqrt{\frac{2}{\Gamma(\tau-1)}} \, \kappa^{\tau-1} \, 
z^{\tau - 3/2} \, e^{-\kappa^2 z^2/2} \, L_n^{\tau-2}(\kappa^2 z^2) 
\en 
and mass spectrum 
\eq\label{mass2_J}
M^2_{n\tau\!J} = 4 \kappa^2 \Big( n + \frac{\tau + J}{2} - 1 \Big) \,.
\en 
Here $L_n^{m}(x)$ is the  
generalized Laguerre polynomial. 

In the case of fermion fields $\Psi_{K_1 \cdots K_{J-1/2}}(x,z)$ 
with spin $J$, the soft-wall AdS/QCD action reads~\cite{Gutsche:2011vb}: 
\eq
\hspace*{-.5cm}
S_{F} &=&  \int d^dx dz \, \sqrt{g} \, e^{-\varphi(z)} \,
g^{K_1N_1} \cdots g^{K_{J-1/2}N_{J-1/2}} \,
\biggl[ \frac{i}{2} \bar\Psi_{K_1 \cdots K_{J-1/2}}(x,z)
\epsilon_a^M \Gamma^a {\cal D}_M \Psi_{N_1 \cdots N_{J-1/2}}(x,z)
\nonumber\\
&-& \frac{i}{2}
({\cal D}_M\Psi_{K_1 \cdots K_{J-1/2}}(x,z))^\dagger
\Gamma^0 \epsilon_a^M \Gamma^a \Psi_{N_1 \cdots N_{J-1/2}}(x,z)
- \bar\Psi_{K_1 \cdots K_{J-1/2}}(x,z) \Big(\mu + V_F(z)\Big)
\Psi_{N_1 \cdots N_{J-1/2}}(x,z) \biggr] \,,
\en
where $V_F(z) = \varphi(z)/R$ is the dilaton potential, 
${\cal D}_M$ is the covariant derivative acting on the spin-tensor
field $\Psi^\pm_{N_1 \cdots N_{J-1/2}}$ as: 
\eq
{\cal D}_M \Psi_{N_1 \cdots N_{J-1/2}} =
\partial_M \Psi_{N_1 \cdots N_{J-1/2}}
- \frac{1}{8} \omega_M^{ab} [\Gamma_a, \Gamma_b]
\Psi_{N_1 \cdots N_{J-1/2}} \,,
\en
where $\omega_M^{ab} = A^\prime(z) \,  
(\delta^a_z \delta^b_M - \delta^b_z \delta^a_M)$ 
is the spin connection term, and 
$\Gamma^a=(\gamma^\mu, - i\gamma^5)$ are the Dirac matrices.

After expanding the fermion field in left- and right-chirality
components $\Psi^{L/R} = (1 \mp \gamma^5)/2 \, \Psi$ and 
performing a KK expansion for the $\Psi^{L/R}(x,z)$ fields 
$\Psi^{L/R}(x,z) = \sum\limits_n
\ \Psi^{L/R}_n(x) \ F_n^{L/R}(z)$,
one can obtain decoupled Schr\"odinger EOMs for the fermion 
bulk profiles $f_n^{L/R}(z) = e^{2 A(z)} \, F_n^{L/R}(z)$: 
\eq
\biggl[ -\partial_z^2
+ \kappa^4 z^2 + 2 \kappa^2 \Big(m \mp \frac{1}{2} \Big)
+ \frac{m (m \pm 1)}{z^2} \biggr] f_{n\tau}^{L/R}(z) 
= M_{n\tau}^2 \, f_{n\tau}^{L/R}(z) \,, 
\en
where $m = \tau - 3/2 = L + 3/2$ and 
\eq\label{fLR_tau}
f_{n\tau}^{L}(z) &=& \sqrt{\frac{2\Gamma(n+1)}{\Gamma(n+\tau)}} \ \kappa^{\tau}
\ z^{\tau-1/2} \ e^{-\kappa^2 z^2/2} \ L_n^{\tau-1}(\kappa^2z^2) \,, 
\nonumber\\
f_{n\tau}^{R}(z) &=& \sqrt{\frac{2\Gamma(n+1)}{\Gamma(n+\tau-1)}} \ \kappa^{\tau-1}
\ z^{\tau-3/2} \ e^{-\kappa^2 z^2/2} \ L_n^{\tau-2}(\kappa^2z^2)
\en
and
\eq
M_{n\tau}^2 = 4 \kappa^2 \Big( n + \tau - 1 \Big) 
            = 4 \kappa^2 \Big( n + L + 2 \Big) \,.  
\en
In order to study electromagnetic properties of hadrons we 
need to obtain the vector bulk-to-boundary propagator $V(q,z)$, dual to 
the $q^2$-dependent electromagnetic current: 
\eq
\partial_z \biggl( \frac{e^{-\varphi(z)}}{z} \,
\partial_z V(-q^2,z)\biggr) + q^2 \frac{e^{-\varphi(z)}}{z} \,
V(-q^2,z) = 0 \,.
\en
The latter equation is solved analytically in terms of 
the gamma $\Gamma(n)$ and Tricomi $U(a,b,z)$ functions, with the result: 
\eq
\label{VInt_q}
V(Q^2,z) = \Gamma(1 + a) \, U(a,0,\kappa^2 z^2) \,, 
\en
where $Q^2 = - q^2$ and $a = Q^2/(4 \kappa^2)$. 
It is convenient to use the integral
representation for $V(Q,z)$~\cite{Grigoryan:2007my}
\eq
\label{VInt}
V(Q^2,z) = \kappa^2 z^2 \int\limits_0^1 \frac{dy}{(1-y)^2}
\, y^{a} \,
e^{- \kappa^2 z^2 \frac{y}{1-y} }\,. 
\en
The expression for the hadron form factors
is given in the soft-wall AdS/QCD by 
\eq
F_{n\tau}(Q^2) = \int\limits_0^\infty dz \, \phi_{n\tau}^2(z) \, V(Q^2,z) \,, 
\en 
where the integrand contains the square of the holographic wave function 
in fifth dimension $z$ (dual to hadron wave function), multiplied with the 
vector bulk-to-boundary propagator $V(Q^2,z)$. 

Next we illustrate the derivation of PDFs in soft-wall AdS/QCD for hadrons 
with arbitrary partonic content (twist)~\cite{Lyubovitskij:2020otz}. 
In the following, for simplicity, we restrict our discussion to a consideration 
of ground states of hadrons with $n=0$. We start 
with the hadronic wave function normalization condition, which
depends on the holographic variable $z$: 
\eq\label{norm_phi}
1 = \int\limits_0^1 dz \, \phi_\tau^2(z) 
\en 
where $\phi_\tau(z)$ is the AdS bulk profile function (for simplicity 
we restrict here to the bosonic case; extension to the fermion 
case is straightforward~\cite{Lyubovitskij:2020otz}). 
In the next step we use the integral representation for unity 
\eq\label{int_unity} 
1 = - e^{\kappa^2 z^2} \, 
\int\limits_0^1 d\biggl[f_\tau(x) \, e^{-\kappa^2 z^2/(1-x)^2} \biggr]
= e^{\kappa^2 z^2} \,   
\int\limits_0^1 dx \, \biggl[ 
\frac{2f_\tau(x) \, \kappa^2 z^2}{(1-x)^3} \, - f'_\tau(x) \biggr] 
\, e^{-\kappa^2 z^2/(1-x)^2} 
\en  
and insert it into Eq.~(\ref{norm_phi}). Here $x$ is the 
light-cone coordinate and $f_\tau(x)$ is the profile function 
with boundary condition $f_\tau(0) = 1$, 
which is specific for a particular hadron and fixed from its PDF. 
The functions $f_\tau(x)$ and $y_\tau(x)$ [see Eq.~(\ref{ytau})] 
are related by: 
\eq 
\Big(1-y_\tau(x)\Big)^{\tau-1} = f_\tau(x) \, (1-x)^{2 (\tau-1)}
\en 
or 
\eq\label{yf_relation} 
y_\tau(x) = 1 - \Big[f_\tau(x)\Big]^{\frac{1}{\tau-1}} \, (1-x)^2 \,. 
\en 
At $x=0$ the functions $y_\tau(x)$ and $f_\tau(x)$  
obey the boundary conditions $y_\tau(0) = 0$ and $f_\tau(0) = 1$. 
At $x=1$ function $f_\tau$ is finite and its value 
depends on the specific choice of twist $\tau$ (see below), 
while $y_\tau(1) = 1$ is independent on twist.

After integration over the variable $z$ one gets 
\eq 
1 = \int\limits_0^1 dx \, (1-x)^{2 \tau - 3} \, 
\biggl[ 2 f_\tau(x) (\tau-1) - f'_\tau(x) (1-x) \biggr] \,. 
\en 
Here and in the following the superscript $(')$ means derivative 
with respect to variable $x$. 
Using the general definition for the hadronic PDF $q_\tau(x)$, 
in the form of an integral representation (zero moment) over $x$  
(here for simplicity we normalize the generic PDF to 1, but 
for specific hadrons the corresponding normalization is 
understood, e.g., 1 for valence PDF in pion, 2 and 1 for valence 
$u$ and $d$ quark PDF in the nucleon respectively, etc.): 
\eq\label{PDF_norm} 
1 = \int\limits_0^1 dx \, q_\tau(x) 
\en 
we get: 
\eq\label{qtau}
q_\tau(x) = (1-x)^{2 \tau - 3} \,
\biggl[ 2 f_\tau(x) (\tau-1) - f'_\tau(x) (1-x) \biggr] 
= \biggl[ - f_\tau(x) (1-x)^{2 \tau - 2} \biggr]' 
\,. 
\en 
We require that the hadronic PDF $q_\tau(x)$ must have the correct 
scaling at large $x$ and this behavior is governed by the profile 
function $f_\tau(x)$. 
In particular, we derived~\cite{Lyubovitskij:2020otz}) 
the following results for the PDFs of pions 
\eq\label{qpi_ADS} 
q_\pi(x) = (1-x)^2 \, \biggl[  \frac{2 f_\pi(x)}{1-x} - f'_\pi(x) \biggr] 
= [-f_\pi(x) (1-x)^2]' 
\en  
and nucleons
\eq\label{udv}
u_v(x) &=& \biggl[ 
- f_u(x) (1-x)^4 \, \Big(1 + 2 \eta_u + (1-x)^2 (1 - 4 \eta_u) 
+ 2 \eta_u (1-x)^4 \Big) \biggr]'  \,, \nonumber\\
d_v(x) &=& \biggl[ 
- f_d(x) (1-x)^4 \, \Big(\frac{1}{2} + 2 \eta_d + (1-x)^2 
\Big(\frac{1}{2} - 4 \eta_d\Big) 
+ 2 \eta_d (1-x)^4 \Big) \biggr]'  \,, \nonumber\\
{\cal E}_v^q(x) &=& k^q \, \Big[ - f_q(x) \, (1-x)^{6} \Big]'  \,. 
\en 
where $f_\pi(x)$, $f_u(x)$, and $f_d(x)$ are the profile 
functions, which define the distribution of quarks in the pion, 
and of $u$ and $d$ quarks in the nucleon, respectively. 
Here $\eta_u = 2 \eta_p + \eta_n$ and $\eta_d = 2 \eta_n + \eta_p$  
are linear combinations of the nucleon couplings with the vector field,
related to nucleon anomalous magnetic moments $k_N$ and fixed 
as~\cite{Abidin:2009hr,Vega:2010ns}: $\eta_N = k_N \kappa/(2 M_N \sqrt{2})$, 
where $M_N$ is the nucleon mass. 

We have shown that the profile functions $f_\pi$, $f_u$, and $f_d$ 
can be fixed using results of extractions of PDFs using world data. 
In particular, using the pQCD prediction for the pion PDF~\cite{Aicher:2010cb}, 
at the initial scale $\mu_0 = 0.63$ GeV  
\eq\label{qpi_QCD} 
q_\pi(x,\mu_0) = N_\pi x^{\alpha-1} \, (1-x)^\beta \, (1+\gamma x^\delta) \,, 
\en 
where $N_\pi$ is the normalization constant, 
$\alpha = 0.70$, $\beta = 2.03$, $\gamma = 13.8$, $\delta = 2$ 
we found~\cite{Lyubovitskij:2020otz} 
\eq\label{fpi_PDF} 
f_\pi(x) (1-x)^2 = 1 - N_\pi \, x^\alpha 
\biggl[ \frac{1}{\alpha} 
- \frac{2x}{\alpha+1} + \frac{x^2}{\alpha+2}  
+ \gamma x^{\delta} \, 
\biggl( \frac{1}{\alpha+\delta} 
- \frac{2x}{\alpha+\delta+1} + \frac{x^2}{\alpha+\delta+2} \biggr) 
\biggr] \,. 
\en 
In the case of nucleons we used 
the Martin-Stirling-Thornem-Watt (MSTW) 2008 LO global analysis of PDFs~\cite{Martin:2009iq}: 
\eq
u_v(x,\mu_0) &=& A_u \, x^{\alpha_u-1} \, (1-x)^{\beta_u} \, 
(1 + \epsilon_u \sqrt{x} + \gamma_u x)\,, 
\nonumber\\
d_v(x,\mu_0) &=& A_d \, x^{\alpha_d-1} \, (1-x)^{\beta_d} \, 
(1 + \epsilon_d \sqrt{x} + \gamma_d x)\,, 
\en
where $\mu_0 = 1$ GeV is the initial scale. 
The normalization constants $A_q$ 
and the constants $\alpha_q$, $\beta_q$, $\epsilon_q$, $\gamma_q$ 
were fixed as 
\eq 
& &A_u = 1.4335\,, \quad A_d = 5.0903\,, 
\nonumber\\
& &\alpha_u = 0.45232\,, \quad \alpha_d = 0.71978\,, 
\nonumber\\
& &\beta_u = 3.0409 \simeq 3\,, \quad \beta_d = 5.1244 \simeq 5\,, 
\\
& &\epsilon_u = -2.3737\,, \quad \epsilon_d = -4.3654\,, 
\nonumber\\
& &\gamma_u = 8.9924\,, \quad \gamma_d = 7.4730 \,.
\nonumber
\en 
Using the MSTW results~\cite{Martin:2009iq} we predict the $u$ and 
$d$ profile functions in the nucleon as~\cite{Lyubovitskij:2020otz}  
\eq 
f_u(x) \, (1-x)^4 &=& 1 - A_u x^{\alpha_u} \, 
\Big[                 B_u(x,0) 
+ \epsilon_u \sqrt{x} B_u(x,1/2)
+ \gamma_u x B_u(x,1) \Big] \,, \nonumber\\
f_d(x) \, (1-x)^6 &=& 1 - A_d x^{\alpha_d} \, 
\Big[                 B_d(x,0) 
+ \epsilon_d \sqrt{x} D_d(x,1/2)
+ \gamma_d x B_d(x,1) \Big] \,, 
\en 
where 
\eq
B_u(x,n) =  \sum\limits_{k=0}^3 \, \frac{C_3^k \, (-x)^k}{\alpha_u+n+k}\,, 
\quad 
B_d(x,n) =  \sum\limits_{k=0}^5 \, \frac{C_5^k \, (-x)^k}{\alpha_d+n+k} \,. 
\en
Here $C_m^k = \frac{m!}{k! (m-k)!}$ are the binomial coefficients. 

TMDs were derived using a normalization condition involving integration over light-cone $x$ 
and transverse momentum $\bfk$ coordinates: 
\eq\label{int_unity2} 
1 &=& - e^{\kappa^2 z^2} \, 
\int\limits_0^1 d\biggl[f_\tau(x) 
e^{-\kappa^2 z^2/(1-x)^2} \biggr] \, 
\int d^2\bfk \, \frac{D_\tau(x)}{\pi \kappa^2} \, 
e^{-\bfk^2 D_\tau(x)/\kappa^2} \nonumber\\ 
&=& \frac{e^{\kappa^2 z^2}}{\pi \kappa^2} \,   
\int\limits_0^1 dx \, \int d^2\bfk \, \biggl[ 
\frac{2f_\tau(x) \, \kappa^2 z^2}{(1-x)^3} \, - f'_\tau(x) \biggr] 
\, D_\tau(x) \, 
e^{-\kappa^2 z^2/(1-x)^2} \, 
e^{-\bfk^2 D_\tau(x)/\kappa^2}  \,, 
\en  
where $D_\tau(x)$ is the longitudinal factor, which is related to 
the profile function $f_\tau(x)$ (or function $y_\tau(x)$ as 
\eq
D_\tau(x) = \frac{\log[1/y_\tau(x)]}{(1-x)^2}  
= - \frac{\log\biggl[1 - \Big[f_\tau(x)\Big]^{\frac{1}{\tau-1}} 
\, (1-x)^2 \biggr]}{(1-x)^2} 
\,. 
\en
For large $x$ the function $D_\tau(x)$ behaves as
\eq
D_\tau(x) = \Big[f_\tau(x)\Big]^{\frac{1}{\tau-1}} \,,
\en
Notice that the explicit form of the function $D_\tau(x)$ was obtained 
via matching the expression for the hadronic form factors in two 
approaches --- soft-wall AdS/QCD and LF QCD. 

As illustration we present the result for the unpolarized quark TMD in the nucleon  
(more details on our results on quark TMDs see in~\cite{Lyubovitskij:2020otz}):
\eq 
f_1^{q_v}(x,\bfk^2) &=& 
\biggl[q_v^+(x) + q_v^-(x) \, 
\frac{\bfk^2 \, D_q(x)}{\kappa^2}\biggr] \, 
\frac{D_q(x)}{2 \pi \kappa^2} \, 
e^{-\bfk^2 D_q(x)/\kappa^2}   
\en 
where $q_v^\pm(x) = q_v(x) \pm \delta q_v(x)$, 
$q_v(x)$ and $\delta q_v(x)$ are the helicity-independent 
and helicity-dependent valence quark parton distributions. 
One can see that in our approach TMDs and PDFs are related. 

Finally, we present results for hadron GPDs ${\cal H}_\tau(y_\tau(x),Q^2)$ 
and form factor $F_\tau(Q^2)$ with arbitrary twist $\tau$~\cite{Gutsche:2013zia}: 
\eq\label{GPD1} 
{\cal H}_\tau(y_\tau(x),Q^2) 
= (\tau - 1) \, 
(1-y_\tau(x))^{\tau-2} \, \Big[y_\tau(x)\Big]^a\,, \quad a = \frac{Q^2}{4 \kappa^2} \,. 
\en 
and 
\eq\label{FF1}  
F_\tau(Q^2) = \int\limits_0^1 dx {\cal H}_\tau(y_\tau(x),Q^2) \,. 
\en 
The GPD can be written in more convenient form  in terms of the PDF: 
\eq\label{GPD2}  
{\cal H}_\tau(x,Q^2) = q_\tau(x) \, \Big[y_\tau(x)\Big]^a = 
q_\tau(x) \, \exp\Big(- a \log\Big[1/y_\tau(x)\Big]\Big) \,. 
\en  
One can see that all parton distributions (PDFs, TMDs, GPDs) and form 
factors are related to each other. 
 
\section{Gluon parton distribution in soft-wall AdS/QCD} 

In this section we extend our formalism for parton distributions from quarks to 
gluons. Such extension is straightforward. We just take into account that 
in case of gluons we should use specific values of twist and the available normalization 
conditions for gluon parton densities. 

\subsection{Gluon PDFs} 

We start with gluon PDFs. We will base our discussion on the QCD predictions for 
the gluon PDFs derived in Refs.~\cite{Brodsky:1989db,Brodsky:1994kg}. 
The results of Refs.~\cite{Brodsky:1989db} and~\cite{Brodsky:1994kg} 
we will call, respectively, as QCDI and QCDII. As stressed in 
Refs.~\cite{Brodsky:1989db,Brodsky:1994kg}, the gluon PDFs in nonexotic  
hadrons (e.g., pion, nucleon, etc.) must fall off at large $x$ 
by at least one power faster than the respective quark PDFs. 
It means that the helicity-nonflip gluon PDFs in pion and nucleon should 
fall off at large $x$ as $(1-x)^3$ and $(1-x)^4$, respectively.  
Hence, the EOM for the bulk wave function of 
the gluon content in a hadron with twist $\tau$ is deduced from 
the general formula~(\ref{qtau}) for the profile of parton density 
with arbitrary twist $f_\tau(x)$ by shifting the twist as $\tau \to \tau + 1/2$: 
\eq\label{Eq_gluon}
\Big[ - \frac{d^2}{dz^2} + \frac{4 (\tau-3/2)^2 - 1}{4z^2} + U_J(z)
\Big] \phi_{n\tau}(z) = M^2_{n\tau\!J} \phi_{n\tau}(z)
\en 
This EOM gives the following analytical solution for the gluon bulk profile 
\eq\label{gluon_WF}
\phi_{n\tau}(z) &=& \sqrt{\frac{2}{\Gamma(\tau-1/2)}} \, \kappa^{\tau-1/2} \, 
z^{\tau - 1} \, e^{-\kappa^2 z^2/2} \, L_n^{\tau-3/2}(\kappa^2 z^2) 
\en 

Using Eq.~(\ref{gluon_WF}) we derive the differential 
equations for the gluon profile functions $f_\tau^G(x)$ and $f_\tau^{\Delta G}(x)$, 
characterizing the unpolarized and polarized gluon PDFs 
$G(x)$ and $\Delta G(x)$, respectively. Using the
corresponding normalization unpolarized and polarized gluon 
PDFs, and QCD predictions for the moments of gluon PDFs~(\ref{gluon_momentsI}) 
and~(\ref{gluon_momentsII}) we derive the following differential equations 
for $f_\tau^G(x)$ and $f_\tau^{\Delta G}(x)$: 

QCDI~\cite{Brodsky:1989db} 
\eq 
& &\biggl[ - f_\tau^G(x) (1-x)^{2\tau-1} \biggr]' = x G(x) 
= N_1 (1-x)^4 (2 + 2x + x^2) \,, 
\nonumber\\
& &\biggl[ - f_\tau^{\Delta G}(x) (1-x)^{2\tau-1} \biggr]' = \Delta G(x) 
= N_1 (1-x)^4 (6 - x) \,,
\en  
where $N_1 = \frac{21}{10} \, \la x_g \ra = \frac{6}{7} \, \Delta G$.  
 
QCDII~\cite{Brodsky:1994kg}
\eq 
& &\biggl[ - f_\tau^G(x) (1-x)^{2\tau-1} \biggr]' = x G(x) 
= N_2 (1-x)^4 (1 + (1 + x)^2) \,, 
\nonumber\\
& &\biggl[ - f_\tau^{\Delta G}(x) (1-x)^{2\tau-1} \biggr]' = \Delta G(x) 
= N_2 (1-x)^4 (2 - x) \,.
\en   
where $N_2 = \frac{35}{12} \, \la x_g \ra = \frac{30}{11} \, \Delta G$.  

Taking $\tau = 3$ as leading twist value of gluon PDF and integrating over $x$ 
with boundary conditions $f_G(0) = \la x_g \ra$ 
and $f_{\Delta G}(0) = \Delta G$ 
one gets:

QCDI~\cite{Brodsky:1989db} 
\eq\label{fGI} 
f_G(x)          &=& \la x_g \ra \, \biggl[ 1 + \frac{4}{5} x + \frac{3}{10} x^2 \biggr] 
\,, \nonumber\\
f_{\Delta G}(x) &=& \Delta G    \, \biggl[ 1 - \frac{x}{7} \biggr] 
\,. 
\en 
 
QCDII~\cite{Brodsky:1994kg}
\eq\label{fGII}  
f_G(x)          &=& \la x_g \ra \, \biggl[ 1 - \frac{5}{6} x + \frac{5}{12} x^2 \biggr] 
\,, \nonumber\\
f_{\Delta G}(x) &=& \Delta G    \, \biggl[ 1 - \frac{5}{11} x \biggr] 
\,. 
\en  
Note that at large $x$ the profile functions approach a constant and 
degenerate for each QCD-based framework version: 

QCDI~\cite{Brodsky:1989db} 
\eq\label{large_QCDI} 
f_g = f_G(1) = f_{\Delta G}(1) = \frac{21}{10} \, \la x_g \ra = N_1 \,.
\en 

QCDII~\cite{Brodsky:1994kg} 
\eq\label{large_QCDII}  
f_g = f_G(1) = f_{\Delta G}(1) = \frac{7}{12} \, \la x_g \ra = \frac{N_2}{5} \,. 
\en 

We derive also the profile function defining the gluon distribution in pions.
Choosing $\tau = 2$ we get the differential equation for the gluon PDF
in a pion:
\eq\label{diff_pion}
\biggl[ - f_\pi^G(x) (1-x)^3 \biggr]' = x G_\pi(x) = N_{G_\pi} \, (1-x)^3
\en
with boundary condition
\eq
f_\pi^G(0) = \int\limits_0^1 dx \biggl[ - f_\pi^G(x) (1-x)^3 \biggr]'
           = \int\limits_0^1 dx \, x \, G_\pi(x) = \la x_g^\pi \ra \,.
\en
As in the nucleon case, the value of the profile function $f_\pi^G(x)$ at $x=0$ 
is related to the first moment of the gluon PDF $G_\pi(x)$. On the other hand, $\la x_g^\pi \ra$
can be uniquely fixed from the energy-momentum sum rule~\cite{Gluck:1991ey,Novikov:2020snp}:
\eq\label{Energy_SR}
\int\limits_0^1 dx \, x \, \Big[ 2 q_\pi(x) + S_\pi(x) + G_\pi(x) \Big] 
= 2 \la x^\pi \ra + \la x^\pi_s \ra + \la x^\pi_g \ra = 1 
\,,
\en
where $\la x^\pi \ra$, $\la x^\pi_s \ra$, and $\la x^\pi_g \ra$ are the 
first moments of pion PDFs (valence, total sea, and gluon contributions).  
Using the xFitter Developers Team~\cite{Novikov:2020snp} parametrization 
for the gluon PDF in the pion 
\eq 
x G_\pi(x) = \la x^\pi_g \ra \, (1 + C_{g}) \, (1-x)^{C_{g}}  
\en  
where $C_g = 3 \pm 1$ at the initial scale squared $\mu_0^2 = 1.9$ GeV$^2$ 
is the parameter determining the large $x$ behavior of the PDF $G_\pi(x)$. 
Solving the differential equation~(\ref{diff_pion}) one gets: 
\eq\label{fpiG} 
f_\pi^G(x) = \la x^\pi_g \ra \, (1-x)^{C_{g}-2}
\en  
Note that the solution~(\ref{fpiG}) obeys the boundary condition $f_\pi^G(0) = \la x^\pi_g \ra$.  

\subsection{Gluon TMDs}
 
Now we are in a position to derive the soft-wall AdS/QCD prediction 
for the T-even gluon TMDs in a hadron. In the following 
we use light-cone kinematics, which is specified by two 
light-light vectors $n_\pm$ 
as~\cite{Mulders:2000sh}:
\eq 
n_+^\mu = (1,0,\bfn)\,, \quad 
n_-^\mu = (0,1,\bfn) 
\en 
obeying the conditions $n_+ \, n_- = 1$ and $n_\pm^2 = 0$. 
Any four momentum $p$ is expanded through $n_\pm$ as 
\eq 
p^\mu = p^+ n_+^\mu + p^- n_-^\mu + p^\mu_\perp = (p^+,p^-,\bfp) \,. 
\en 
We work in the hadron rest frame, where the nucleon $P$ 
and parton $k$ momenta are specified as  
\eq 
P^\mu &=& \Big(P^+, P^-, \bfn \Big) \,, \quad P^- = \frac{M^2}{2P^+} \,, 
\nonumber\\
k^\mu &=& \Big(xP^+, \frac{k^2 + \bfk^2}{2xP^+}, \bfk\Big) \,, 
\en 
where $M$ is a hadron mass, $x = k^+/P^+$ is the longitudinal 
momentum fraction carried by the gluon.  
The spin vector of the hadron is expanded into one-dimensional 
longitudinal $S_L$ (helicity) and two-dimensional transverse ${\bf S}_T$ components 
in a manifestly covariant way as
\eq 
S^\mu = S_L^\mu + S_T^\mu \,, 
\en 
where 
\eq 
S_L^\mu &=& S_L \, \frac{P n_-}{M} \, n_+^\mu 
        - S_L \, \frac{P n_+}{M} \, n_-^\mu 
        = S_L \, \biggl(\frac{P^+}{M}, - \frac{P^-}{M}, {\bf 0}\biggr) \,, 
\nonumber\\
S_T^\mu &=& \Big(0, 0, {\bf S}_T\Big)      
\en 
with $S_L^2 + {\bf S}_T^2  = 1$. 
Note that in the infinite momentum frame $P^+ \to \infty$ the $P^-$ component and 
components of all four-vectors proportional to $P^-$ vanish. 

Next we specify the two symmetric $g^{\mu\nu}_T$ and $\eta^{\mu\nu}_T$ 
and antisymmetric $\epsilon^{\mu\nu}_T$ transverse tensors, 
using the $n_\pm^\mu$ vectors~\cite{Mulders:2000sh}: 
\eq 
g^{\mu\nu}_T &=& g^{\mu\nu} - n^\mu_+ n^\nu_- - n^\mu_- n^\nu_+ = {\rm diag}(0,0,-1,-1) \,, 
\nonumber\\[3mm]
\eta^{\mu\nu}_T &=& g^{\mu\nu}_T + \frac{2 \bfk^\mu \bfk^\nu}{\bfk^2} 
= \left( 
\begin{array}{cc}
\mbox{\large 0} & \mbox{\large 0} \\[1mm]
\mbox{\large 0} & \begin{array}{cr}
    \cos 2\phi_k &   \sin 2\phi_k \\
    \sin 2\phi_k & - \cos 2\phi_k \\
    \end{array}
\end{array}
\right)
\,, 
\nonumber\\[3mm]
\epsilon^{\mu\nu}_T &=& \epsilon^{\alpha\beta\mu\nu} \, n_{+ \alpha} \, n_{- \beta} 
= \epsilon^{n_+n_-\mu\nu} = \epsilon^{- + \mu\nu} 
= \left( 
\begin{array}{cc}
\mbox{\large 0} & \ \ \ \mbox{\large 0} \\[1mm]
\mbox{\large 0} & \begin{array}{rc}
    0 & 1 \\
   -1 & 0 \\
    \end{array}
\end{array}
\right)
\,, 
\en 
where $\bfk^\mu = (0,0,\bfk) = \sqrt{\bfk^2} \, (0,0,\cos\phi_k,\sin\phi_k)$ 
and $\phi_k$ is the azimuthal angle, defining an orientation of $\bfk$ in the 
transverse plane.   
These three tensors play a fundamental role 
in the classification of TMDs~\cite{Mulders:2000sh} and obey the 
following normalization and orthogonality conditions: 
\eq 
& &
     g^{\mu\nu}_T        \  g_{\mu\nu, T}         = 
     \eta^{\mu\nu}_T     \  \eta_{\mu\nu, T}      = 
     \epsilon^{\mu\nu}_T \  \epsilon_{\mu\nu, T}  = 2\,,  \nonumber\\
& & 
g^{\mu\nu}_T \, \eta_{\mu\nu, T}        = 
g^{\mu\nu}_T \, \epsilon_{\mu\nu, T}    = 
\eta^{\mu\nu}_T \, \epsilon_{\mu\nu, T} = 0 \,. 
\en 
Gluon polarization vectors $\epsilon^\mu_{\lambda}$ read 
\eq 
\epsilon^\mu_{\pm} = (0,0,{\bm\epsilon}_\pm) = 
\frac{1}{\sqrt{2}} \, (0,0,\mp 1,-i) 
\quad  
\en  
in the case of circular polarization and 
\eq 
\epsilon^\mu_{x} 
= (0,0,{\bm\epsilon}_x) = (0,0,1,0)\,, \quad 
\epsilon^\mu_{y} = (0,0,{\bm\epsilon}_y) = (0,0,0,1)\,.  
\en  
in the case of linear polarization. 
The two sets of polarization vectors are related as: 
\eq 
\epsilon^\mu_\pm = \mp \frac{1}{\sqrt{2}} \, 
\biggl[ \epsilon^\mu_x \pm i \epsilon^\mu_y 
\biggr] 
\,. 
\en 
In both cases the polarization vectors obey the completeness and orthonormality
conditions:
\eq
- g^{\mu\nu}_T = \sum\limits_{\lambda}
\, \epsilon^\mu_\lambda  \, \epsilon^{\dagger\nu}_\lambda \,, \qquad
\epsilon^{\dagger\mu}_{\lambda} \, \epsilon_{\mu\lambda'} = - \delta_{\lambda\lambda'} \,.
\en
The antisymmetric tensor $\epsilon^{\mu\nu}_T$ is expressed in terms of
polarization vectors for the case of linear and circular polarizations as
\eq
- i \epsilon^{\mu\nu}_T = - i \, \Big[
\epsilon^{\mu}_x \, \epsilon^{\dagger\nu}_y -
\epsilon^{\mu}_y \, \epsilon^{\dagger\nu}_x \Big]
= \epsilon^{\mu}_+ \, \epsilon^{\dagger\nu}_+
- \epsilon^{\mu}_- \, \epsilon^{\dagger\nu}_- \,.
\en
The unpolarized gluon TMD in the nucleon is defined in analogy with the corresponding quark TMD: 
\eq\label{f1_TMD} 
f_1^{g}(x,\bfk^2) &=& 
\biggl[ G(x) + G^-(x) \, \alpha_+(x) 
\biggl(\frac{\bfk^2 \, D_g^2(x)}{\kappa^2} - 1 \biggr) 
\biggr] \, \frac{D_g(x)}{\pi \kappa^2} \, e^{-\bfk^2 D_g(x)/\kappa^2}   
\,, 
\en 
where $G^\pm = (G \pm \Delta G)/2$ are the helicity-aligned 
and helicity-antialigned gluon PDFs, $D_{g}(x) > 0$ is the profile function, and  
\eq 
\alpha_\pm(x) = \frac{1 \pm  (1-x)^2}{(1-x)^2} \,. 
\en 
It is clear that the integration 
of the $f_1^{g}(x,\bfk)$ over $\bfk$ gives the unpolarized gluon PDF: 
\eq 
f_1^g(x) = \int d^2\bfk \, f_1^{g}(x,\bfk^2) = G(x) \,. 
\en 

Notice that that Gaussian form of the $\bfk^2$ dependence of TMD in our approach 
[see, e.g., Eq.~(\ref{f1_TMD}) 
is dictated by the Gaussian dependence of the light-front wave functions (LFWFs),
defined below in Eq.~(\ref{LFWFs_symmetric}). The reason is that the
Gaussian ansatz for the LFWFs is dictated by the Gaussian $Q^2$-dependence of 
the form factors [see Eq.(\ref{FF_ADSQCD})], 
calculated using holographic wave functions of hadrons. Since
all parton densities and form factors in our approach are related, we can
guarantee that LFWFs produce the same functional behavior of form factors.

Following Refs.~\cite{Brodsky:2000ii} and~\cite{Lu:2016vqu,Kaur:2020pvc}
one can set up the light-front wave function (LFWF) for the nucleon as a composite 
system of a spin-1 gluon and spin-$\frac{1}{2}$ three-quark spectator. This is the
so-called spectator model, which was applied also 
in Refs.~\cite{Meissner:2007rx,Bacchetta:2008af,Bacchetta:2020vty} 
for the derivation of quark and gluon TMDs. 
In our case we derive LFWF model without refering to additional 
parameters such as spectator mass (a similar model was considered by us 
for the nucleon as bound state of an active quark and a scalar diquark 
spectator~\cite{Gutsche:2013zia,Gutsche:2014yea,Gutsche:2016gcd,%
Lyubovitskij:2020otz}. We introduce the notation of LFVW 
$\psi^{\lambda_N}_{\lambda_g\lambda_X}(x,\bfk)$, describing the
bound state of a gluon $(g)$ and a three-quark spectator $(X)$, 
where $\lambda_N$, $\lambda_g$, and $\lambda_X$ are the helicities 
of nucleon, gluon, and three-quark spectator, respectively. 
For a nucleon with up helicity $\lambda_N = + \frac{1}{2}$ (we use shorten notation $+$), 
four possible LFWFs are available, taking into account the different helicities 
$\lambda_g = \pm 1$ 
and $\lambda_X = \pm \frac{1}{2}$ 
\eq 
\psi^+_{+1 +\frac{1}{2}}(x,\bfk) &=& 
\frac{k^1-ik^2}{M_N} \, \varphi^{(2)}(x,\bfk^2) 
\,, 
\nonumber\\
\psi^+_{+1 -\frac{1}{2}}(x,\bfk) &=&   \varphi^{(1)}(x,\bfk^2) \,, 
\nonumber\\
\psi^+_{-1 +\frac{1}{2}}(x,\bfk) &=& - \frac{k^1+ik^2}{M_N} \, (1-x) 
\, \varphi^{(2)}(x,\bfk^2) 
\,, 
\nonumber\\
\psi^+_{-1 -\frac{1}{2}}(x,\bfk) &=& 0 \,. 
\en 
The LFWFs with down helicity $\lambda_N = - \frac{1}{2}$ (we use shorten notation $-$) are 
\eq\label{LFWF_in} 
\psi^-_{+1 +\frac{1}{2}}(x,\bfk) &=& 0 \,, \nonumber\\
\psi^-_{+1 -\frac{1}{2}}(x,\bfk) &=& 
- \Big[\psi^+_{-1 +\frac{1}{2}}(x,\bfk)\Big]^* =
  \frac{k^1-ik^2}{M_N} \, (1-x) \, \varphi^{(2)}(x,\bfk) 
\,, 
\nonumber\\
\psi^-_{-1 +\frac{1}{2}}(x,\bfk) &=& 
+ \Big[\psi^+_{+1 -\frac{1}{2}}(x,\bfk)\Big]^* = 
\varphi^{(1)}(x,\bfk^2) \,, 
\nonumber\\
\psi^-_{-1 -\frac{1}{2}}(x,\bfk) &=& 
- \Big[\psi^+_{+1 +\frac{1}{2}}(x,\bfk)\Big]^* = 
- \frac{k^1+ik^2}{M_N} \, \varphi^{(2)}(x,\bfk^2) 
\,. 
\en 
where the functions $\varphi^{(1)}(x,\bfk)$ and 
$\varphi^{(2)}(x,\bfk)$ can be expressed through the 
gluon PDF functions $G^\pm(x)$ as 
\eq\label{LFWFs_symmetric}
\varphi^{(1)}(x,\bfk^2) &=& 
\frac{4 \pi}{\kappa} \, \sqrt{G^+(x)} \, \beta(x) \, 
\sqrt{D_g(x)}  \, 
\exp\biggl[- \frac{\bfk^2}{2 \kappa^2} \, D_g(x) \biggr] \,,
\nonumber\\
\frac{1}{M_N} \,
\varphi^{(2)}(x,\bfk^2) &=& \frac{4 \pi}{\kappa^2} \, 
\sqrt{G^-(x)} \, \frac{D_g(x)}{1-x} \,
\exp\biggl[- \frac{\bfk^2}{2 \kappa^2} \, D_g(x) \biggr] \,,
\en 
where 
\eq\label{beta_def} 
\beta(x) = \sqrt{1 - \frac{G^-(x)}{G^+(x) (1-x)^2}} \,  
\,.
\en 
One can see that the $\varphi^{(2)}(x,\bfk^2)$ function is expressed 
in terms of the $G^-(x)$ PDF, while the 
$\varphi^{(1)}(x,\bfk^2)$ is expressed through the combination of 
functions $G^+(x)$ and $G^-(x)$. Using Eq.~(\ref{LFWFs_symmetric}) 
we get useful formula:  
\eq 
                       \Big[\varphi^{(1)}(x,\bfk^2)\Big]^2 
+ \frac{\bfk^2}{M_N^2} \, \Big[\varphi^{(2)}(x,\bfk^2)\Big]^2 
= \frac{16 \pi^2}{\kappa^2} \, D_g(x) \, 
\biggl[G^+(x) + \frac{G^-(x)}{(1-x)^2} 
\, \biggl( \frac{\bfk^2}{\kappa^2} D_g(x) - 1 \biggr) \biggr] 
\exp\biggl[- \frac{\bfk^2}{\kappa^2} \, D_g(x) \biggr] 
\,. 
\en  
Note that in Eq.~(\ref{LFWFs_symmetric}) the LFWFs $\varphi^{(i)}(x,\bfk^i)$ 
are defined in terms of a single profile function $D_g(x)$.    
A generalization to the case when the LFWFs $\varphi^{(i)}(x,\bfk^i)$ 
are accompanied by two different functions $D_{g_1}(x)$ 
and $D_{g_2}(x)$, respectively, is straightforward. 
{\it Generalized version} is considered in 
Appendix~{\ref{app_TMD}}, where we present the corresponding 
list of the LFWFs and TMDs. 

Note that from the positivity of the argument of the square root in the 
definition of the $\varphi^{(1)}(x,\bfk^2)$ function, it follows 
that the gluon PDF $G^+(x)$ and $G^-(x)$ should 
satisfy the bound 
\eq\label{positivity_bound} 
\frac{G^+(x) (1-x)^2}{G^-(x)} \ge 1
\en 
i.e. $\beta(x)$ defined in Eq.~(\ref{beta_def}) must be real. 
The positivity bound~(\ref{positivity_bound}) is fulfilled 
in both versions of the QCD derivation of gluon PDFs~\cite{Brodsky:1989db,Brodsky:1994kg}. 
In particular, QCDI~\cite{Brodsky:1989db,Brodsky:1994kg} and 
QCDII~\cite{Brodsky:1994kg} give, respectively  
\eq 
\frac{G^+(x) (1-x)^2}{G^-(x)} = 1 + 4x \ge 1
\en 
and 
\eq 
\frac{G^+(x) (1-x)^2}{G^-(x)} =  1 \,. 
\en 

Note that our wave functions $\varphi^{(1)}(x,\bfk^2)$ and  
$\varphi^{(2)}(x,\bfk^2)$ are generalizations of the 
LFWF $\varphi(x,\bfk^2)$ 
used in Refs.~\cite{Brodsky:2000ii} and~\cite{Lu:2016vqu,Kaur:2020pvc}: 
\eq 
\varphi^{(1)}(x,\bfk^2) \longrightarrow 
\varphi(x,\bfk^2) \, \sqrt{2} \, \frac{M_N (1-x) - M_X}{1-x} \,, 
\qquad 
\varphi^{(2)}(x,\bfk^2) \longrightarrow 
\varphi(x,\bfk^2) \, \sqrt{2} \, \frac{M_N}{x (1-x)} 
\,, 
\en 
where $M_X$ is the mass of the three-quark spectator.  
Also, the functions  $\varphi^{(1)}(x,\bfk^2)$ and
$\varphi^{(2)}(x,\bfk^2)$ can be considered as extensions 
of the quark-gluon coupling in the quark target 
model~\cite{Meissner:2007rx} and 
the form factor $g_1(k^2)$, 
parametrizing the minimal nucleon-gluon coupling in 
the spectator model proposed in Ref.~\cite{Bacchetta:2020vty}   
\eq
& &\varphi^{(1)}(x,\bfk^2) \longrightarrow  
2 g \sqrt{\frac{2}{3x}} \, \frac{m x^2}{m^2 x^2 + \bfk^2} 
\,, \nonumber\\ 
& &\varphi^{(2)}(x,\bfk^2) \longrightarrow  
2 g \sqrt{\frac{2}{3x}} \, \frac{M_N}{m^2 x^2 + \bfk^2} 
\,. 
\en 
and 
\eq 
& &\varphi^{(1)}(x,\bfk^2) \longrightarrow  
\frac{g_1(k^2)}{k^2} \, \sqrt{2x} \, \frac{M_N (1-x) - M_X}{1-x} 
\,, \nonumber\\ 
& &\varphi^{(2)}(x,\bfk^2) \longrightarrow  
\frac{g_1(k^2)}{k^2} \, \sqrt{2x}  \, \frac{M_N}{x (1-x)} 
\,. 
\en  
It is clear that LFWF $\varphi(x,\bfk^2)$ in Refs.~\cite{Brodsky:2000ii,Lu:2016vqu,Kaur:2020pvc}  
is related with form factor $g_1(k^2)$ from Ref.~\cite{Bacchetta:2020vty} as 
\eq 
\varphi(x,\bfk^2) = \frac{g_1(k^2)}{k^2} \, \sqrt{x} \,. 
\en  
In terms of LFWFs~(\ref{LFWF_in}) the unpolarized
gluon TMD in the nucleon reads (see also Refs.~\cite{Lu:2016vqu,Kaur:2020pvc}) 
\eq\label{f1g_def} 
f_1^g(x,\bfk^2) &=& 
- \frac{1}{16 \pi^3} \, \frac{1}{2} \, g^{\mu\nu}_T \, 
\sum\limits_{\lambda_N  \lambda_g \lambda_g' \lambda_X} 
\, \epsilon^{\dagger\lambda_g'}_\mu \, \epsilon^{\lambda_g}_\nu \, 
\, \psi^{*\lambda_N}_{\lambda_g'\lambda_X}(x,\bfk) 
\, \psi^{\lambda_N}_{\lambda_g\lambda_X}(x,\bfk) 
\nonumber\\
&=&\frac{1}{16 \pi^3} \, \frac{1}{2} \, 
\sum\limits_{\lambda_N  \lambda_g \lambda_X} \, 
|\psi^{\lambda_N}_{\lambda_g\lambda_X}(x,\bfk)|^2 
\nonumber\\
&=& 
\frac{1}{16 \pi^3} \, \biggl[ 
  |\psi^{+}_{+1 +\frac{1}{2}}(x,\bfk)|^2
+ |\psi^{+}_{+1 -\frac{1}{2}}(x,\bfk)|^2
+ |\psi^{+}_{-1 +\frac{1}{2}}(x,\bfk)|^2
\biggr] 
\nonumber\\
&=& \frac{1}{16 \pi^3} \, \biggl[
\Big(\varphi^{(1)}(x,\bfk^2)\Big)^2
+ \frac{\bfk^2}{M_N^2} \, \Big[1+(1-x)^2\Big] 
\, \Big(\varphi^{(2)}(x,\bfk^2)\Big)^2 \biggr]
\nonumber\\
&=& \frac{D_g(x)}{\pi \kappa^2} \, 
\biggl[
G(x)  + G^-(x) \, \alpha_+(x) \, \biggl( 
\frac{\bfk^2}{\kappa^2} \, D_g(x)  - 1 \biggr) \biggr] \, 
\exp\biggl[- \frac{\bfk^2}{\kappa^2} \, D_g(x) \biggr] 
\,.
\en 
The LFWF describing the nucleon, with both transverse 
and longitudinal polarizations, is constructed as a superposition 
of two LFWFs with definite nucleon helicity $\lambda_N = + 1/2$ 
and $\lambda_N = - 1/2$~\cite{Diehl:2005jf}: 
\eq\label{generic_TL} 
\psi^{{\bf S},S^z}_{\lambda_g\lambda_X}(x,\bfk) &=& 
\cos\frac{\theta}{2} \, \psi^{+}_{\lambda_g\lambda_X}(x,\bfk) + 
\sin\frac{\theta}{2} \, e^{i\phi} \, \psi^{-}_{\lambda_g\lambda_X}(x,\bfk) 
\,. 
\en 
Here the three-dimensional nucleon spin 
is expressed in terms of two spherical angles $\phi$ and $\theta$ as 
\eq 
\vec{S} = ({\bf S},S^z)\,, 
\quad {\bf S}=(\sin\theta \cos\phi, \sin\theta \sin\phi)\,, 
\quad S^z = \cos\theta \,. 
\en 
The limiting cases are: 

(1) Nucleon has only transverse polarization 
with ${\bf S}_T = (\cos\phi, \sin\phi)$ and $S_L = 0$,  
i.e. $\psi^{{\bf S},S^z}_{\lambda_g\lambda_X} \to 
\psi^{{\bf S}_T}_{\lambda_g\lambda_X}$ at $\theta=90^{\circ}$; 

(2) Nucleon has only longitudinal polarization 
with ${\bf S}_T = 0$ and $S_L = \pm 1$, 
i.e. $\psi^{{\bf S},S^z}_{\lambda_g\lambda_X} \to 
\psi^{S_L}_{\lambda_g\lambda_X} \equiv 
\psi^{\lambda_N}_{\lambda_g\lambda_X}$ 
at $\theta=0^{\circ}$ or $180^{\circ}$.  

The helicity TMD $g_{1L}^g(x,\bfk^2)$, describing a distribution of a
circularly polarized gluon in a longitudinally polarized nucleon, 
is defined by analogy with the quark case~\cite{Bacchetta:2008af} as 
\eq\label{g1L_def}  
g_{1L}^g(x,\bfk^2) &=& 
- \frac{1}{16 \pi^3} \, \frac{1}{S_L}
i \epsilon^{\mu\nu}_T \, 
\sum\limits_{\lambda_g \lambda_g' \lambda_X}
\, \epsilon^{\dagger\lambda_g'}_\mu \, \epsilon^{\lambda_g}_{\nu} \, 
\psi^{*S_L}_{\lambda_g'\lambda_X}(x,\bfk) \, 
\psi^{S_L}_{\lambda_g\lambda_X}(x,\bfk) 
\nonumber\\
&=& 
\frac{1}{16 \pi^3} \, \frac{1}{S_L}
\sum\limits_{\lambda_X}
\biggl[ 
  |\psi^{S_L}_{+1 \lambda_X}(x,\bfk)|^2 
- |\psi^{S_L}_{-1 \lambda_X}(x,\bfk)|^2 
\biggr]
\nonumber\\
&=& 
  \frac{1}{16 \pi^3} \, \biggl[ 
  |\psi^{+}_{+1 +\frac{1}{2}}(x,\bfk)|^2
+ |\psi^{+}_{+1 -\frac{1}{2}}(x,\bfk)|^2
- |\psi^{+}_{-1 +\frac{1}{2}}(x,\bfk)|^2
\biggr] 
\nonumber\\
&=& \frac{1}{16 \pi^3} \, 
\biggl[ (\varphi^{(1)}(x,\bfk)\Big)^2
+ \frac{\bfk^2}{M_N^2} \, \Big[1-(1-x)^2\Big] 
\, \Big(\varphi^{(2)}(x,\bfk)\Big)^2 \biggr]
\nonumber\\
&=& \frac{D_g(x)}{\pi \kappa^2} \, 
\biggl[
\Delta G(x)  + G^-(x) \, \alpha_-(x) \, \biggl( 
\frac{\bfk^2}{\kappa^2} \, D_g(x)  - 1 \biggr) \biggr] \, 
\exp\biggl[- \frac{\bfk^2}{\kappa^2} \, D_g(x) \biggr] 
\,.
\en 

The T worm-gear TMD $g_{1T}^g(x,\bfk^2)$, describing the distribution of a
circularly polarized gluon in a transversally polarized nucleon, 
is defined as 
\eq\label{g1T_def}  
g_{1T}^g(x,\bfk^2) &=& - \frac{M_N}{{\bf S}_T \, {\bfk}}
\, \frac{1}{16 \pi^3} \, i \epsilon^{\mu\nu}_T \, 
\sum\limits_{\lambda_g \lambda_g' \lambda_X}
\, \epsilon^{\dagger\lambda_g'}_\mu \, \epsilon^{\lambda_g}_\nu \, 
\psi^{*{\bf S}_T}_{\lambda_g'\lambda_X}(x,\bfk) \, 
\psi^{{\bf S}_T}_{\lambda_g\lambda_X}(x,\bfk)  
\nonumber\\
&=& \frac{M_N}{{\bf S}_T \, {\bfk}} \, 
\frac{1}{16 \pi^3} 
\sum\limits_{\lambda_X}
\biggl[ 
  |\psi^{{\bf S}_T}_{+1 \lambda_X}(x,\bfk)|^2 
- |\psi^{{\bf S}_T}_{-1 \lambda_X}(x,\bfk)|^2 
\biggr]
\nonumber\\
&=& \frac{M_N}{{\bf S}_T \, {\bfk}} \, 
\frac{1}{16 \pi^3} \, {\bf S_T} \, {\bf M_T}(x,\bfk)  
= \frac{1}{8 \pi^3} \, 
\varphi^{(1)}(x,\bfk^2) \, \varphi^{(2)}(x,\bfk^2) \, (1-x) 
\nonumber\\
&=& \frac{D_g^{3/2}(x) M_N}{\pi \kappa^3} \, 
\sqrt{G^2(x)-\Delta G^2(x)} \ \beta(x) 
\, \exp\biggl[- \frac{\bfk^2}{\kappa^2} \, D_g(x) \biggr] 
\,,  
\en 
where ${\bf M_T} = ({\bf M}^x, {\bf M}^y)$ and 
\eq 
{\bf M}^x(x,\bfk)  &=& 
  \psi^{*+}_{+1 -\frac{1}{2}}(x,\bfk) \psi^{-}_{+1 -\frac{1}{2}}(x,\bfk) 
+ {\rm H.c.} = 
  \frac{2 k^1}{M_N} \, 
\varphi^{(1)}(x,\bfk^2) \, \varphi^{(2)}(x,\bfk^2) \, (1-x) \,, 
\nonumber\\
{\bf M}^y(x,\bfk)  &=& 
i \, \psi^{*+}_{+1 -\frac{1}{2}}(x,\bfk) \psi^{-}_{+1 -\frac{1}{2}}(x,\bfk) 
+ {\rm H.c.} = 
  \frac{2 k^2}{M_N} \, \varphi^{(1)}(x,\bfk^2) \, \varphi^{(2)}(x,\bfk^2) \, (1-x)
\,. 
\en 

The Boer-Mulders TMD $h_{1\perp}^g(x,\bfk)$, describing a
linearly polarized gluon inside an unpolarized nucleon, are 
\eq\label{h1g_def} 
h_1^{\perp g}(x,\bfk^2) &=&
\frac{1}{16 \pi^3}   \, 
\frac{M_N^2}{\bfk^2} \, 
\eta^{\mu\nu}_T \, 
\sum\limits_{\lambda_N \lambda_g \neq \lambda_g' \lambda_X}
\, \epsilon^{\dagger\lambda_g'}_\mu \, \epsilon^{\lambda_g}_\nu \, 
\psi^{* \lambda_N}_{\lambda_g' \lambda_X}(x,\bfk) \, 
\psi^{\lambda_N}_{\lambda_g \lambda_X}(x,\bfk) \nonumber\\
&=& - \frac{1}{16 \pi^3}   \, 
\frac{M_N^2}{\bfk^4} \, 
\sum\limits_{\lambda_N \lambda_X} \, \biggl[ 
(k_-)^2 \, \psi^{* \lambda_N}_{+ \lambda_X}(x,\bfk) \, 
         \psi^{\lambda_N}_{- \lambda_X}(x,\bfk) + 
(k_+)^2 \, \psi^{* \lambda_N}_{- \lambda_X}(x,\bfk) \, 
         \psi^{\lambda_N}_{+ \lambda_X}(x,\bfk) \biggr] 
\nonumber\\
&=& \frac{1}{4 \pi^3}  
\Big[\varphi^{(2)}(x,\bfk^2)\Big]^2 \, (1-x) 
= \frac{2 D_g^2(x) M_N^2}{\pi \kappa^4}  
\, \frac{G(x)-\Delta G(x)}{1-x} \  
\, \exp\biggl[- \frac{\bfk^2}{\kappa^2} \, D_g(x) \biggr] 
\,,
\en 
where $k_\pm = k^1 \pm i k^2$. 

From decomposition of our gluon TDMs in terms of LFWFs one can see that they 
obey the sum rule: 
\eq\label{SR}
\biggl[f_1^g(x,\bfk^2)\biggr]^2  =
\biggl[g_{1L}^g(x,\bfk^2)\biggr]^2 
\,+\, \biggl[\frac{|\bfk|}{M_N} \, g_{1T}^g(x,\bfk^2)\biggr]^2 
\,+\, \biggl[ \frac{\bfk^2}{2M_N^2} h_1^g(x,\bfk^2)\biggr]^2 
\,.
\en
This sum rule is very important 
because it establishes the relation between square of
unpolarized TMD [left-hand side (lhs) of Eq.~(\ref{SR})] 
and the superposition of squares of
3 polarized TMDs [right-hand side (rhs) of Eq.~(\ref{SR})]. 
From this sum rule immediately follow 
the model-independent Mulders-Rodrigues 
inequalities (positivity bounds)~\cite{Mulders:2000sh}, without 
referring to a specific functional form of LFWFs $\varphi^{(i)}(x,\bfk)$, 
i.e. independent on a choice of $\varphi^{(i)}(x,\bfk)$:   
\eq\label{str_in}
& &
\sqrt{\biggl[g_{1L}^g(x,\bfk^2)\biggr]^2 
\,+\, \biggl[\frac{|\bfk|}{M_N} \, g_{1T}^g(x,\bfk^2)\biggr]^2}
\le f_1^g(x,\bfk^2) \,, \nonumber\\
& &
\sqrt{\biggl[g_{1L}^g(x,\bfk^2)\biggr]^2 \,+\,
      \biggl[\frac{\bfk^2}{2M_N^2} \, h_1^{\perp g}(x,\bfk^2)\biggr]^2}
\le f_1^g(x,\bfk^2) \,, \\
& &
\sqrt{\biggl[\frac{|\bfk|}{M_N} \, g_{1T}^g(x,\bfk^2)\biggr]^2 \,+\,
\biggl[\frac{\bfk^2}{2M_N^2} \, h_1^{\perp g}(x,\bfk^2)\biggr]^2}
\le f_1^g(x,\bfk^2) \,. \nonumber
\en
From above inequalities follow less stringent MR bounds
involving unpolarized and one of the polarized TMDs~\cite{Mulders:2000sh}:
\eq 
& &|g_{1L}^g(x,\bfk^2)| \le f_1^g(x,\bfk^2) \,, 
\label{g1Lf1g}\\[1mm] 
& &|g_{1T}^g(x,\bfk^2)| \le \frac{M_N}{|\bfk|} \, 
f_1^g(x,\bfk^2) \,, \label{g1Tf1g}\\[1mm] 
& &|h_{1}^{\perp g}(x,\bfk^2)| \le \frac{2 M_N^2}{\bfk^2} \, 
f_1^g(x,\bfk^2) \label{h1gf1g} \,. 
\en 
In particular, for any choice of $\varphi^{(i)}(x,\bfk^2)$ and for all 
values of $x$ and $\bfk$ variables 
the inequalities~(\ref{g1Lf1g})-(\ref{h1gf1g}) follow respectively from
\eq 
& &\Big[\varphi^{(2)}(x,\bfk^2)\Big]^2 \, \frac{\bfk^2}{M_N^2} \, (1-x)^2  \ge 0 \,, 
\nonumber\\[1mm]
& &\biggl[\varphi^{(1)}(x,\bfk^2) - \frac{|\bfk|}{M_N} \, (1-x) \, 
\varphi^{(2)}(x,\bfk^2)\biggr]^2  
+ \frac{\bfk^2}{M_N^2} \, \biggl[\varphi^{(2)}(x,\bfk^2)\biggr]^2  
\ge 0 \,, 
\nonumber\\[1mm]
& &\Big[\varphi^{(1)}(x,\bfk^2)\Big]^2 
+  \frac{\bfk^2}{M_N^2} \, \Big[x \, \varphi^{(2)}(x,\bfk^2)\Big]^2 \ge 0 \,. 
\en 

Note that our approach is similar to three frameworks 
developed in Refs.~\cite{Meissner:2007rx,Bacchetta:2020vty,Kaur:2020pvc} as on the level LFWFs,  
as on the level of gluon TMDs. 
In particular, one can see that expressions for the gluon TMDs in terms 
of LFWFs are the same in four approaches (in our and in the ones proposed 
in Refs.~\cite{Meissner:2007rx,Bacchetta:2020vty,Kaur:2020pvc}). 
One should stress that the sum rule derived by us in Eq.~(\ref{SR}) also 
holds in Ref.~\cite{Bacchetta:2020vty} (if to restrict to the minimal 
coupling of gluon with three-quark spectator) and in Ref.~\cite{Kaur:2020pvc}. 
In case of Ref.~\cite{Meissner:2007rx} a slightly different sum rule occurs 
(the nucleon mass $M_N$ is replaced by the quark mass $m$): 
\eq 
\biggl[f_1^g(x,\bfk^2)\biggr]^2  =
\biggl[g_{1L}^g(x,\bfk^2)\biggr]^2 
\,+\, \biggl[\frac{|\bfk|}{m} \, g_{1T}^g(x,\bfk^2)\biggr]^2 
\,+\, \biggl[ \frac{\bfk^2}{2m^2} h_1^g(x,\bfk^2)\biggr]^2 
\,.
\en 
Also it important to stress that our consideration of the quark and gluon TMDs 
is based on ideas of the quark-scalar diquark model and the gluon-three-quark 
spectator model proposed one of us in collaboration with Brodsky, Hwang, 
and Ma in Refs.~\cite{Brodsky:2002cx,Brodsky:2000ii} and was successfully 
used, e.g., in Refs.~\cite{Meissner:2007rx,Gutsche:2016gcd,Lu:2016vqu,%
Kaur:2020pvc,Lyubovitskij:2020otz}. The similar approaches to the quark 
and gluon TMDs use based on quark target model~\cite{Meissner:2007rx} and 
spectator model~\cite{Bacchetta:2020vty} allow to get description 
of the TMDs in very good agreement with us. Note that agreement between 
our formalism using light-front picture and spectator model for the case 
of quark parton densities was proved in Ref.~\cite{Bacchetta:2008af}. For 
the case of gluon parton densities we make a conclusion about argeement 
in the present paper. 
The main difference is that 
we also guarantee the correct scaling of TMDs and other parton densities at 
at large $x$ (see detailed discussion in Sec.IIIc) in consistency 
with constraints imposed by QCD, while it is not the case in the calculations 
presented in Refs.~\cite{Meissner:2007rx,Bacchetta:2020vty,Kaur:2020pvc}.  
One should stress that an implementation of large $x$ behavior in 
frameworks used in Refs.~\cite{Meissner:2007rx,Bacchetta:2020vty,Kaur:2020pvc} 
can be done easily --- by taking into account a specific 
$x$-dependence of the couplings/form factors. Such idea was 
proposed in Ref.~\cite{Jakob:1997wg}, where the spectator model was proposed. 
In particular, it was suggested that the multipole form factors of
spectator diquark contain a free parameter $\alpha$, which indicates the power of 
form factors and can be clearly fixed
to fulfill large $x$ counting rules. 

The four T-even gluon TMDs, upon integration over $d^2\bfk$, are normalized as 
\eq\label{gluon_PDFs} 
f_1^g(x)    &=& \int d^2\bfk \, f_1^g(x,\bfk^2)    = G(x)        \,, 
\nonumber\\ 
g_{1L}^g(x) &=& \int d^2\bfk \, g_{1L}^g(x,\bfk^2) = \Delta G(x) \,, 
\nonumber\\ 
g_{1T}^g(x) &=& \int d^2\bfk \, g_{1T}^g(x,\bfk^2) = 
\frac{D_g^{1/2}(x) M_N}{\kappa} \, 
\sqrt{G^2(x)-\Delta G^2(x)} \ \beta(x) \,, 
\nonumber\\  
h_1^{\perp g}(x) &=& \int d^2\bfk \, h_1^{\perp g}(x,\bfk^2) = 
\frac{2 D_g(x) M_N^2}{\kappa^2} \, \frac{G(x)-\Delta G(x)}{1-x} 
\,.
\en  
One can see that they obey the condition 
\eq 
\frac{\Big[g_{1T}^g(x)\Big]^2}{\Big[f_1^g(x)+g_{1L}^g(x)\Big] \, h_1^{\perp g}(x)} 
= \frac{1-x}{2} \, \beta^2(x) \,. 
\en 
For completeness we now discuss how to construct the T-even gluon correlators 
(unpolarized $\Gamma^{\mu\nu}_U$, longitudinally polarized $\Gamma^{\mu\nu}_L$, and 
transversally polarized $\Gamma^{\mu\nu}_T$) derived in Ref.~\cite{Mulders:2000sh} 
in terms of our LFWFs. 
First, we split 
the product of the gluon polarization vectors into a sum of 
antisymmetric $(A B)^{[\mu\nu]}   = A^\mu B^\nu - A^\nu B^\mu$ 
and symmetric $(A B)^{\{\mu\nu\}} = A^\mu B^\nu + A^\nu B^\mu$ 
combinations under exchange of Lorenz indices $\mu$ and $\nu$: 
\eq 
  \epsilon^{\dagger\mu}_{\lambda_g'} \, \epsilon^\nu_{\lambda_g} 
= \frac{1}{2} \Big(\epsilon^{\dagger}_{\lambda_g'} \, \epsilon_{\lambda_g}\Big)^{[\mu\nu]} 
+ \frac{1}{2} \Big(\epsilon^{\dagger}_{\lambda_g'} \, \epsilon_{\lambda_g}\Big)^{\{\mu\nu\}} 
\,. 
\en 
In case of T-even gluon TMDs symmetric combinations of polarization vectors contribute 
to the unpolarized $f_1^g(x,\bfk^2)$ and $h_1^{\perp g}(x,\bfk^2)$ TMDs, 
while antisymmetric combinations to the polarized $g_{1L}^g$ and $g_{1T}^g$ TMDs.  

The unpolarized tensor $\Gamma^{\mu\nu}_U$ is given by 
\eq\label{GammaU_f1_h1_1}  
\Gamma_U^{\mu\nu}(x,\bfk) &=& - g^{\mu\nu}_T \, f_1^g(x,\bfk^2) 
+ \frac{\bfk^2}{2 M_N^2} \, \eta^{\mu\nu}_T \, h_1^{\perp g}(x,\bfk^2) 
\nonumber\\
&=& 
\frac{1}{16 \pi^3} \, \frac{1}{2} \, \sum\limits_{\lambda_N \lambda_g \lambda_g' \lambda_X} \, 
\Big(\epsilon^{\dagger}_{\lambda_g'} \, \epsilon_{\lambda_g}\Big)^{\{\mu\nu\}}   \,
\psi^{*\lambda_N}_{\lambda_g'\lambda_X} \, \psi^{\lambda_N}_{\lambda_g\lambda_X} \,. 
\en 
Next we split the sum over $\lambda_g$ and $\lambda_g'$ in two terms: 
$\lambda_g = \lambda_g'$ (it produces the $f_1^g(x,\bfk^2)$ TMD) 
and $\lambda_g \neq \lambda_g'$ (it produces the $h_1^{\perp g}(x,\bfk^2)$ TMD): 
\eq\label{GammaU_f1_h1_2}  
\Gamma_U^{\mu\nu}(x,\bfk) = \Gamma_{U; f_1^g}^{\mu\nu}(x,\bfk) + 
\Gamma_{U; h_1^{\perp g}}^{\mu\nu}(x,\bfk)\,, 
\en  
where 
\eq\label{GammaU_f1_h1_3} 
\Gamma_{U; f_1^g}^{\mu\nu}(x,\bfk) &=& - g^{\mu\nu}_T \, f_1^g(x,\bfk^2) 
\nonumber\\
&=& - \frac{1}{16 \pi^3} \, \frac{1}{2} \, g^{\mu\nu}_T \, 
\sum\limits_{\lambda_N \lambda_g \lambda_X} \, 
|\psi^{\lambda_N}_{\lambda_g\lambda_X}(x,\bfk)|^2 \,, \nonumber\\
\Gamma_{U; h_1^{\perp g}}^{\mu\nu}(x,\bfk) &=& 
\frac{\bfk^2}{2 M_N^2} \, \eta^{\mu\nu}_T \, h_1^{\perp g}(x,\bfk^2) 
\nonumber\\
&=& \frac{1}{16 \pi^3} \, \frac{1}{2} \, \sum\limits_{\lambda_N \lambda_g\neq\lambda_g'  \lambda_X} \, 
\Big(\epsilon^{\dagger}_{\lambda_g'} \, \epsilon_{\lambda_g}\Big)^{\{\mu\nu\}} \, 
\psi^{* \lambda_N}_{\lambda_g'\lambda_X}(x,\bfk) \, \psi^{\lambda_N}_{\lambda_g\lambda_X}(x,\bfk)  
\,.  
\en 
From Eqs.~(\ref{GammaU_f1_h1_1})-(\ref{GammaU_f1_h1_3}) and using the
normalization and orthogonality conditions for the tensors $g^{\mu\nu}_T$ and 
$\eta^{\mu\nu}_T$ the master formulas 
for the $f_1^g(x,\bfk^2)$~(\ref{f1g_def}) and $h_1^{\perp g}(x,\bfk^2)$~(\ref{h1g_def}) 
follow:  
\eq 
f_1^g(x,\bfk^2) &=& - \frac{1}{2} \, g_{\mu\nu; T} \, \Gamma_U^{\mu\nu}(x,\bfk) 
=  - \frac{1}{2} \, g_{\mu\nu; T} \, \Gamma_{U; f_1^g}^{\mu\nu}(x,\bfk) \,, \nonumber\\
h_1^{\perp g}(x,\bfk^2) &=& \frac{M_N^2}{\bfk^2} \, \eta_{\mu\nu; T} \, 
\Gamma_U^{\mu\nu}(x,\bfk) 
=  \frac{M_N^2}{\bfk^2} \, \eta_{\mu\nu; T} \, 
\Gamma_{U; h_1^{\perp g}}^{\mu\nu}(x,\bfk) \,.
\en 

By analogy, we construct longitudinal $\Gamma^{\mu\nu}_L$ and
transversal $\Gamma^{\mu\nu}_T$ polarized tensors. 
Corresponding gluon correlator tensors read: 
\eq\label{Gamma_LT} 
\Gamma^{\mu\nu}_L(x,\bfk) &=& - i \epsilon^{\mu\nu}_T \, 
S_L \, g_{1L}^g(x,\bfk^2) \nonumber\\
&=& - \frac{1}{16 \pi^3} \, 
\sum\limits_{\lambda_g \lambda_g' \lambda_X} \, 
\Big(\epsilon^{\dagger}_{\lambda_g'} \, \epsilon_{\lambda_g}\Big)^{[\mu\nu]} 
\, \psi^{*S_L}_{\lambda_g'\lambda_X}(x,\bfk) 
\, \psi^{S_L}_{\lambda_g\lambda_X}(x,\bfk) 
\,, \nonumber\\
\Gamma^{\mu\nu}_T(x,\bfk) &=& - i \epsilon^{\mu\nu}_T \, 
\frac{{\bf S}_T \, {\bfk}}{M_N} \, g_{1T}^g(x,\bfk^2) \nonumber\\
&=& - \frac{1}{16 \pi^3} \, 
\sum\limits_{\lambda_g \lambda_g' \lambda_X} \, 
\Big(\epsilon^{\dagger}_{\lambda_g'} \, \epsilon_{\lambda_g}\Big)^{[\mu\nu]} 
\, \psi^{*{\bf S}_T}_{\lambda_g'\lambda_X}(x,\bfk) 
\, \psi^{{\bf S}_T}_{\lambda_g\lambda_X}(x,\bfk) \,.
\en 
From Eq.~(\ref{Gamma_LT}) follow master formulas  
for the TMDs $g_{1L}^g(x,\bfk^2)$~(\ref{g1L_def}) 
and $g_{1T}^g(x,\bfk^2)$~(\ref{g1T_def}):
\eq 
S_L \, g_{1L}^g(x,\bfk^2) = 
\frac{i}{2} \epsilon_{\mu\nu; T} \, \Gamma_L^{\mu\nu}(x,\bfk)\,, \qquad 
\frac{{\bf S}_T \, {\bfk}}{M_N} \, 
g_{1T}^g(x,\bfk^2) = \frac{i}{2} \epsilon_{\mu\nu; T} \, \Gamma_T^{\mu\nu}(x,\bfk)\,. 
\en

Now we present the results for the T-even gluon TMDs in 
transverse impact space $\bfb$, performing the Fourier transform 
of the TMDs with respect to the  $\bfk$. For completeness we also display 
the decomposition for the gluon impact TMDs in terms of the impact LFWF functions 
$\tilde\varphi(x,\bfb)$ and $\tilde\varphi(x,\bfb)$  
\eq
\tilde\varphi^{(1)}(x,\bfb^2) &=&  \frac{1}{2\pi} 
\, \sqrt{G^+(x)} \ \beta(x) \, 
\exp\Big[ - \frac{\bfb^2 \kappa^2}{8 D_g(x)}\Big]\,, 
\nonumber\\
\tilde\varphi^{(2)}(x,\bfb^2) &=&  \frac{1}{2\pi} \, \sqrt{G^-(x)} \  
\frac{\sqrt{D_g(x)}}{1-x} \, 
\exp\Big[ - \frac{\bfb^2 \kappa^2}{8 D_g(x)}\Big]\,. 
\en

The list of the T-even gluon TMDs in impact space reads: 

Unpolarized gluon TMD in impact space $\tilde f_{1}^g(x,\bfb^2)$
\eq 
\tilde f_1^g(x,\bfb^2) &=& \int \frac{d^2\bfk}{(2\pi)^2} 
\, e^{i\bfb\bfk} \, f_1^g(x,\bfk^2)
\nonumber\\ 
&=& \Big[\tilde\varphi^{(1)}(x,\bfb^2)\Big]^2 
+   \Big[\tilde\varphi^{(2)}(x,\bfb^2)\Big]^2 \, 
\frac{1 + (1-x)^2}{D_g(x)} \, 
\biggl[ 1 + \frac{\bfb^2 \kappa^2}{4 D_g(x)} \biggr] 
\nonumber\\ 
&=& \frac{1}{4\pi^2} \, \biggl[ G(x) + G^-(x) \, \alpha_+(x) \, 
\frac{\bfb^2 \kappa^2}{4 D_g(x)} \biggr] 
\, \exp\Big[-\frac{\bfb^2 \kappa^2}{4 D_g(x)}\Big] \,.
\en 

Helicity gluon TMD $\tilde g_{1L}^g(x,\bfb^2)$
\eq
\tilde g_{1L}^g(x,\bfb^2) &=& 
\int \frac{d^2\bfk}{(2\pi)^2} \, e^{i\bfb\bfk} \, g_{1L}^g(x,\bfk^2) 
\nonumber\\
&=& 
  \Big[\tilde\varphi^{(1)}(x,\bfb^2)\Big]^2 
+ \Big[\tilde\varphi^{(2)}(x,\bfb^2)\Big]^2 \, 
\frac{1 - (1-x)^2}{D_g(x)} \, 
\biggl[ 1 + \frac{\bfb^2 \kappa^2}{4 D_g(x)} \biggr] 
\nonumber\\ 
&=& \frac{1}{4\pi^2} \, \biggl[ \Delta G(x) + G^-(x) \,  \alpha_-(x) \, 
\frac{\bfb^2 \kappa^2}{4 D_g(x)} \biggr] 
\, \exp\Big[-\frac{\bfb^2 \kappa^2}{4 D_g(x)}\Big] \,.
\en  

T worm-gear gluon TMD $\tilde g_{1T}^g(x,\bfb^2)$
\eq
\tilde g_{1T}^g(x,\bfb^2) &=& 
\int \frac{d^2\bfk}{(2\pi)^2} \, e^{i\bfb\bfk} \, g_{1T}^g(x,\bfk^2) 
\nonumber\\
&=& \frac{2M_N}{\kappa} \, 
\tilde\varphi^{(1)}(x,\bfb^2) 
\tilde\varphi^{(2)}(x,\bfb^2) (1-x)  
\nonumber\\
&=& \frac{1}{4 \pi^2} \, g_{1T}^g(x) 
\, \exp\Big[-\frac{\bfb^2 \kappa^2}{4 D_g(x)}\Big] \,.
\en  

Boer-Mulders gluon TMD $\tilde h_1^{\perp g}(x,\bfb^2)$
\eq
\tilde h_1^{\perp g}(x,\bfb^2) &=& \int \frac{d^2\bfk}{(2\pi)^2}  \, e^{i\bfb\bfk} 
\, h_1^{\perp g}(x,\bfk^2) \nonumber\\
&=& \frac{4M_N^2}{\kappa^2} \, 
\Big[\tilde\varphi^{(2)}(x,\bfb^2)\Big]^2 \,  (1-x)  
\nonumber\\
&=& \frac{1}{4 \pi^2} \, h_1^{\perp g}(x) 
\, \exp\Big[-\frac{\bfb^2 \kappa^2}{4 D_g(x)}\Big]
\en  

Gluon TMDs in momentum and impact space obey the following relations: 
\eq
{\rm TMD}^g(x) = \int d^2\bfk {\rm TMD}^g(x,\bfk^2) = 4\pi^2 \, {\widetilde{\rm TMD}}^g(x,0)\,, 
\en  
where ${\rm TMD} = f_1, g_{1L}, g_{1T}, h_1^\perp$. 

The gluon TMDs in the impact space obey the following sum rule: 
\eq\label{SR_impact} 
\biggl[\tilde f_1^g(x,\bfb^2)\biggr]^2 = 
\biggl[\tilde g_{1L}^g(x,\bfb^2)\biggr]^2 + 
\biggl[S(x,\bfb^2) \, 
\tilde g_{1T}^g(x,\bfb^2)\biggr]^2 + 
\biggl[\frac{S^2(x,\bfb^2)}{2} \, \tilde h_1^{\perp g}(x,\bfb^2)\biggr]^2 \,,
\en 
where 
\eq 
S(x,\bfb^2) = \frac{\kappa}{M_N \sqrt{D_g(x)}} \, \sqrt{1 + \frac{\bfb^2 \kappa^2}{4 D_g(x)}} \,. 
\en 
From this sum rule one can deduce the following inequalities: 
\eq 
& &
\sqrt{\biggl[\tilde g_{1L}^g(x,\bfb^2)\biggr]^2 \,+\, 
\biggl[S(x,\bfb^2) \, 
\tilde g_{1T}^g(x,\bfb^2)\biggr]^2}  
\le \tilde f_1^g(x,\bfb^2) \,, \\
& &\sqrt{\biggl[\tilde g_{1L}^g(x,\bfb^2)\biggr]^2 \,+\, 
\biggl[\frac{S^2(x,\bfb^2)}{2} \, 
\tilde h_1^{\perp g}(x,\bfb^2)\biggr]^2} 
\le \tilde f_1^g(x,\bfb^2) \,, \\
& &\sqrt{\biggl[S(x,\bfb^2) \, \tilde g_{1T}^g(x,\bfb^2)\biggr]^2 + 
\biggl[\frac{S^2(x,\bfk^2)}{2} \, 
\tilde h_1^{\perp g}(x,\bfb^2)\biggr]^2} 
\le \tilde f_1^g(x,\bfb^2)
\en 
and 
\eq 
& &\tilde g_{1L}^g(x,\bfb^2) \leq \tilde f_1^g(x,\bfb^2) 
\,, \label{ineq1_bp} \\[2mm] 
& &\tilde g_{1T}^g(x,\bfb^2) \leq 
\frac{M_N}{\kappa} \, \sqrt\frac{D_g(x)}{1+\frac{\bfb^2 \kappa^2}{4 D_g(x)}} 
\, \tilde f_1^g(x,\bfb^2)
\leq 
\frac{M_N}{\kappa} \, \sqrt{D_g(x)} \, \tilde f_1^g(x,\bfb^2)
\,, \label{ineq2_bp} \\[2mm] 
& &\tilde h_1^{\perp g}(x,\bfb^2) 
\leq \frac{2 M_N^2}{\kappa^2} 
\, \frac{D_g(x)}{1+\frac{\bfb^2 \kappa^2}{4 D_g(x)}}
\, \tilde f_1^g(x,\bfb^2) 
\leq \frac{2 M_N^2}{\kappa^2} 
\, D_g(x) \, \tilde f_1^g(x,\bfb^2) 
\,. \label{ineq3_bp} 
\en 
One can see that the above inequalities involve model parameters $\kappa$ and $D_g(x)$. 
The nucleon mass and the parameter $\kappa$ are related in our approach, as 
$M_N = 2 \sqrt{2} \, \kappa$~\cite{Gutsche:2011vb,Gutsche:2019jzh}. 
The profile function $D_g(x)$ can be also related to the profile function 
$f_g(x)$, using the relations between form factors, GPDs, and TMDs 
(see discussion in Sec.~\ref{GPD_FF}). The inequalities between 
impact gluon TMDs for generalized case are shown 
in Appendix~\ref{app_TMD}. 

\subsection{Gluon GPDs and form factors} 
\label{GPD_FF}

In this section we derive the GPDs and form factors 
describing the distribution of gluons in hadrons. 
Using Eqs.~(\ref{GPD1})-(\ref{GPD2}) for specific 
twist (we shift $\tau \to \tau + 1/2$ and put $\tau = 3$)  
we get four gluon GPDs corresponding to the four PDFs/TMDs  
[see Eq.~(\ref{gluon_PDFs})]:  
\eq\label{FF_ADSQCD} 
{\cal H}_G(x,Q^2) &=& \ \ G(x)                       \ \exp\Big[-a \, D_{G}(x) \, (1-x)^2 \Big] \,, 
\nonumber\\ 
{\cal H}_{\Delta G}(x,Q^2) &=& \Delta G(x)           \ \exp\Big[-a \, D_{\Delta G}(x) \, (1-x)^2 \Big] \,, 
\nonumber\\
{\cal H}_{g_{1T}^g}(x,Q^2) &=& g_{1T}(x)               \ \exp\Big[-a \, D_{g_{1T}^g}(x) \, (1-x)^2 \Big] \,, 
\nonumber\\
{\cal H}_{h_1^{\perp g}}(x,Q^2) &=& f_1^{\perp g}(x) \ \exp\Big[-a \, D_{h_1^{\perp g}}(x) \, (1-x)^2 \Big] \,, 
\en
where $a = Q^2/(4 \kappa^2)$ and $D_g$ $(g=G, \Delta G, g_{1T}^g, h_1^{\perp g})$ is the set   
of the profile functions relevant for specific GPD.  
The corresponding form factors are given by 
\eq 
F_g(Q^2) = \int\limits_0^1 dx \, {\cal H}_g(x,Q^2)\,, \qquad 
g = G, \Delta G, g_{1T}^g, f_1^{\perp g} \,. 
\en 
$D_g(x)$ is related to the function $f_g(x)$~\cite{Lyubovitskij:2020otz}   
\eq   
D_g(x) = - \frac{\log[1- [f_g(x)]^{2/5} (1-x)^2]}{(1-x)^2} \,.  
\en 
On the other hand, the profile function $f_g(x)$ is fixed from gluon PDF 
by solving differential equation with respect to $x$ variable. 
Note, within {\it symmetric version} ($D_{g_1}=D_{g_2}=D_{g}$) and 
{\it generalized version} ($D_{g_1} \neq D_{g_2}$) of 
our approach the profile functions $D_G(x)$ and $D_{\Delta G}(x)$ 
degenerate $D_G(x) = D_{\Delta G}(x) = D_g(x) = D_{g_1}$ [see 
Eqs.~(\ref{f1g_full}) and~(\ref{g1L_full}) in Appendix~\ref{app_TMD}].
Such a scenario can be realized. In particular, 
as we showed in Eqs~(\ref{large_QCDI}) and~(\ref{large_QCDII}) 
in both QCD-based approaches the profile functions for unpolarized $G(x)$ 
and polarized $\Delta G(x)$ gluon PDFs degenerate at large $x$ and become constant: 
$D_g = [f_g]^{2/5}$, where 
$f_g = \frac{21}{10} \, \la x_g \ra$ (QCDI~\cite{Brodsky:1989db})
and $f_g = \frac{7}{12}  \, \la x_g \ra$ (QCDII~\cite{Brodsky:1994kg}).
Using the central value of the latest lattice result 
$\la x_g \ra = 0.427$~\cite{Alexandrou:2020sml} one gets: 
$D_{g_1}=D_g = 0.957 \simeq 1$ (QCDI~\cite{Brodsky:1989db}) 
and $D_{g_1}=D_g =0.574 \simeq 0.6$ (QCDII~\cite{Brodsky:1994kg}).
In case of the ${\cal H}_{g_{1T}^g}(x,Q^2)$ and ${\cal H}_{h_1^{\perp g}}(x,Q^2)$ 
GPDs the profile functions $D_{g_{1T}^g}(x)$ and $D_{h_1^{\perp g}}(x)$ 
degenerate with profile functions $D_G(x) = D_{\Delta G}(x)$ in the 
{\it symmetric version} of our approach and deviate from them 
in {\it generalized version}. In the latter case we get the 
following relations between two sets $D_{g_{1T}}(x)$, $D_{h_1^{\perp g}}(x)$ 
and $D_{g_1}$, $D_{g_1}$: 
\eq 
D_{g_{1T}^g}(x) = \frac{D_{g_1}(x) + D_{g_2}(x)}{2}\,, \qquad 
D_{f_1^{\perp g}}(x) = D_{g_2}(x) \,. 
\en 
Finally, our profile functions $D_G(x)$, $D_{\Delta G}(x)$,  
$D_{g_{1T}^g}(x)$, $D_{f_1^{\perp g}}(x)$ obey the relations: 
\eq 
& &D_{g_1}(x) = D_G(x) = D_{\Delta G}(x) \,, \nonumber\\
& &D_{g_2}(x) = D_{f_1^{\perp g}}(x) \,, \nonumber\\
& &\frac{D_{g_1}(x)+D_{g_2}(x)}{2} = D_{g_{1T}^g}(x) \,, \nonumber\\
& &2 D_{g_{1T}^g}(x) = D_G(x) + D_{f_1^{\perp g}}(x) 
= D_{\Delta G}(x) +  D_{f_1^{\perp g}}(x) \,. 
\en 
The main conclusion from above discussion is that the profile functions 
in the T-even gluon TMDs/GPDs are related with each other and become  
constants at large $x$. It is supported by the corresponding parametrizations 
of $\bfk$ dependence of the gluon TMDs with constant scale parameters 
(see, e.g., Refs.~\cite{Anselmino:2001js,Schweitzer:2010tt,Boer:2011kf}). 
On the other hand, we can also explain why 
the scale parameters are not the same for all gluon TMDs. 
In particular, in Ref.~\cite{Boer:2011kf} two different scales in 
the Gaussians parametrizing the $\bfk^2$ dependence of the 
$f_1^g(x,\bfk^2)$ and $h_1^{\perp g}(x,\bfk^2)$ TMDs have been considered: 
\eq\label{Boer_TMDs} 
f_1^g(x,\bfk^2) &=& \frac{G(x)}{\pi \la \bfk^2 \ra} \, 
\exp\biggl[ - \frac{\bfk^2}{\la \bfk^2 \ra}\biggr]\,, \nonumber\\
h_1^{\perp g}(x,\bfk^2) &=& \frac{M_N^2 G(x)}{\pi \la \bfk^2 \ra^2} \, 
\, \frac{2 e (1-r)}{r} \, 
\exp\biggl[ - \frac{\bfk^2}{r \la \bfk^2 \ra}\biggr]\,, 
\en  
where $\la \bfk^2 \ra$ is the width, $r$ is a free parameter fixed as $2/3$. 
The above ansatz satisfies the model-independent 
Mulders-Rodrigues inequality~(\ref{h1gf1g}) due to holding the inequality 
$e^{s-1} \ge s$ (consequence of the well-known Bernoulli inequality), 
where $s = \frac{\bfk^2}{\la \bfk^2 \ra} \, \frac{1-r}{r}$.  
Also two TMDs in Eq.~(\ref{Boer_TMDs}) 
contain two different scale parameters, parametrized by $\la \bfk^2 \ra$ and $r$. 
As we showed our formalism supports existence of two different scales 
in the $f_1^g(x,\bfk^2)$ and $h_1^{\perp g}(x,\bfk^2)$ TMDs.  
Two scales considered in Ref.~\cite{Boer:2011kf} are related to our 
parameters $D_{g_1}$, $D_{g_2}$, and $\kappa$ as 
\eq 
\la \bfk^2 \ra = \frac{\kappa^2}{D_{g_1}} \,, \qquad 
r = \frac{D_{g_1}}{D_{g_2}} \,. 
\en
Here we neglect by $x$ dependence of the $D_{g_1}$ and $D_{g_2}$ and 
by the second subleading term 
in our expression for the $f_1^g(x,\bfk^2)$~(\ref{f1g_full}) proportional to $G^-(x)$.  

Now we are on the position to derive the large $x$ and large $Q^2$ behavior of 
the gluon quantities (LFWFs, PDFs, TMDs, GPDs, and form factors) based 
on two version QCD-based approaches to gluon PDFs: 
QCDI~\cite{Brodsky:1989db} and QCDII~\cite{Brodsky:1994kg}. 
In both approaches the gluon PDFs scale at large $x$ as 
\eq 
G(x) \sim \Delta G(x) \sim G^+(x) \sim (1-x)^4\,, \qquad 
G^-(x) \sim (1-x)^6\,. 
\en  
The only difference is in large $x$ scaling of the 
$\beta(x)$ defined through $G^+(x)$ and $G^-(x)$ in 
Eq.~(\ref{beta_def}):   
\eq
\beta(x) \sim
\left\{
\begin{array}{cl}
          1,   & \ \ \ {\rm  QCDI}  \\ 
{\cal O}(1-x), & \ \ \ {\rm  QCDII} \\ 
\end{array}
\right.
\en
We start with large $x$ scaling of the LFWFs: 
\eq 
& &\varphi^{(1)}(x,\bfk^2) \sim \tilde\varphi^{(1)}(x,\bfb^2) 
\sim \sqrt{G^+(x)} \, \beta(x) \sim 
\left\{ 
\begin{array}{cl}  
(1-x)^2, & \ \ \ {\rm QCDI}  \nonumber\\
(1-x)^3, & \ \ \ {\rm QCDII} \nonumber\\
\end{array}
\right. \,, \nonumber\\ 
& &\varphi^{(2)}(x,\bfk^2) \sim \tilde\varphi^{(2)}(x,\bfb^2) 
\sim \frac{\sqrt{G^-(x)}}{1-x}  \sim (1-x)^2 \,. 
\en 
Gluon PDFs, TMDs, and GPDs scale as 
\eq 
& &G(x) \sim f_1^g(x,\bfk^2) \sim \tilde f_1^g(x,\bfb^2) 
\sim {\cal H}_G(x,Q^2) \sim (1-x)^4\,, 
\nonumber\\[4mm]
& &\Delta G(x) 
\sim g_{1L}^g(x,\bfk^2) \sim 
\tilde g_{1L}^g(x,\bfb^2) \sim 
{\cal H}_{\Delta G}(x,Q^2) \sim (1-x)^4 \,,
\nonumber\\[4mm]
& &g_{1T}^g(x) \sim 
g_{1T}^g(x,\bfk^2) \sim 
\tilde g_{1T}^g(x,\bfb^2) \sim 
{\cal H}_{g_{1T}^g}(x,Q^2) \nonumber\\[2mm]
& &\hspace*{1.1cm}  \sim
\sqrt{G^+(x) G^-(x)} \, \beta(x) \sim (1-x)^5 \, \beta(x) 
\sim 
\left\{ 
\begin{array}{cl}   
(1-x)^5,   & \ \ \ {\rm QCDI}   \\ 
(1-x)^6,   & \ \ \ {\rm QCDII}  \\ 
\end{array}
\right.  \,, \nonumber\\[4mm]
& &h_1^{\perp g}(x) \sim 
h_1^{\perp g}(x,\bfk^2) \sim 
\tilde h_1^{\perp g}(x,\bfb^2) \sim 
{\cal H}_{f_1^{\perp g}}(x,Q^2) 
\sim \frac{G^-(x)}{1-x}  \sim (1-x)^5 \,. 
\en 
Finally the gluon form factors scale at large $Q^2$ as 
\eq 
& &F_G(Q^2) \sim F_{\Delta G}(x) \sim \int\limits_0^1 dx \, (1-x)^4 
\, \exp\Big[-Q^2 (1-x)^2 \Big] \sim \frac{1}{Q^5} \,, 
\nonumber\\
& &F_{g_{1T}^g}(Q^2) \sim \int\limits_0^1 dx \, (1-x)^5 \, \beta(x)  
\, \exp\Big[-Q^2 (1-x)^2 \Big] \sim 
\left\{ 
\begin{array}{cl}   
\frac{1}{Q^6}\,,   & \ \ \ {\rm QCDI}  \\[2mm]
\frac{1}{Q^7}\,,   & \ \ \ {\rm QCDII} \\ 
\end{array}
\right.  \,, \nonumber\\
& &F_{f_1^{\perp g}}(Q^2) \sim \int\limits_0^1 dx \, (1-x)^5 
\, \exp\Big[-Q^2 (1-x)^2 \Big] \sim \frac{1}{Q^6} \,.  
\en 

\section{Numerical applications} 

In this section we present numerical applications of our analytical results 
for gluon parton densities. We concentrate on the T-even gluon TMDs, which were 
analyzed recently in similar approach in Ref.~\cite{Bacchetta:2020vty}.  
The main point is that we derive our analytical formulas 
for the gluon parton densities in terms of unpolarized $G(x)$ and polarized $\Delta G(x)$ 
gluon PDFs, which we take from present-day world data analysis. 
Recently, in Ref.~\cite{Sufian:2020wcv} such set was derived based on 
ideas of QCD approaches~\cite{Brodsky:1989db} and~\cite{Brodsky:1994kg} 
and using world data analysis performed 
by the NNPDF Collaboration in Refs.~\cite{Ball:2017nwa,Nocera:2014gqa}. 
In particular, the authors of Ref.~\cite{Sufian:2020wcv} derived three 
parametrization based on the NNPDF global analysis. They are more or less are 
equivalent and we take the one (so-called ``ansatz 3''), which is direct 
extension of QCD formalism developed in Refs.~\cite{Brodsky:1989db,Brodsky:1994kg} 
and produce the factor $\beta \neq 0$ [see definition in Eq.~(\ref{beta_def})]. 

The ansatz 3 for the gluon PDFs of Ref.~\cite{Sufian:2020wcv} reads: 
\eq
x G^+(x) &=& x^\alpha \, \Big[ A (1-x)^{4+\beta} + B (1-x)^{5+\beta} \,  
\Big(1 + \gamma \sqrt{x} + \delta x\Big)\,, 
\nonumber\\
x G^-(x) &=& x^\alpha \, \Big[ A (1-x)^{6+\beta'} + B (1-x)^{7+\beta'} \,  
\Big(1 + \gamma' \sqrt{x} + \delta' x\Big) 
\,, 
\en
where $A$ and $B$ are the normalization parameters determined from the 
moments of the gluon PDFs $\la x_g  \ra$ and $\Delta G$, the set of 
the parameters $(\alpha,\beta,\gamma,\delta)$ was fixed using 
the NNPDF global analysis as 
\eq\label{NNPDF_data} 
& &
\alpha  =  0.034 \pm 0.064\,, \quad 
\beta   =  0.54  \pm 1.30 \,, \quad 
\gamma  = -2.63  \pm 0.60 \,, \quad 
\delta  =  2.54  \pm 1.01 \,, \nonumber\\
& &
\beta'  =  0.54  \pm 1.30 \,, \quad 
\gamma' = -2.63  \pm 0.60 \,, \quad 
\delta' =  2.54  \pm 1.01 \,. 
\en 

\begin{figure}
\begin{center}
\epsfig{figure=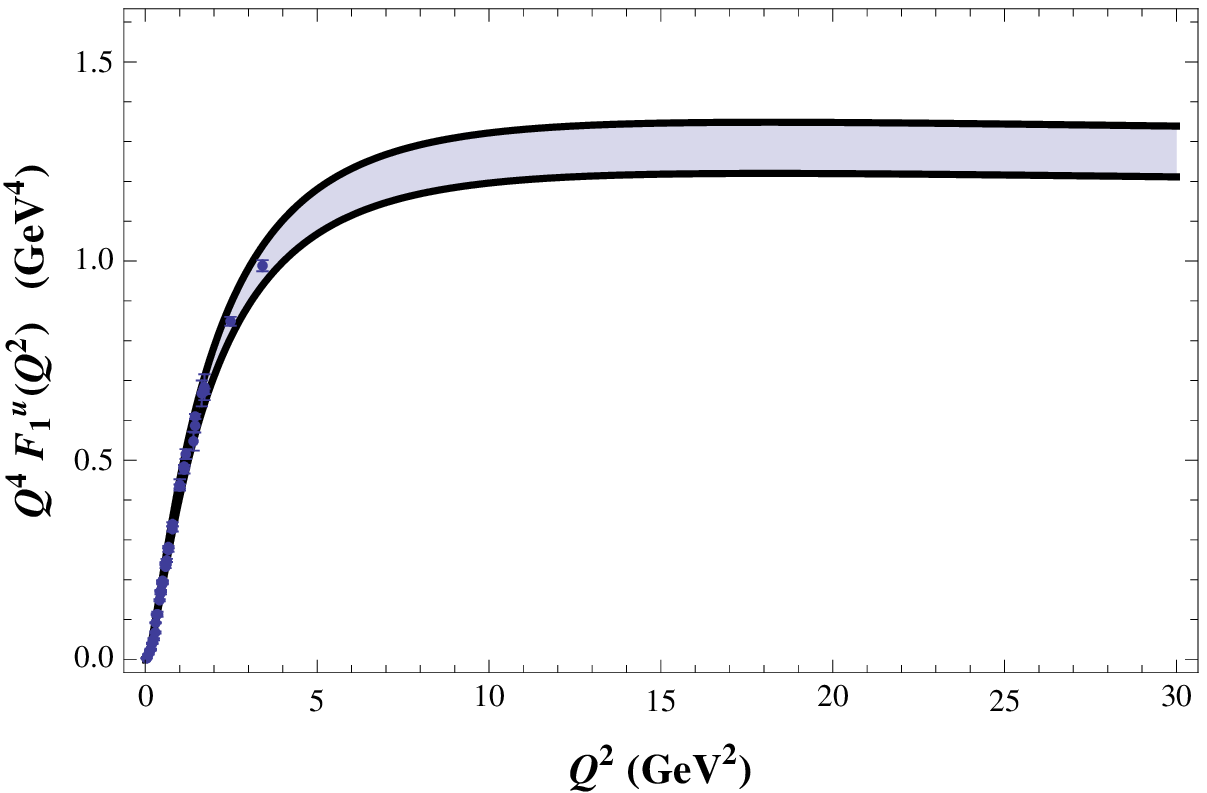,scale=.575}
\hspace*{1cm}
\epsfig{figure=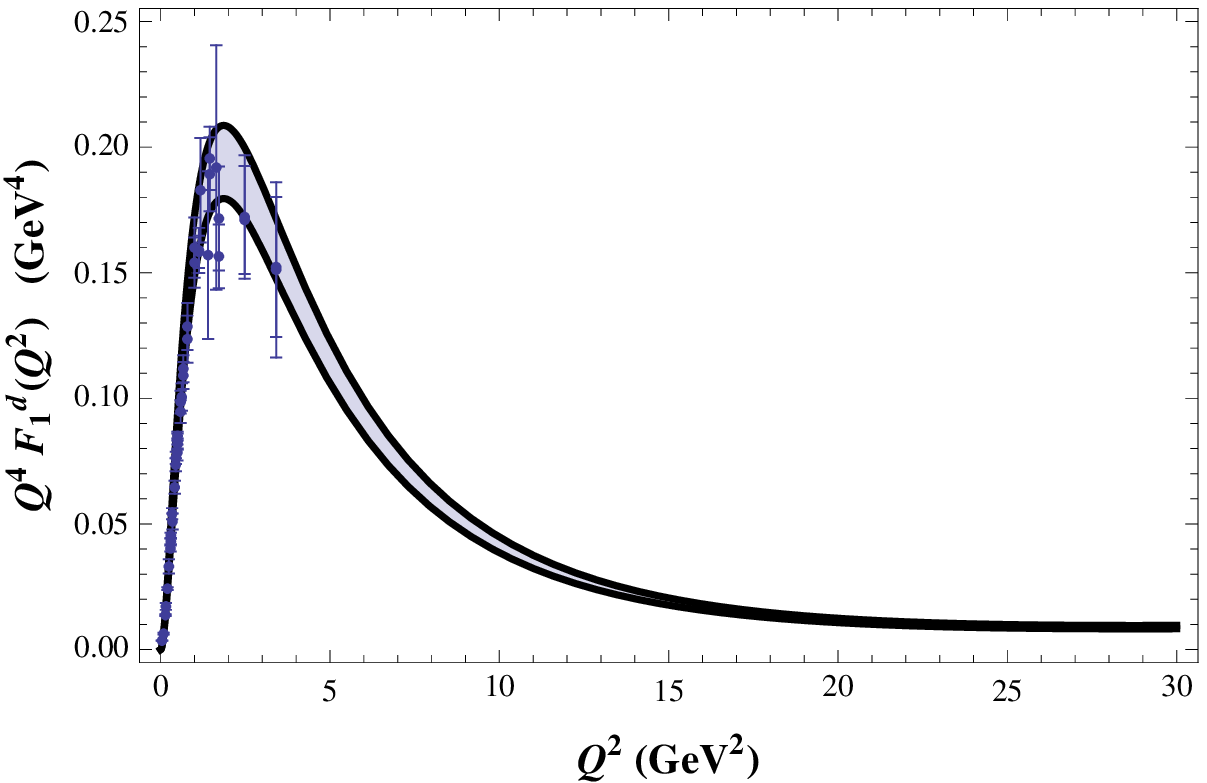,scale=.575}
\end{center}
\vspace*{-.6cm}
\noindent
\caption{Dirac $u$ and $d$ quark form factors.  
\label{fig1}}

\begin{center}
\epsfig{figure=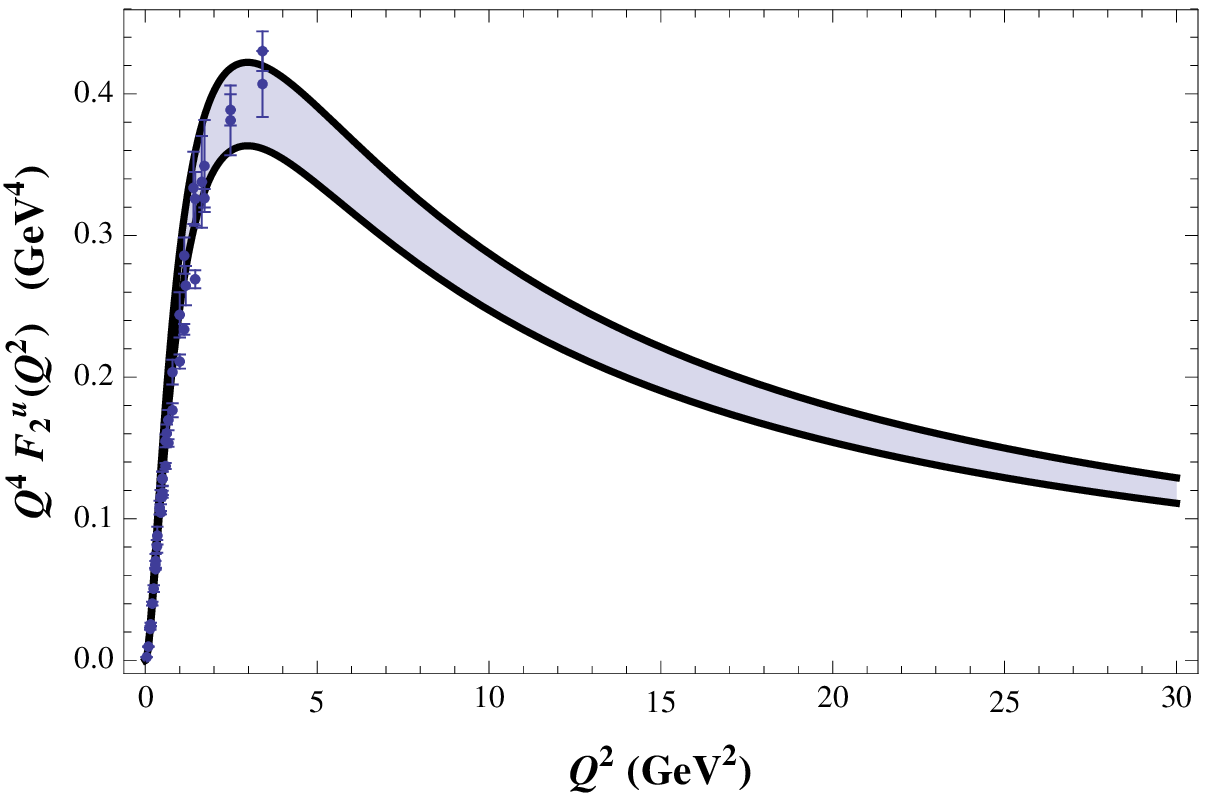,scale=.575}
\hspace*{1cm}
\epsfig{figure=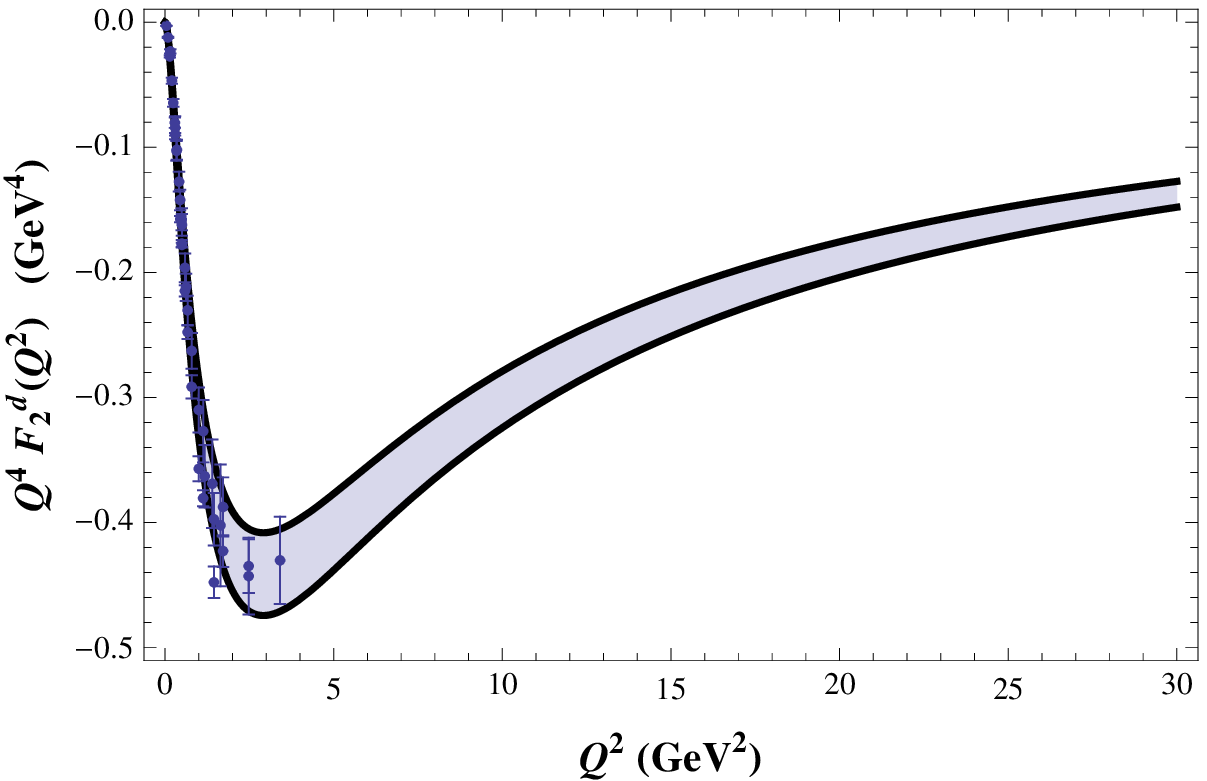,scale=.575}
\end{center}
\vspace*{-.6cm}
\noindent
\caption{Pauli $u$ and $d$ quark form factors.  
\label{fig2}} 

\begin{center}
\epsfig{figure=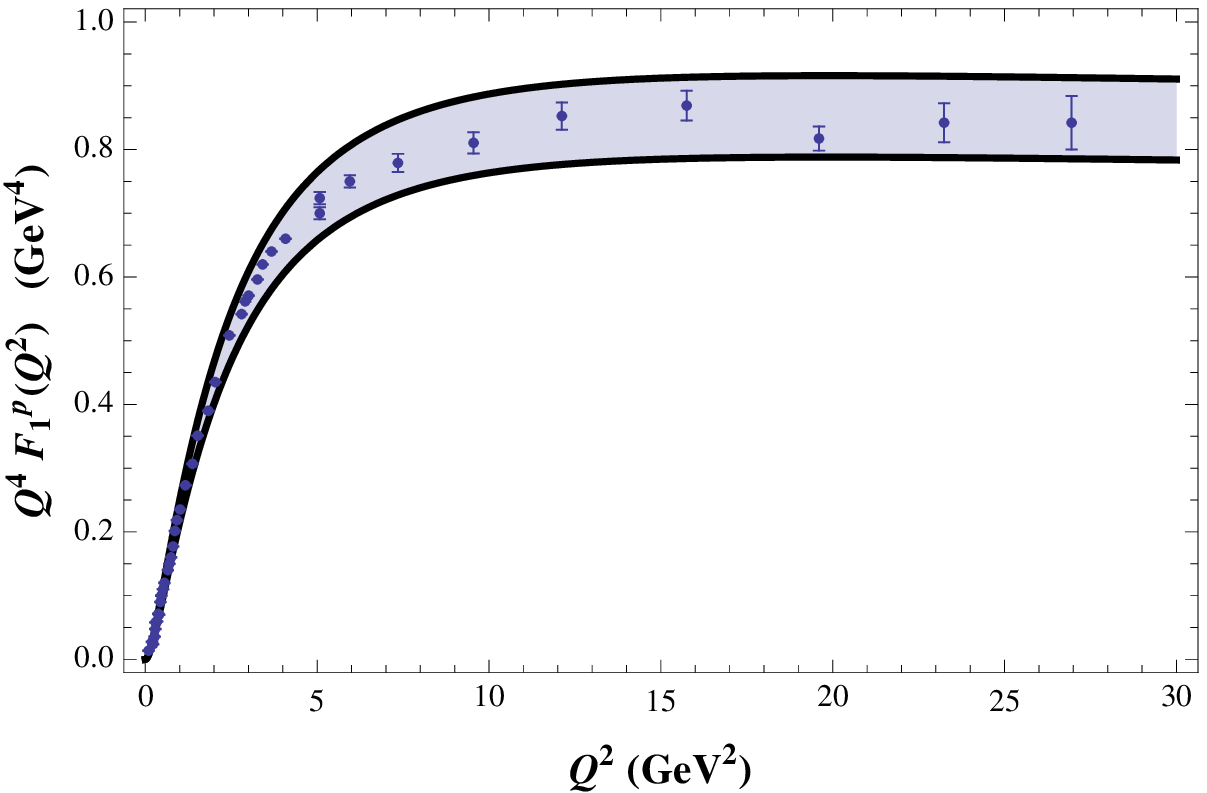,scale=.575}
\hspace*{1cm}
\epsfig{figure=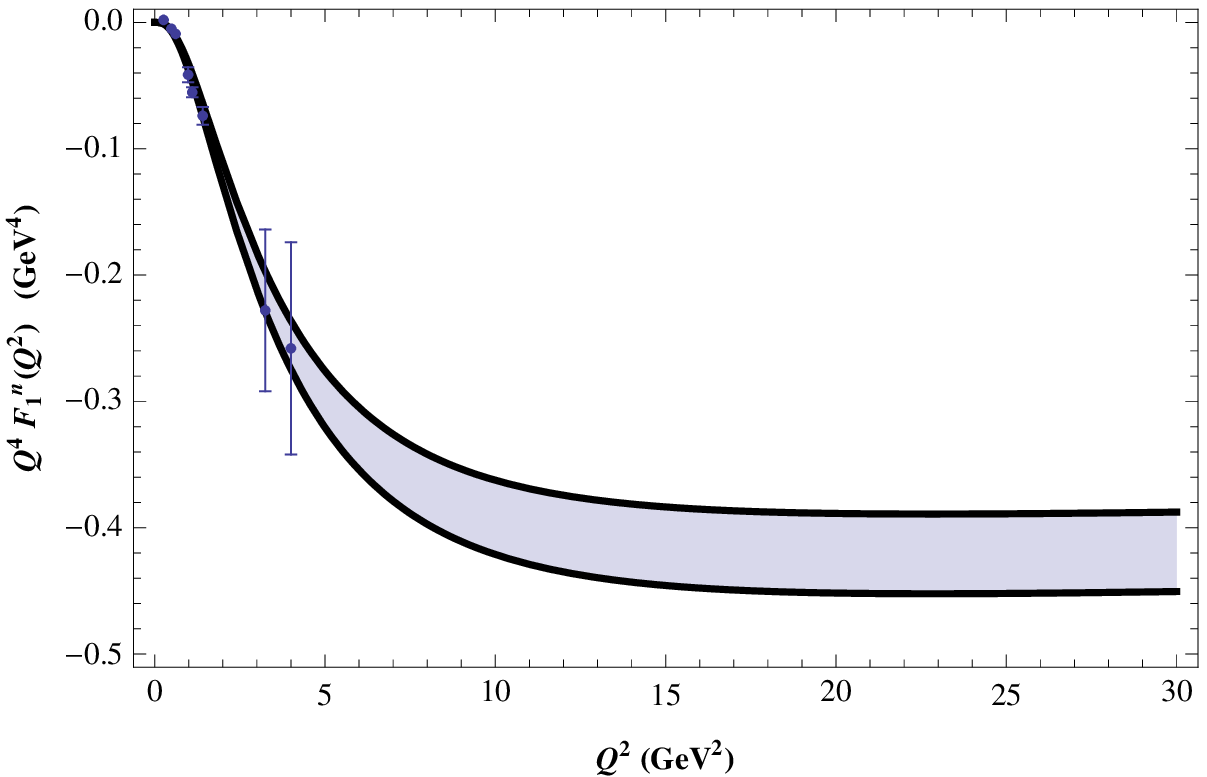,scale=.575}
\end{center}
\vspace*{-.6cm}
\noindent
\caption{Proton and neutron Dirac form factors.   
\label{fig3}}

\begin{center}
\epsfig{figure=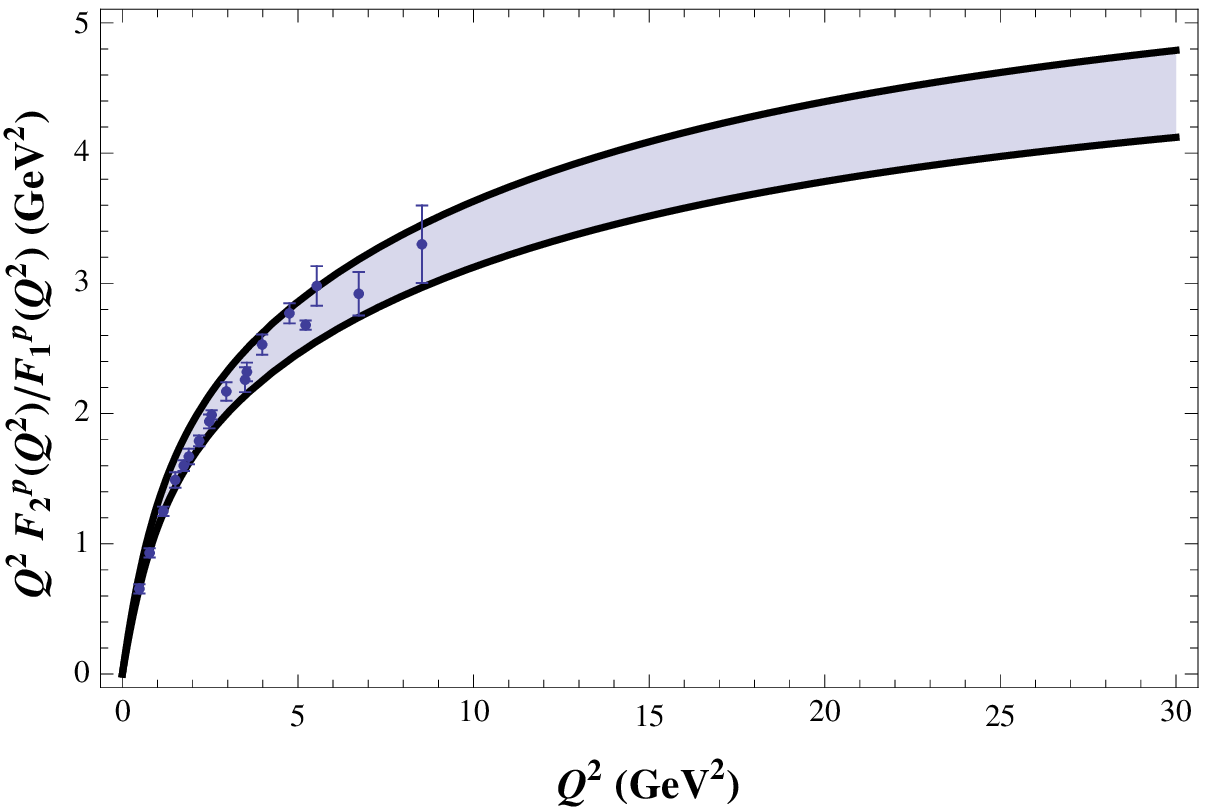,scale=.575}
\end{center}
\vspace*{-.6cm}
\noindent
\caption{Ratio $Q^2 F_2^p(Q^2)/F_1^p(Q^2)$ for proton.  
\label{fig4}}
\end{figure}

Note, that approach developed in Ref.~\cite{Bacchetta:2020vty} 
also used the NNPDF Collaboration input for the gluon PDFs. 
Therefore, it will be easy to compare the predictions of two 
approaches for gluon TMDs. Before we turn to discussion 
of the numerical results for the gluon TMDs we would like 
to present numerical results for the electromagnetic 
form factors of nucleons induced by the set of quark 
PDFs. All analytical formulas for quark densities were derived 
in our recent paper~\cite{Lyubovitskij:2020otz} in the same 
formalism (soft-wall AdS/QCD approach). 
In particular, using the GRV/GRSV world data analysis 
of the quark PDFs~\cite{Gluck:1998xa,Gluck:2000dy} 
we calculate the electromagnetic form factors of nucleons  
induced by valence quark contributions. 
The results are given in Figs.~\ref{fig1}-\ref{fig4}. 
In particular, in Figs.~\ref{fig1} and~\ref{fig2} we 
present the results for the Dirac and Pauli 
form factors of $u$ and $d$ quarks multiplied with $Q^4$ 
in the Euclidean region up to 30 GeV$^2$. Data points for the quark 
decomposition of the nucleon form factors are
taken from Refs.~\cite{Diehl:2013xca,Cates:2011pz}. 
In Figs.~\ref{fig1}-\ref{fig4} the shaded bands correspond to 
a variation of single model parameter $\kappa = 383 \pm 30$ MeV. 
In Figs.~\ref{fig3} and~\ref{fig4} we 
present the results for the Dirac and Pauli nucleon form factors 
multiplied with $Q^4$ and 
the ratio of the Pauli and Dirac proton form factors multiplied with 
$Q^2$ in the Euclidean region up to 30 GeV$^2$. 
Note that preliminary analysis of quark parton densities 
and quark/nucleon electromagnetic form factors 
was performed by us in Refs.~\cite{Gutsche:2014yea,Gutsche:2016gcd} 
using GRV/GRSV~\cite{Gluck:1998xa,Gluck:2000dy} and 
MSTW~\cite{Martin:2009iq} parametrizations for the quark PDFs as input 
for our analytical formulas. In particular, we did not find 
a sufficient preference of one or another parametrization 
for the quark PDFs. 

\begin{figure}
\begin{center}
\epsfig{figure=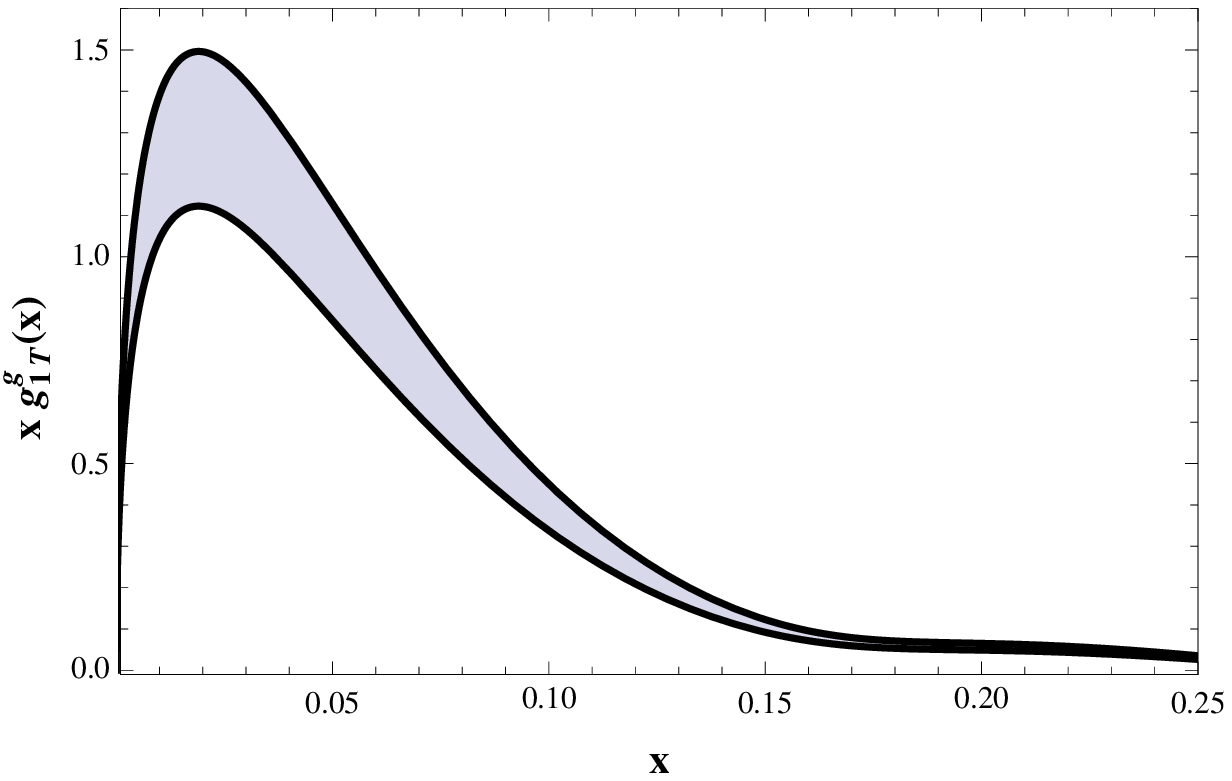,scale=.57}
\hspace*{1cm}
\epsfig{figure=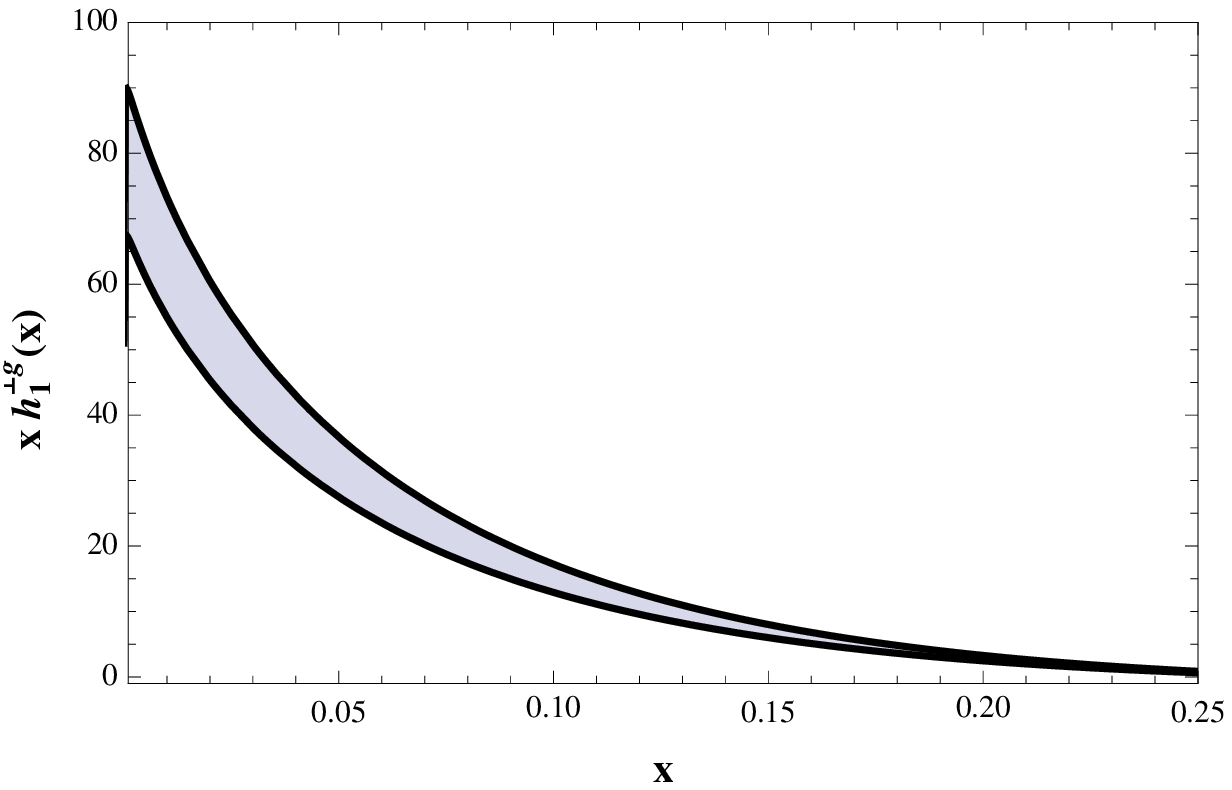,scale=.57}
\end{center}
\vspace*{-.6cm}
\noindent
\caption{Gluon PDFs $x g_{1T}^g(x)$ and $x h_1^{\perp g}(x)$.  
\label{fig5}}

\begin{center}
\epsfig{figure=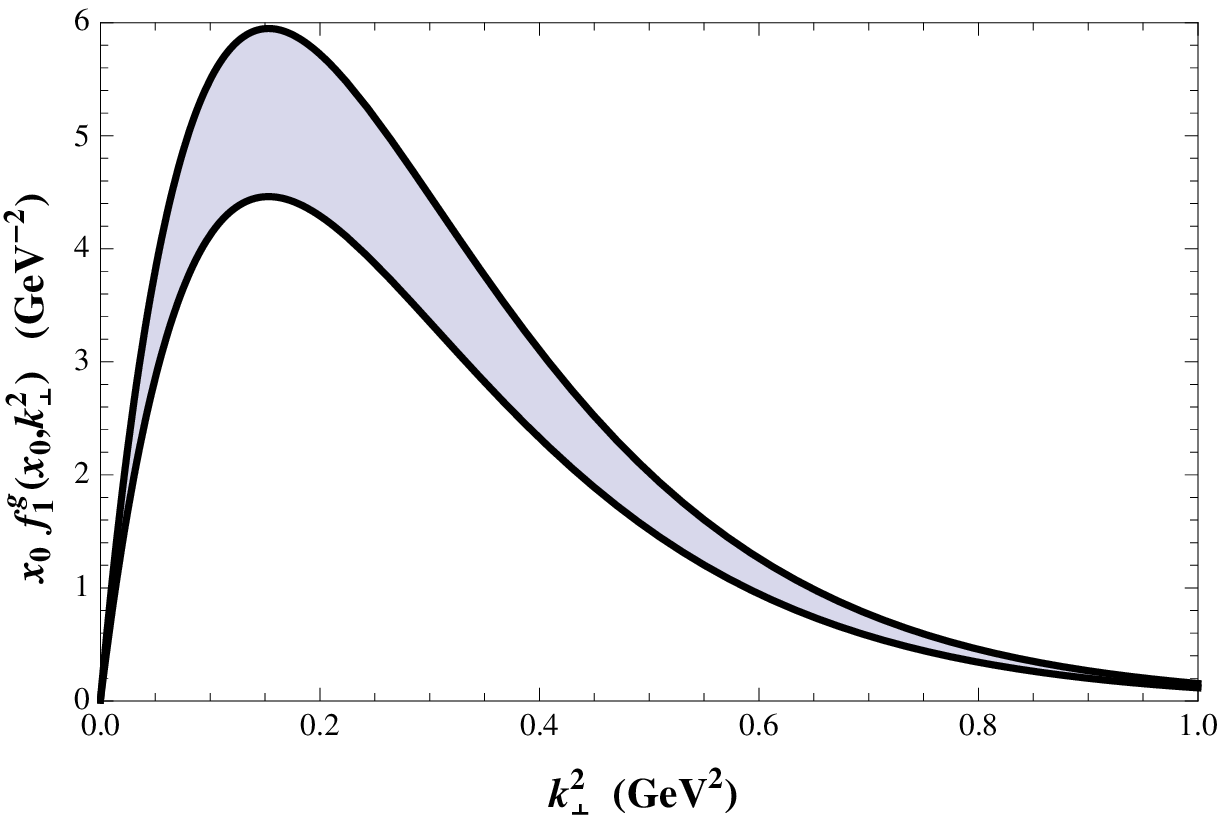,scale=.57}
\hspace*{1cm}
\epsfig{figure=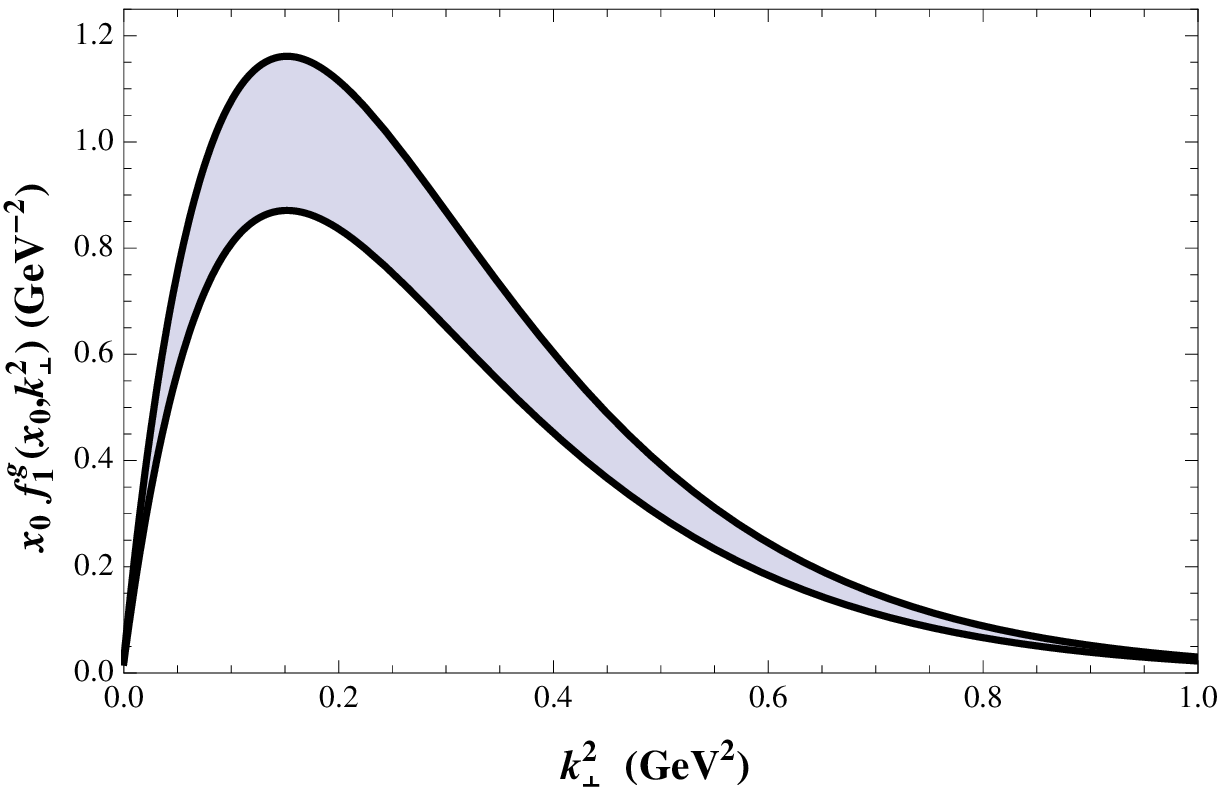,scale=.57}
\end{center}
\vspace*{-.32cm}
\noindent
\caption{Gluon TMD $x_0 f_1^g(x_0,\bfk^2)$ 
for $x_0=0.001$ (left panel) and 
for $x_0=0.1$ (right panel).  
\label{fig6}}

\begin{center}
\epsfig{figure=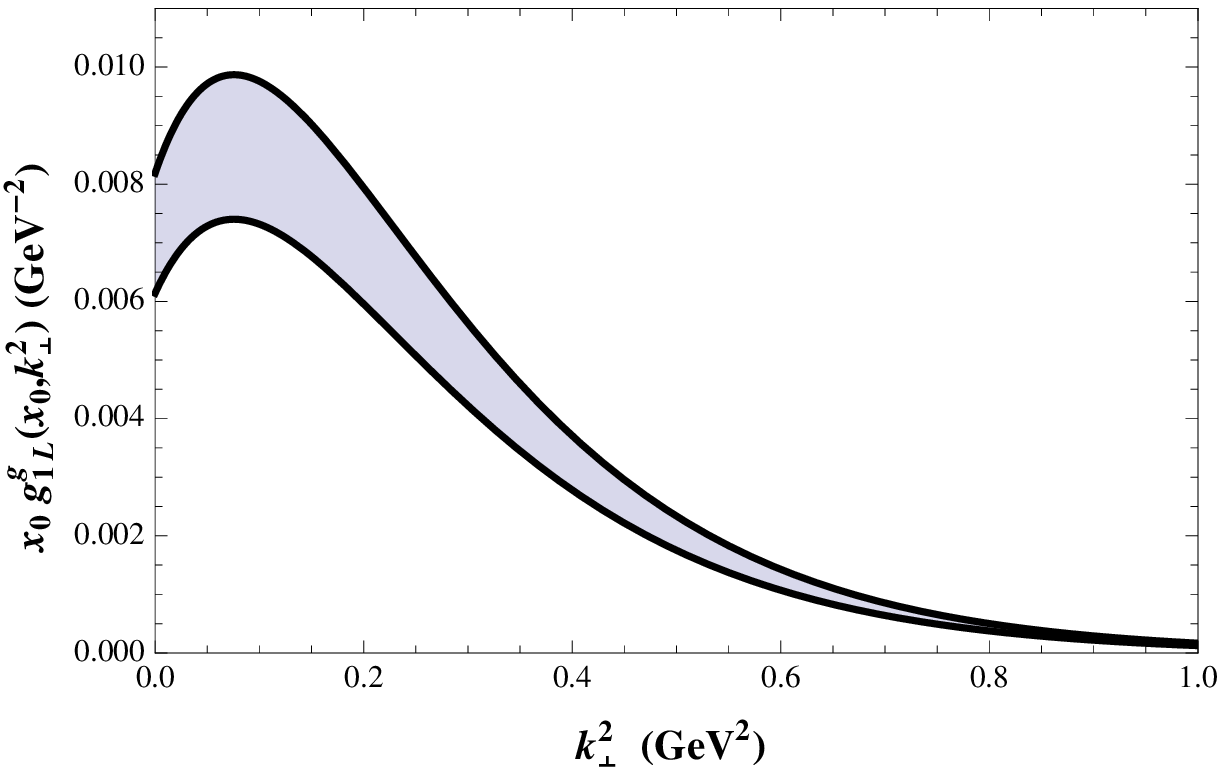,scale=.57}
\hspace*{1cm}
\epsfig{figure=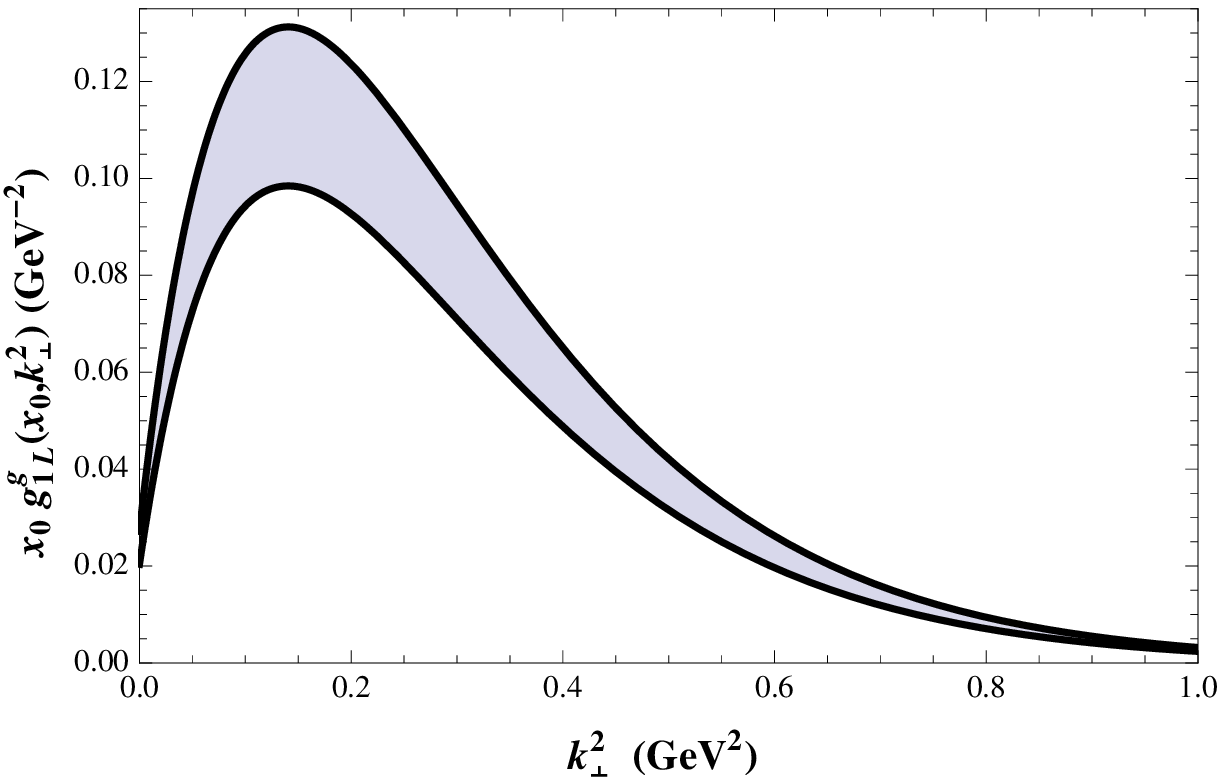,scale=.57}
\end{center}
\vspace*{-.32cm}
\noindent
\caption{Gluon TMD $x_0 g_{1L}^g(x_0,\bfk^2)$ 
for $x_0=0.001$ (left panel) and 
for $x_0=0.1$ (right panel).  
\label{fig7}}

\begin{center}
\epsfig{figure=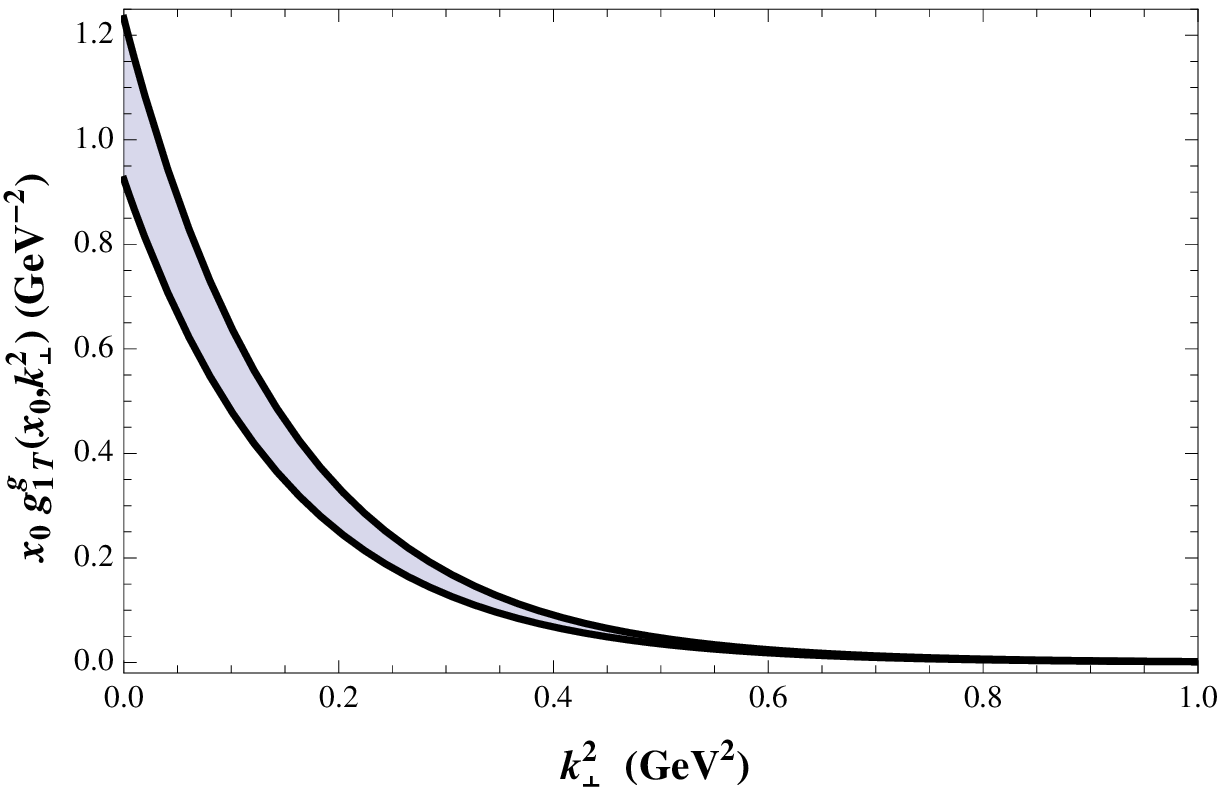,scale=.57}
\hspace*{1cm}
\epsfig{figure=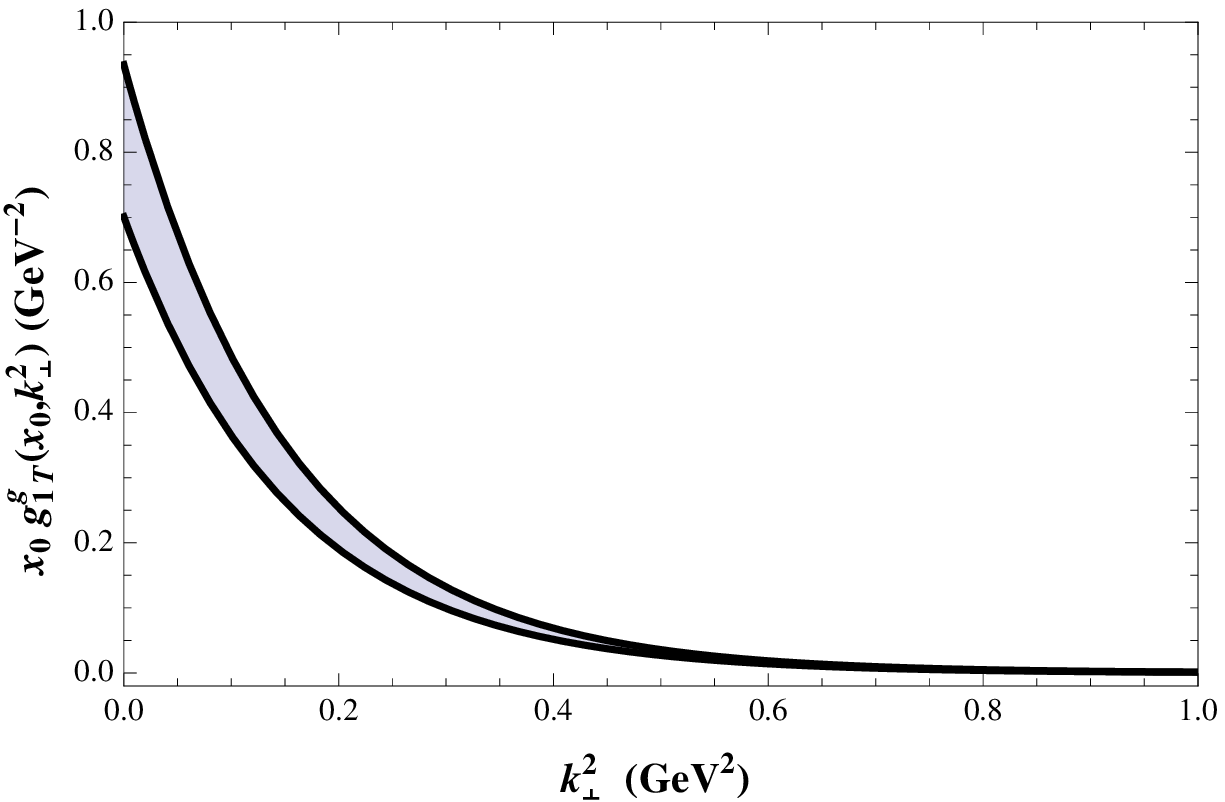,scale=.57}
\end{center}
\vspace*{-.32cm}
\noindent
\caption{Gluon TMD $x_0 g_{1T}^g(x_0,\bfk^2)$ 
for $x_0=0.001$ (left panel) and 
for $x_0=0.1$ (right panel).  
\label{fig8}}
\end{figure}

\begin{figure}
\begin{center}
\epsfig{figure=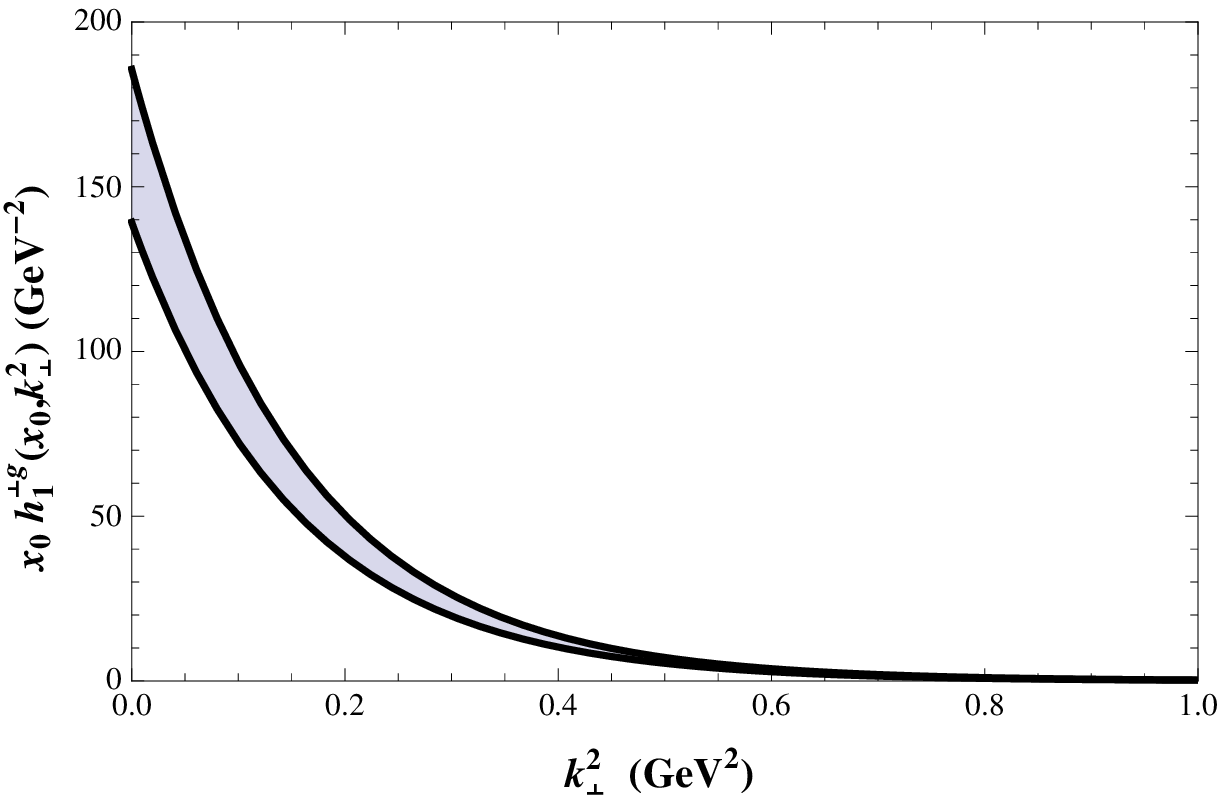,scale=.57}
\hspace*{1cm}
\epsfig{figure=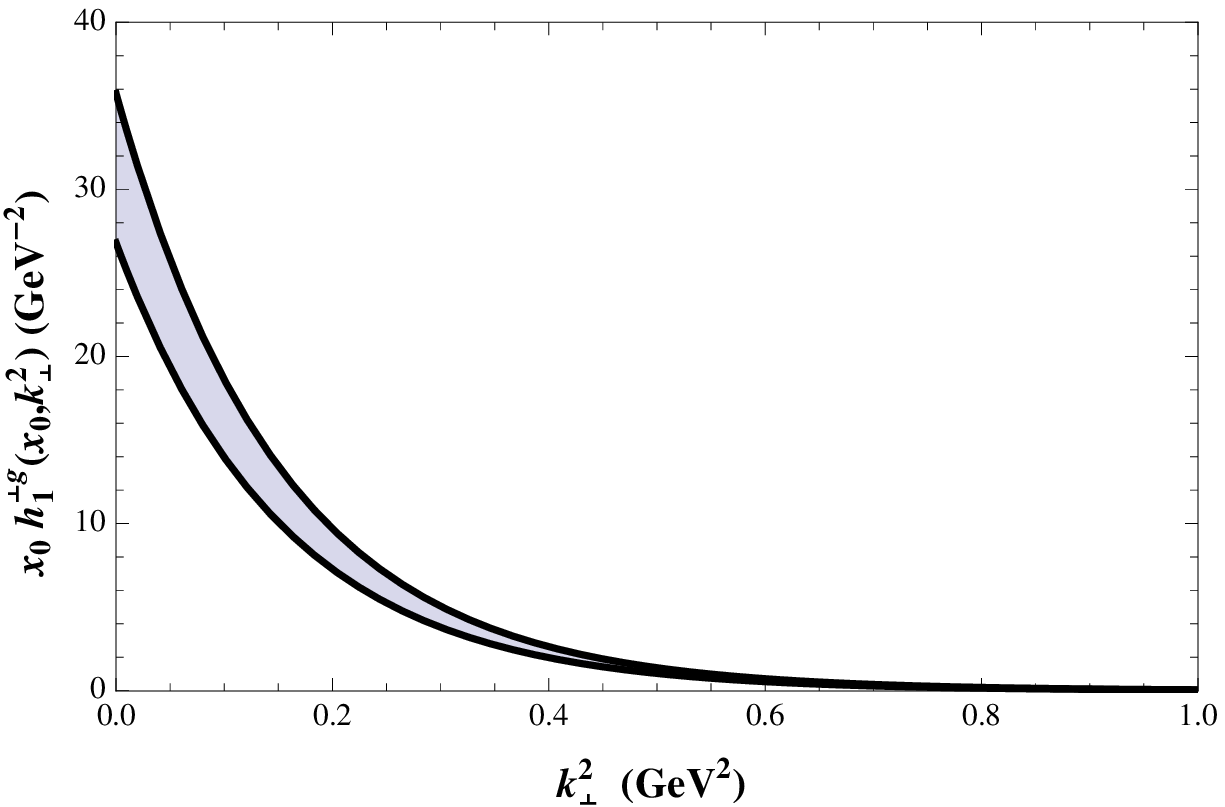,scale=.57}
\end{center}
\vspace*{-.3cm}
\noindent
\caption{Gluon TMD $x_0 h_1^{\perp g}(x_0,\bfk^2)$ 
for $x_0=0.001$ (left panel) and 
for $x_0=0.1$ (right panel).  
\label{fig9}}

\begin{center}
\epsfig{figure=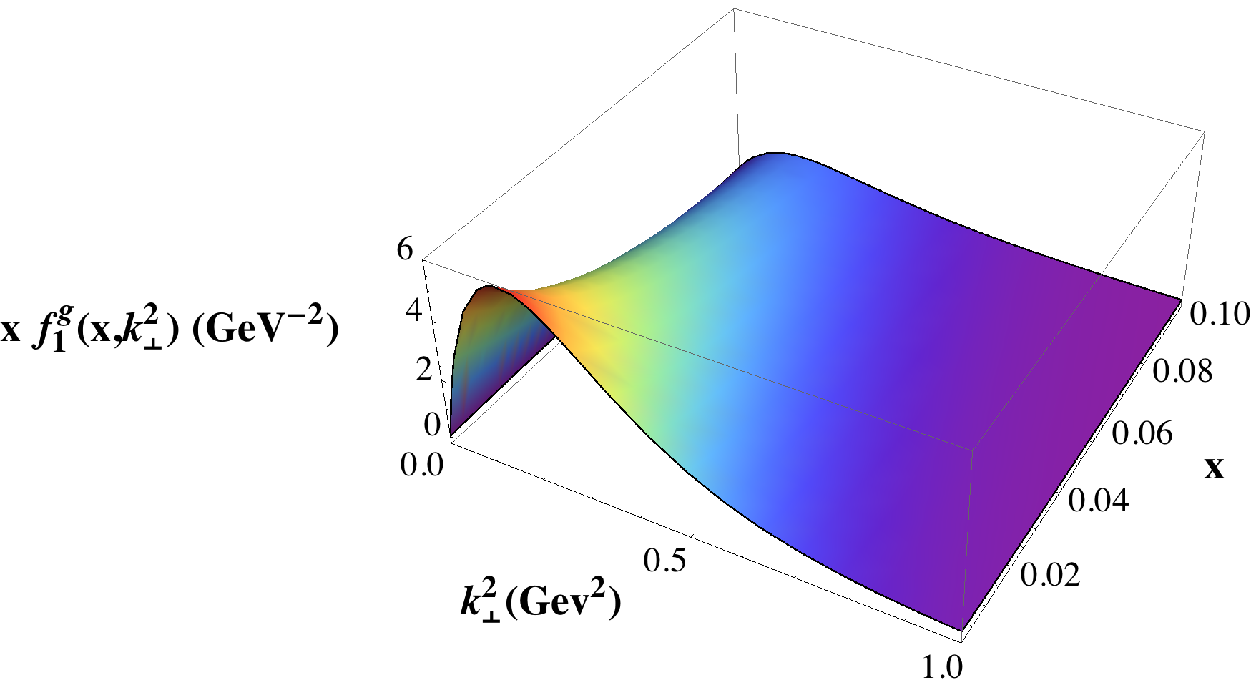,scale=.57}
\hspace*{1cm}
\epsfig{figure=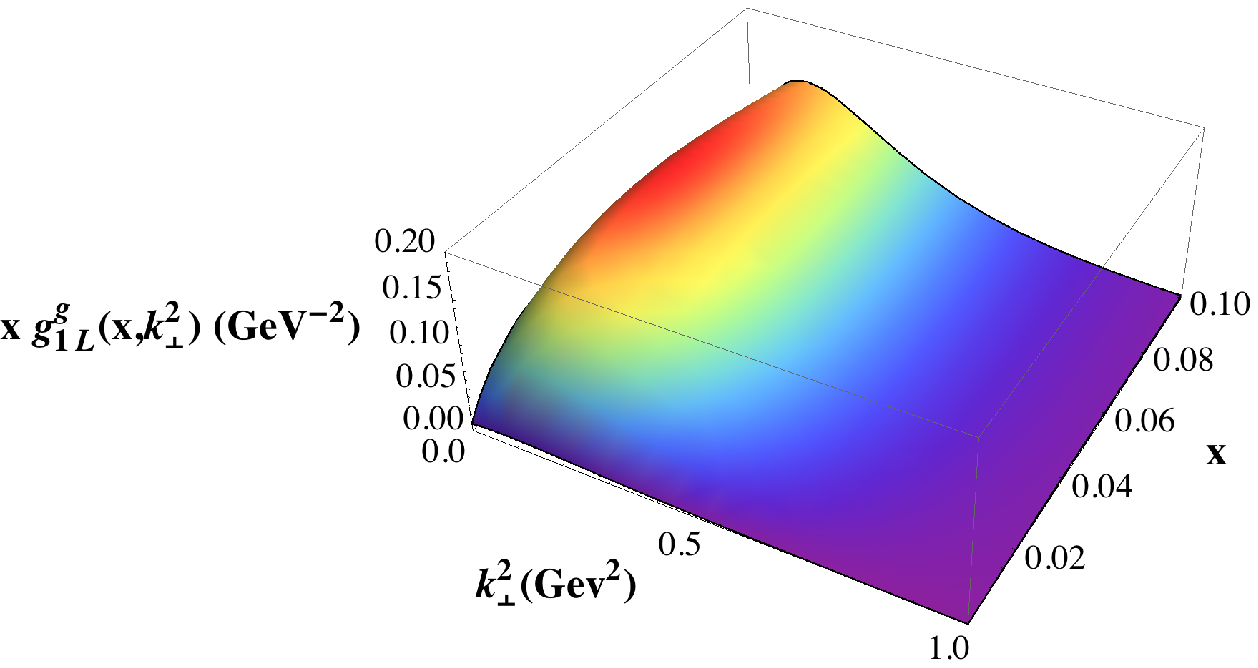,scale=.57}
\end{center}
\vspace*{-.3cm}
\noindent
\caption{3D plots of gluon TMDs $x f_1^g(x,\bfk^2)$ (left panel) 
and $x g_{1L}^g(x,\bfk^2)$ 
(right panel). 
\label{fig10}}

\begin{center}
\epsfig{figure=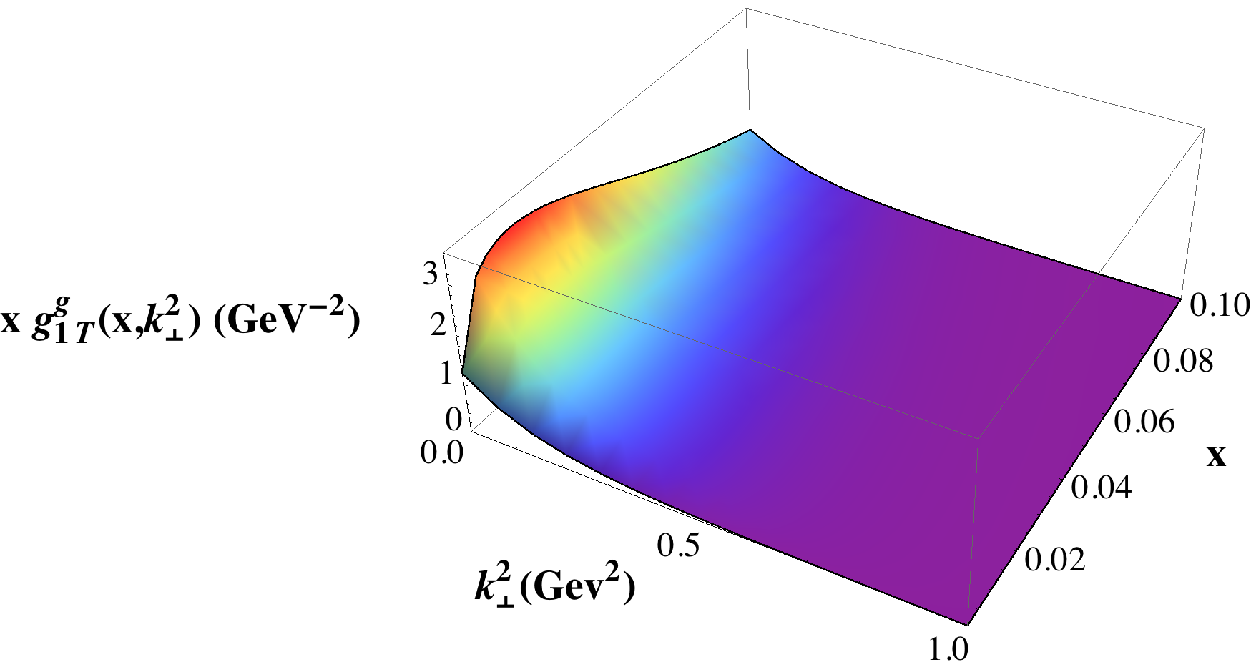,scale=.57}
\hspace*{1cm}
\epsfig{figure=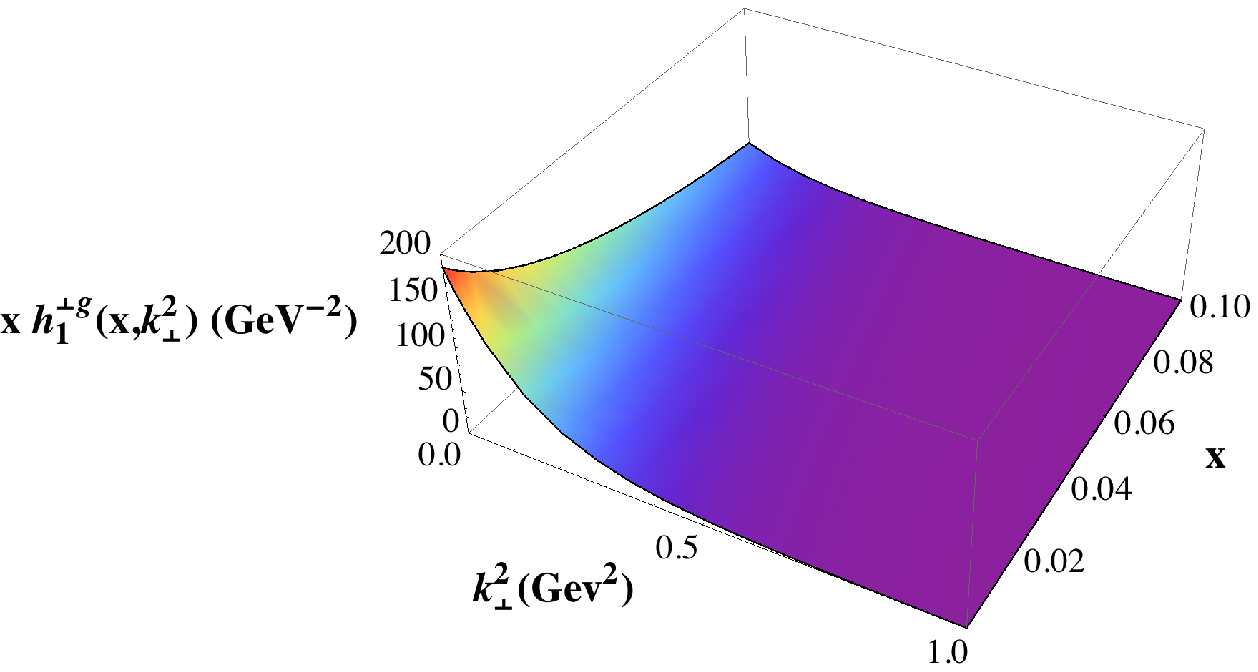,scale=.57}
\end{center}
\vspace*{-.3cm}
\noindent
\caption{3D plots of gluon TMDs $x g_{1T}^g(x,\bfk^2)$ (left panel) 
and $x h_1^{\perp g}(x,\bfk^2)$ 
(right panel). 
\label{fig11}}

\begin{center}
\epsfig{figure=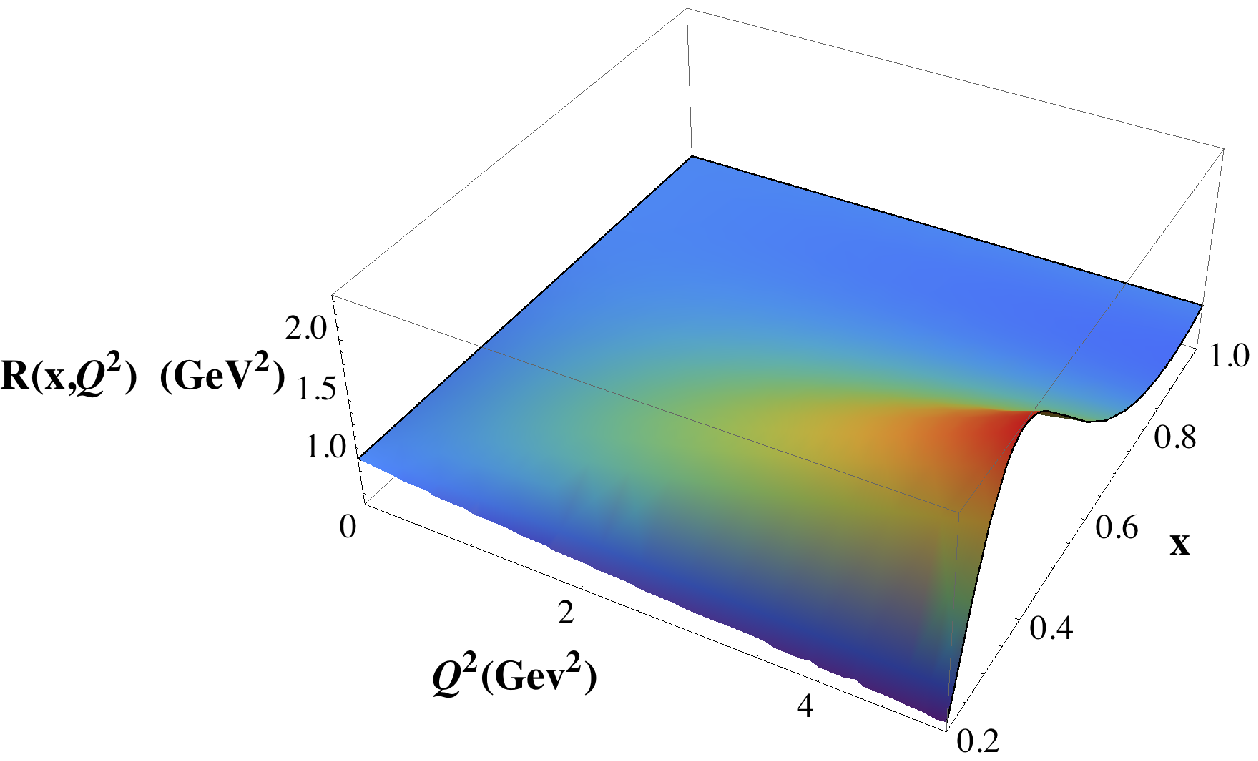,scale=.57}
\hspace*{1cm}
\epsfig{figure=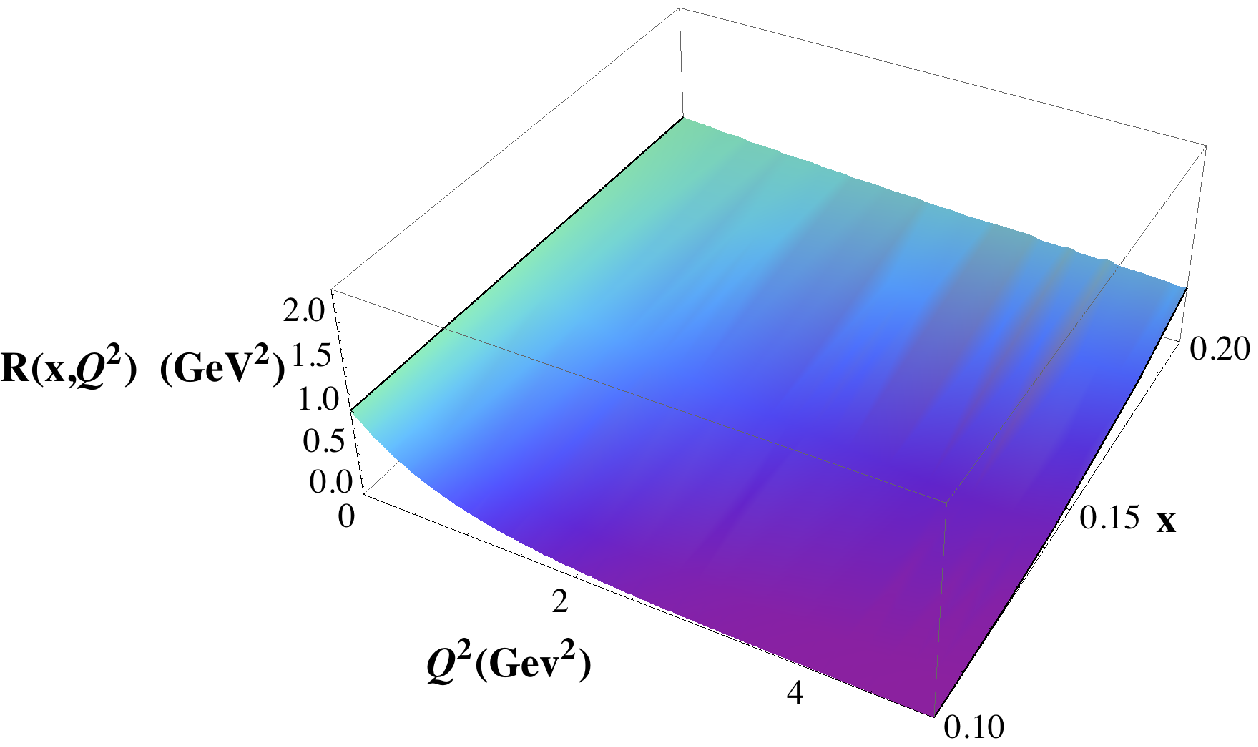,scale=.57}
\end{center}
\vspace*{-.3cm}
\noindent
\caption{3D plots of the ratio $R(x,Q^2)$ for gluon GPDs 
for $0.2 \le x \le 1$ (left panel) and $0.1 \le x \le 0.2$ (right panel).  
\label{fig12}}
\end{figure}

Now we are in the position to discuss numerical results for the T-even 
gluon TMDs and compare them with results of similar approach developed 
in Ref.~\cite{Bacchetta:2020vty}. First of all, for completeness 
in Fig.~\ref{fig5} we plot the $x g_{1T}^g(x)$ and $x h_1^{\perp g}(x)$ 
gluon PDFs. Here and in the following the shaded band corresponds to 
the variation of the parameters defining the set of 
the gluon PDFs~(\ref{NNPDF_data}) fixed from world data analysis 
by the NNPDF Collaboration in Refs.~\cite{Ball:2017nwa,Nocera:2014gqa}.  
The results for the T-even gluon TMDs are presented in
Figs.~\ref{fig6}-\ref{fig11}. In particular, 
in Figs.~\ref{fig6}-\ref{fig9} we display the results 
when $x_0=0.001$ (left panels) and $x_0=0.1$ at $\bfk^2$ varied 
from 0 to 1 GeV$^2$. One should stress that all results 
(except prediction for $x_0 g_{1T}^g(x_0,\bfk^2)$ at $x_0=0.001$) in 
very good agreement with results of Ref.~\cite{Bacchetta:2020vty}. 
One can suppose that the difference of two approaches for plot in Fig.~\ref{fig8}(a)  
can be influenced by a use of an admixutre of minimal and nonminimal
couplings of gluon with three-quark core in Ref.~\cite{Bacchetta:2020vty}. 
In our case the results for all four TMDs are cross-checked by 
Mulders-Rodrigues positivity bounds~\cite{Mulders:2000sh} and 
using the new sum rule~(\ref{SR}) between four T-even gluon TMDs 
derived in this paper for the first time in literature. 
Also we note that we use the sign convention for the $g_{1T}(x,\bfk^2)$ TMD 
as in Ref.~\cite{Mulders:2000sh}, while Ref.~\cite{Bacchetta:2020vty} 
uses opposite sign. 

\begin{figure}
\begin{center}
\epsfig{figure=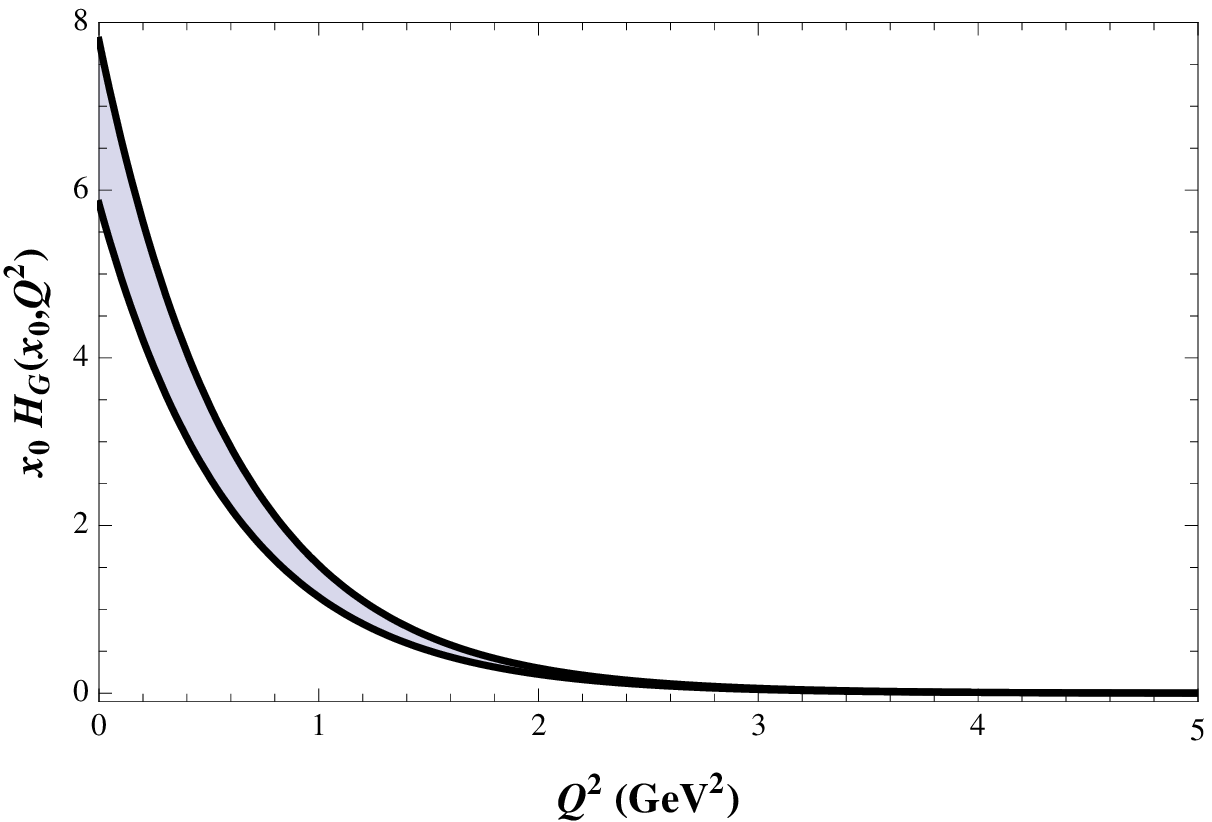,scale=.57}
\hspace*{1cm}
\epsfig{figure=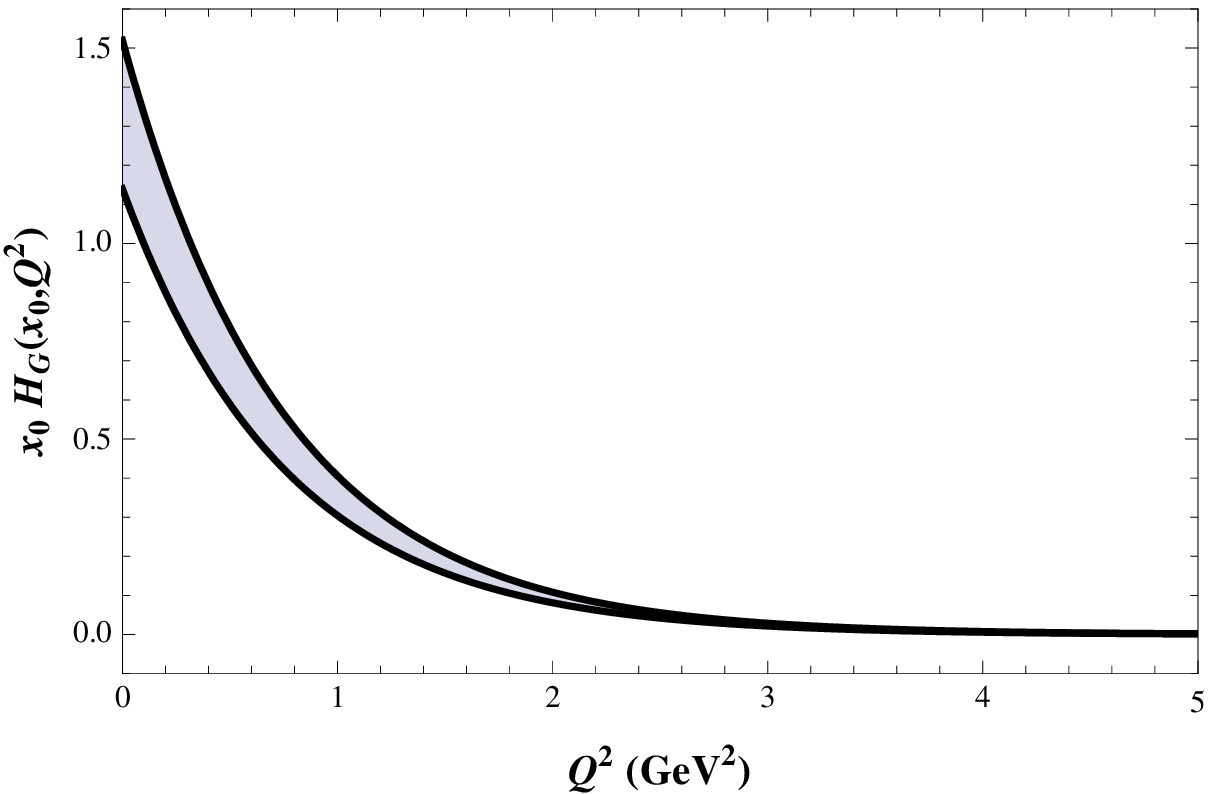,scale=.57}
\end{center}
\vspace*{-.452cm}
\noindent
\caption{$Q^2$ dependence of the gluon GPD 
$x_0 {\cal H}_G(x_0,Q^2)$ 
for $x_0=0.001$ (left panel) and 
for $x_0=0.1$ (right panel).  
\label{fig13}}

\begin{center}
\epsfig{figure=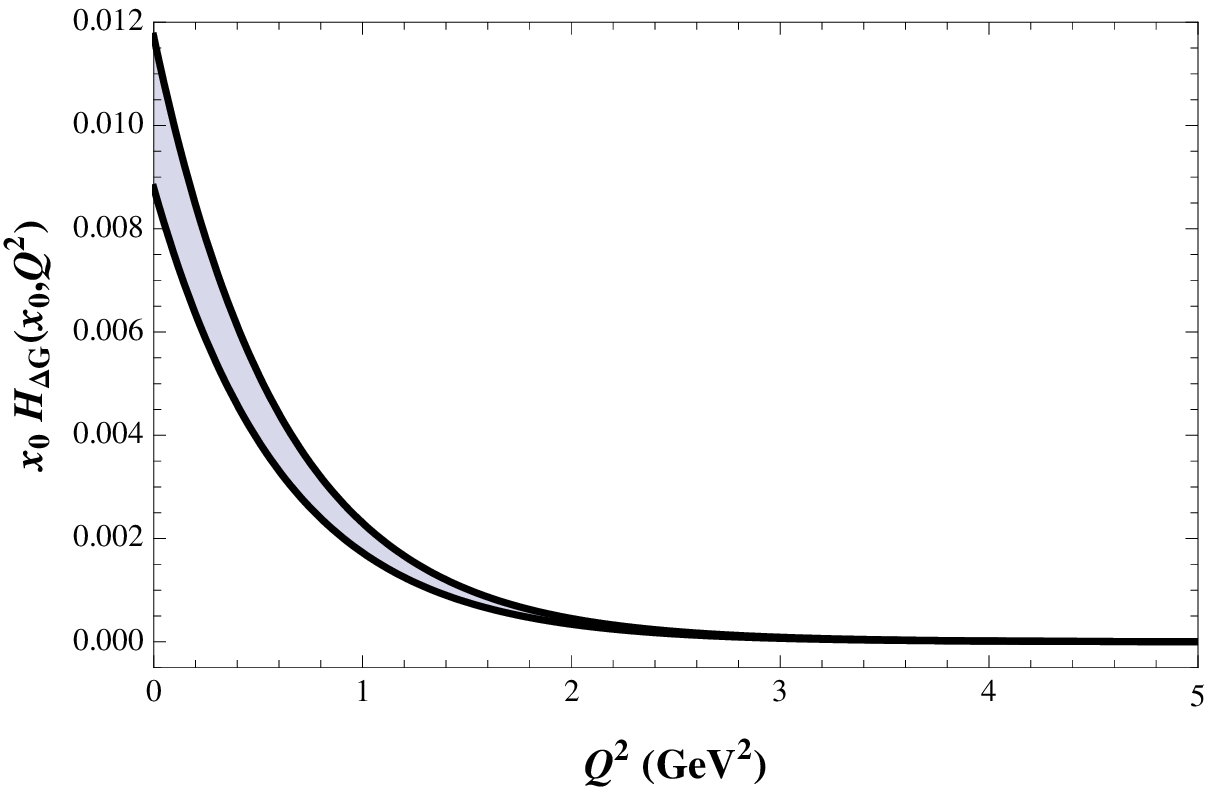,scale=.57}
\hspace*{1cm}
\epsfig{figure=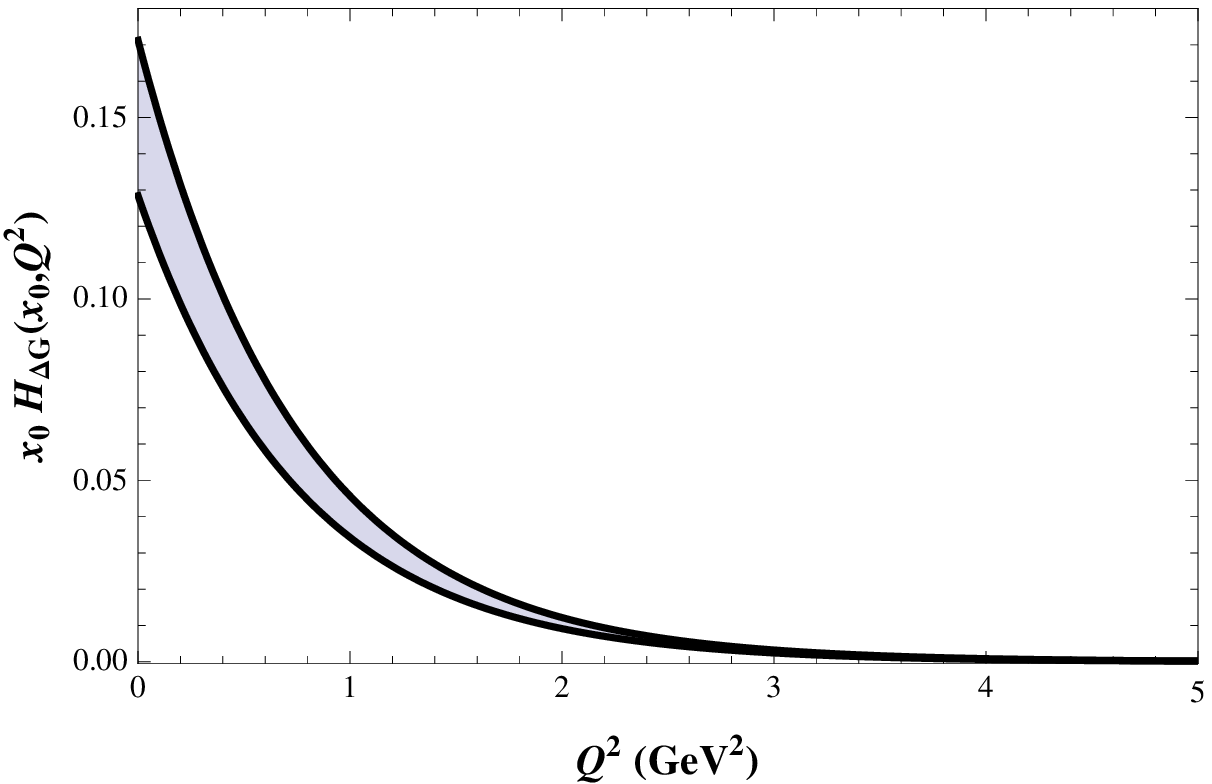,scale=.57}
\end{center}
\vspace*{-.452cm}
\noindent
\caption{$Q^2$ dependence of the gluon GPD 
$x_0 {\cal H}_{\Delta G}(x_0,Q^2)$ 
for $x_0=0.001$ (left panel) and 
for $x_0=0.1$ (right panel).  
\label{fig14}}

\begin{center}
\epsfig{figure=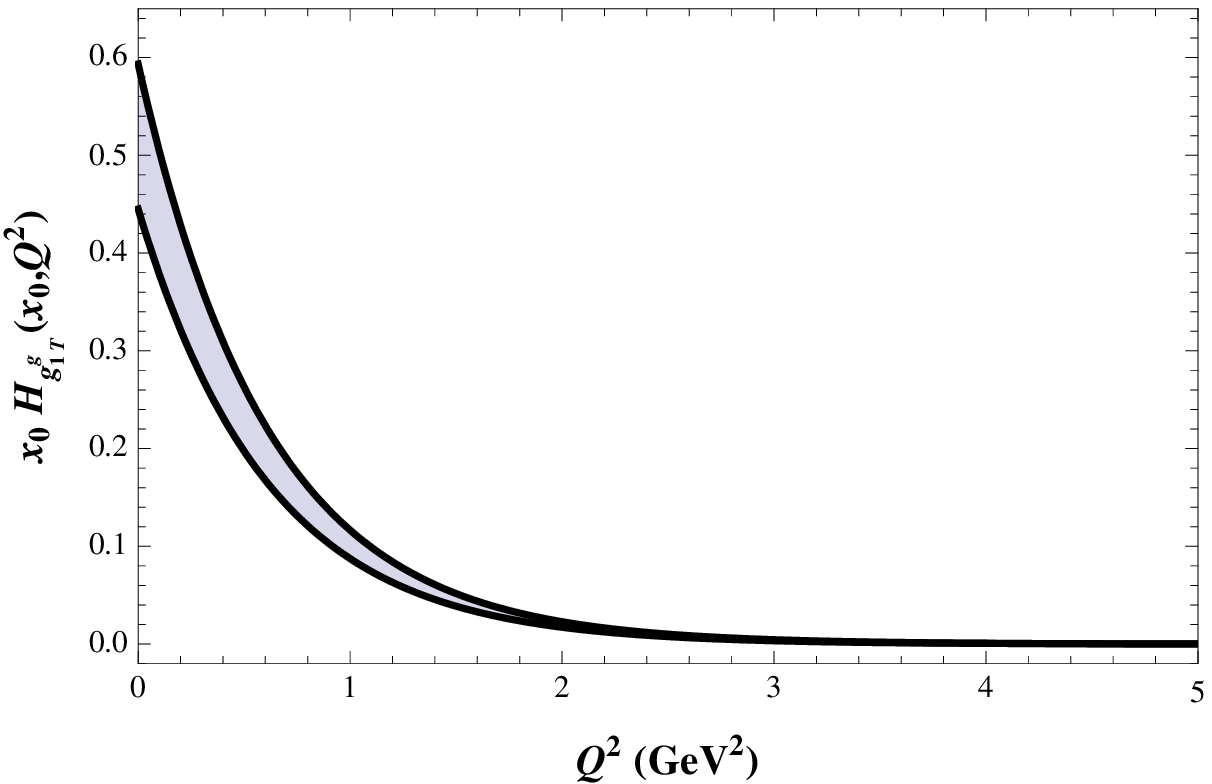,scale=.57}
\hspace*{1cm}
\epsfig{figure=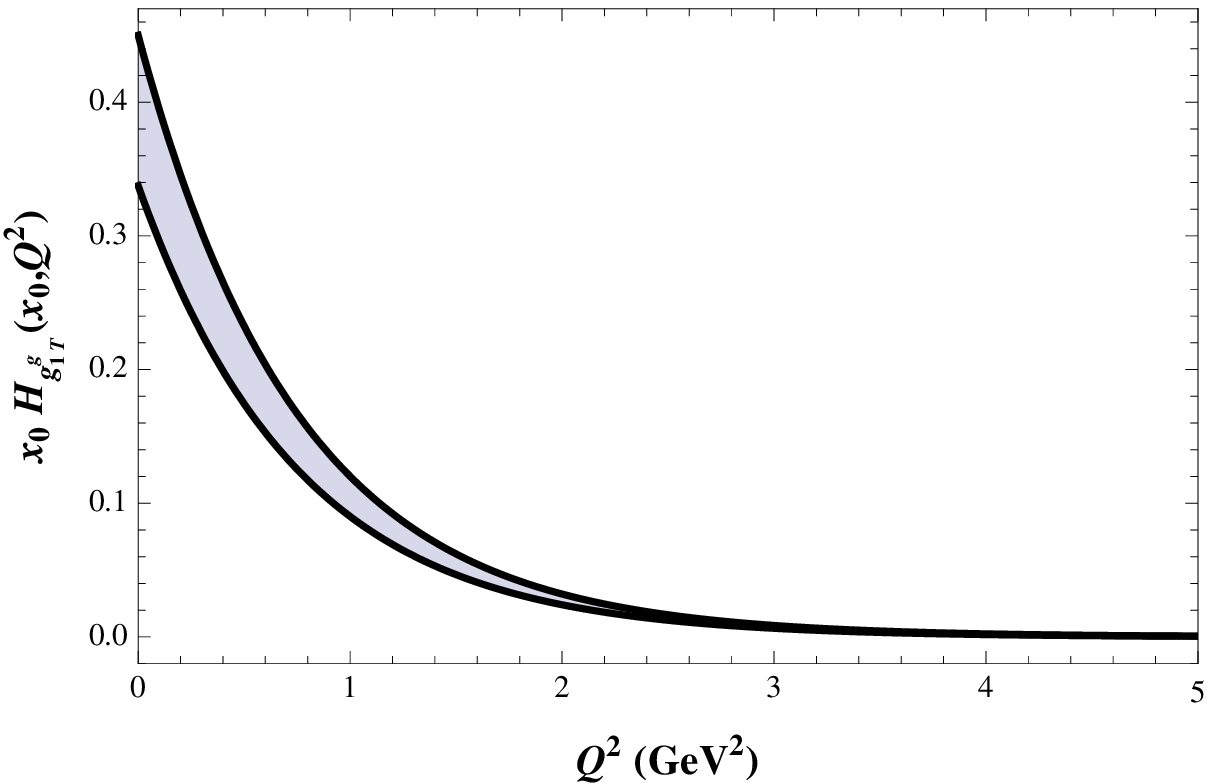,scale=.57}
\end{center}
\vspace*{-.452cm}
\noindent
\caption{$Q^2$ dependence of the gluon GPD 
$x_0 {\cal H}_{g_{1T}^g}(x_0,Q^2)$ 
for $x_0=0.001$ (left panel) and 
for $x_0=0.1$ (right panel).  
\label{fig15}}

\begin{center}
\epsfig{figure=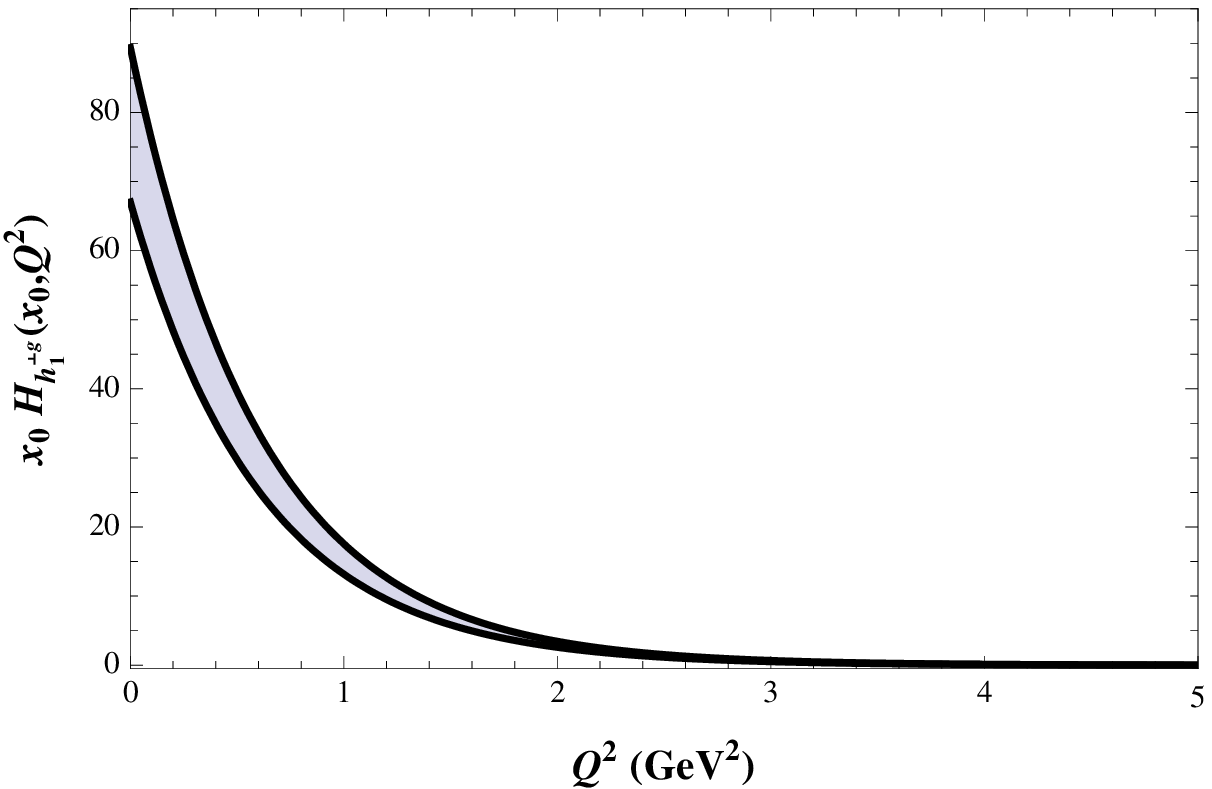,scale=.57}
\hspace*{1cm}
\epsfig{figure=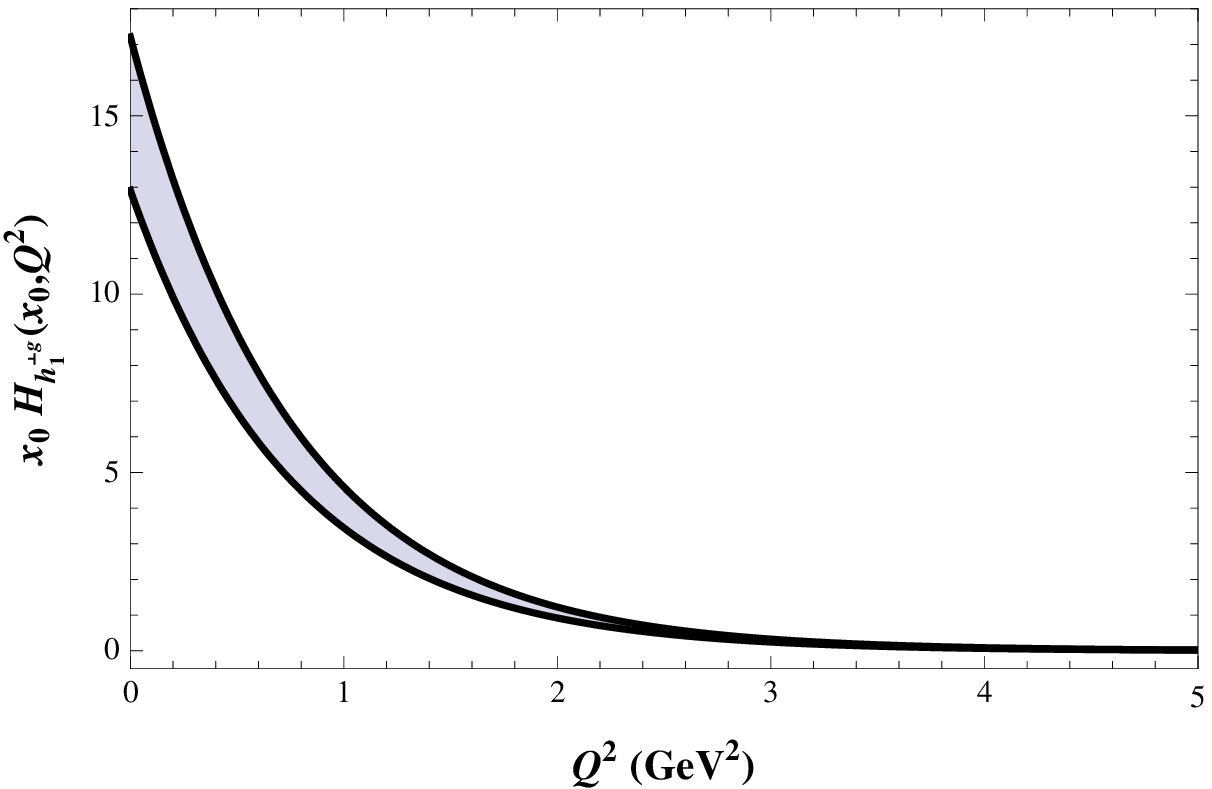,scale=.57}
\end{center}
\vspace*{-.452cm}
\noindent
\caption{$Q^2$ dependence of the gluon GPD 
$x_0 {\cal H}_{f_1^{\perp g}}(x_0,Q^2)$ 
for $x_0=0.001$ (left panel) and 
for $x_0=0.1$ (right panel).  
\label{fig16}}
\end{figure}

\begin{figure}
\begin{center}
\epsfig{figure=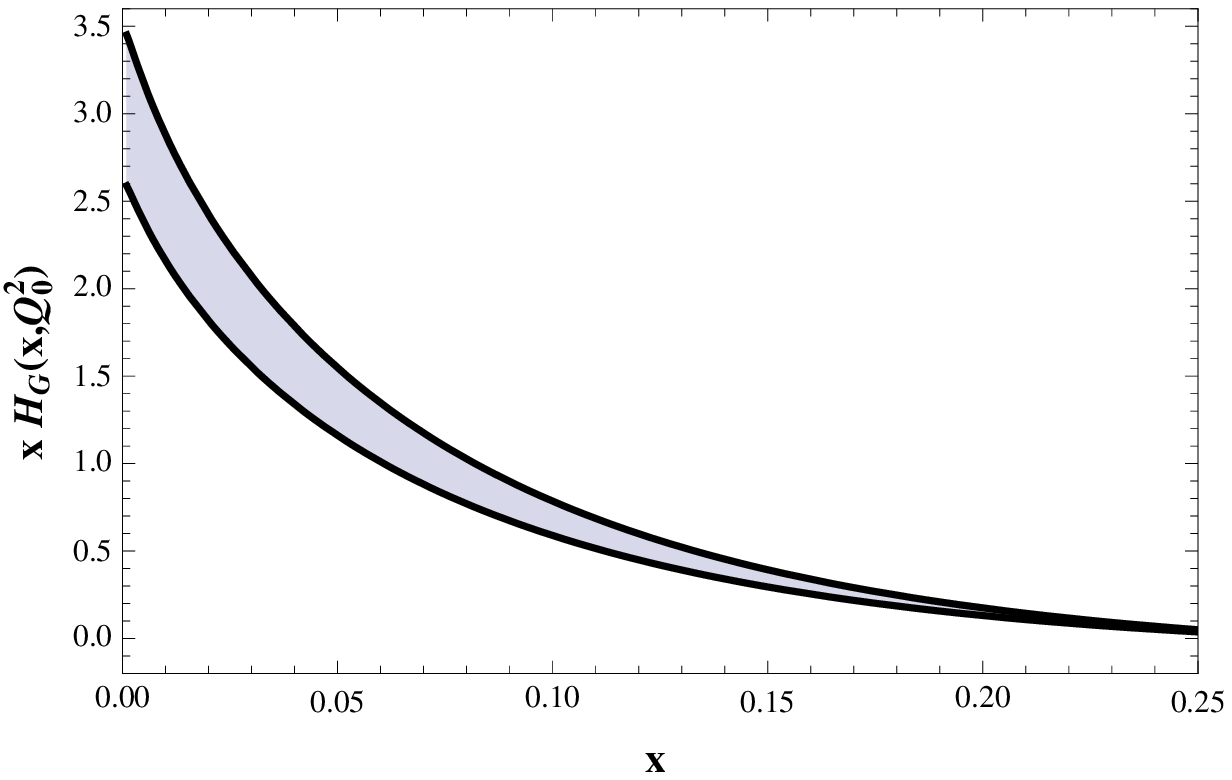,scale=.57}
\hspace*{1cm}
\epsfig{figure=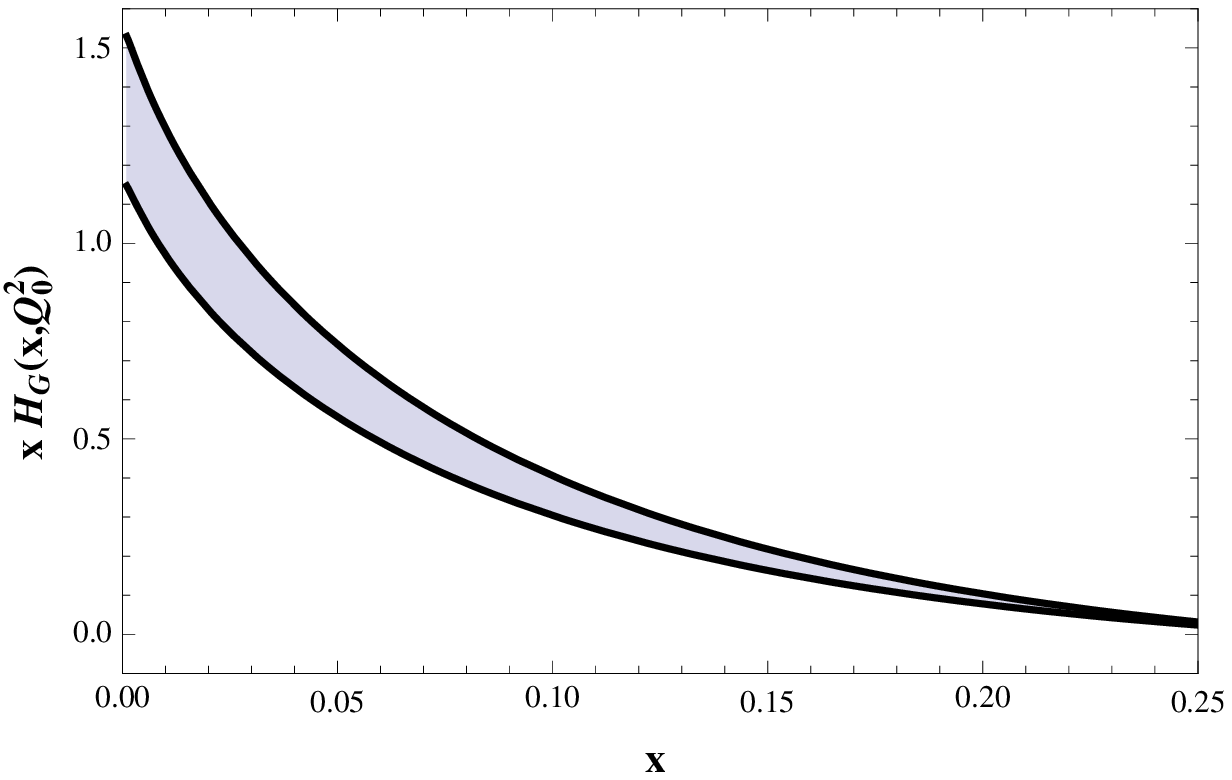,scale=.57}
\end{center}
\vspace*{-.4cm}
\noindent
\caption{$x$ dependence of the gluon GPD 
$x {\cal H}_G(x,Q_0^2)$ 
for $Q_0^2=0.5$ GeV$^2$ (left panel) and 
for $Q_0^2=1$ GeV$^2$   (right panel).  
\label{fig17}}

\begin{center}
\epsfig{figure=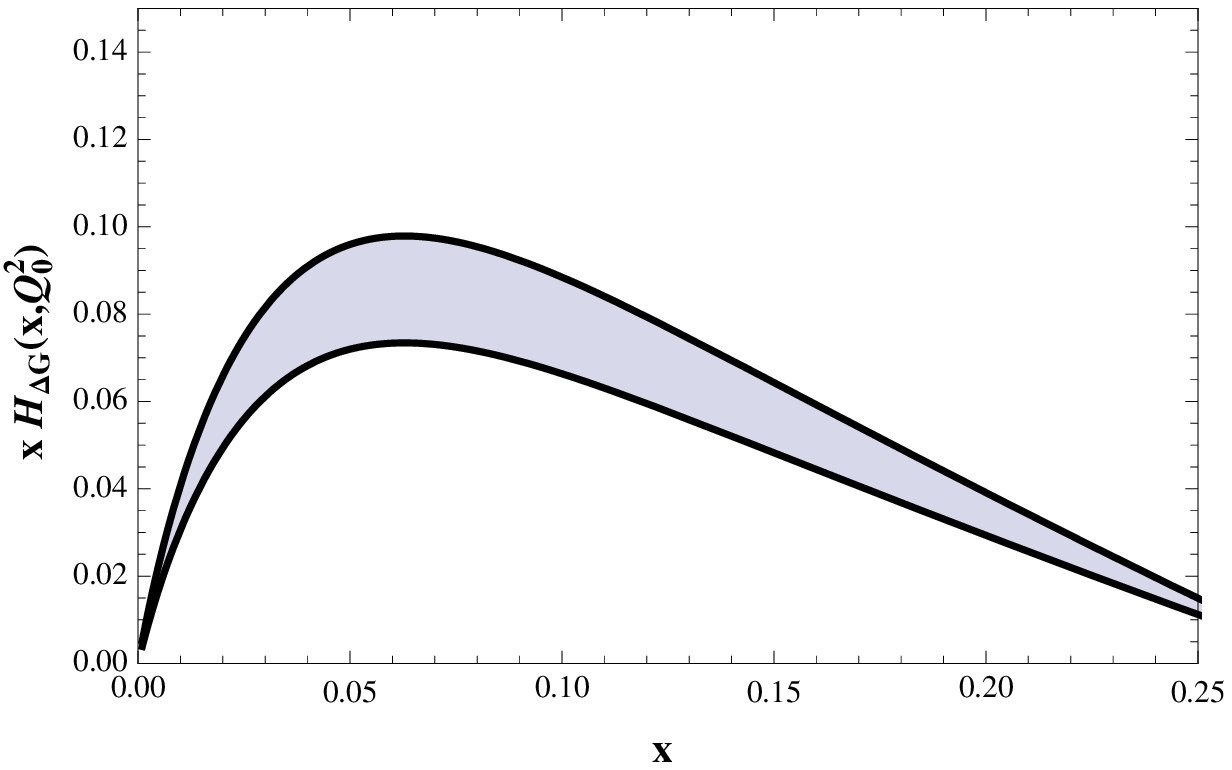,scale=.57}
\hspace*{1cm}
\epsfig{figure=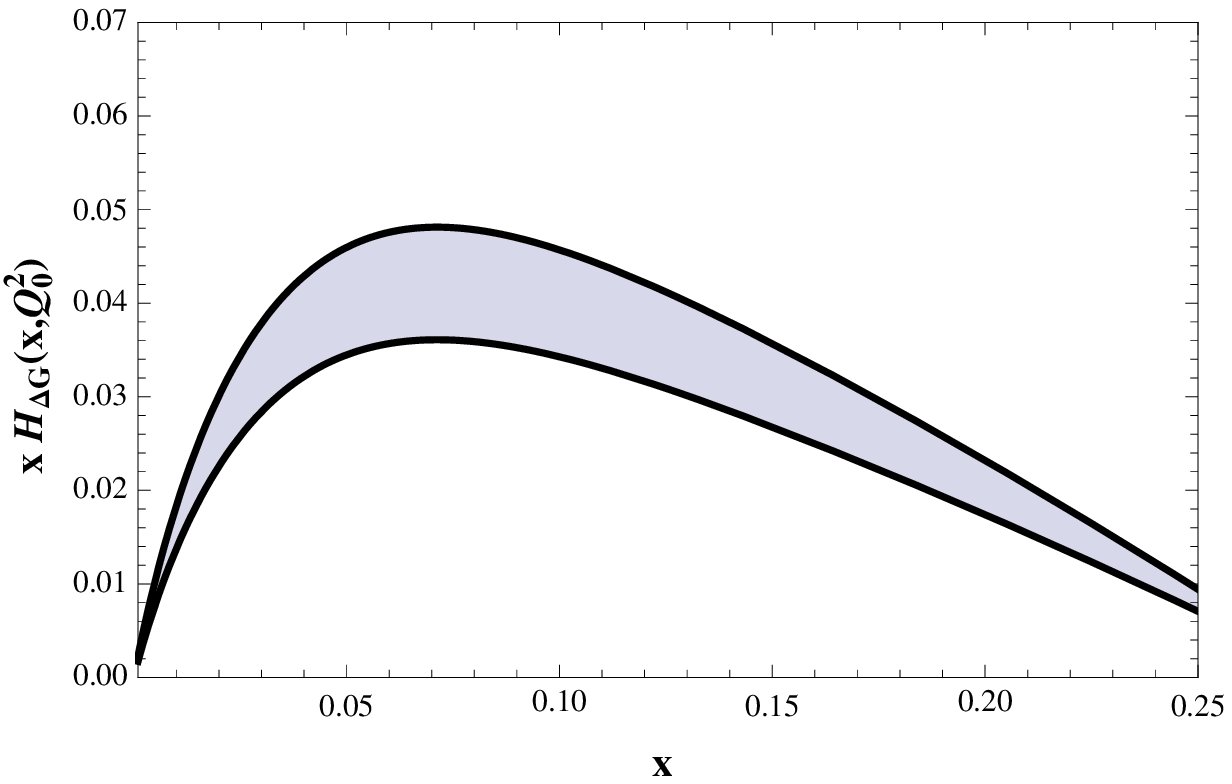,scale=.57}
\end{center}
\vspace*{-.4cm}
\noindent
\caption{$x$ dependence of the gluon GPD 
$x {\cal H}_{\Delta G}(x,Q_0^2)$ 
for $Q_0^2=0.5$ GeV$^2$ (left panel) and 
for $Q_0^2=1$ GeV$^2$   (right panel).  
\label{fig18}}

\begin{center}
\epsfig{figure=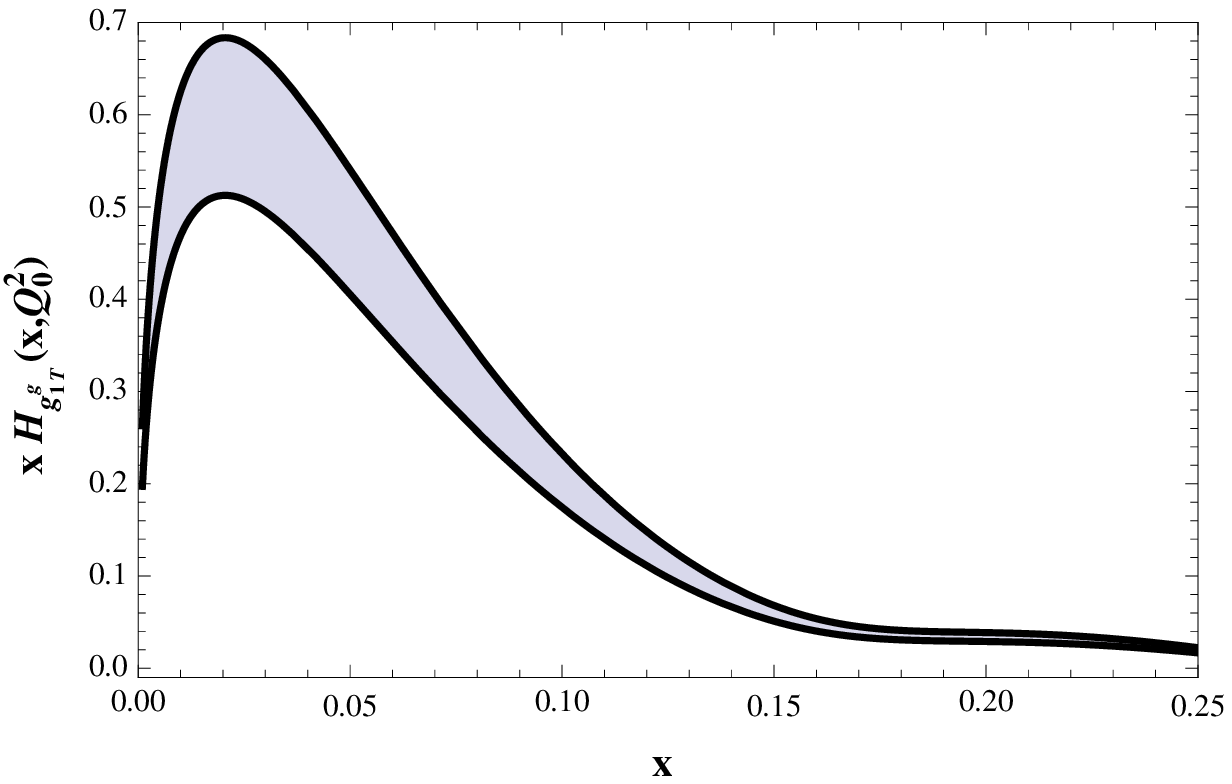,scale=.57}
\hspace*{1cm}
\epsfig{figure=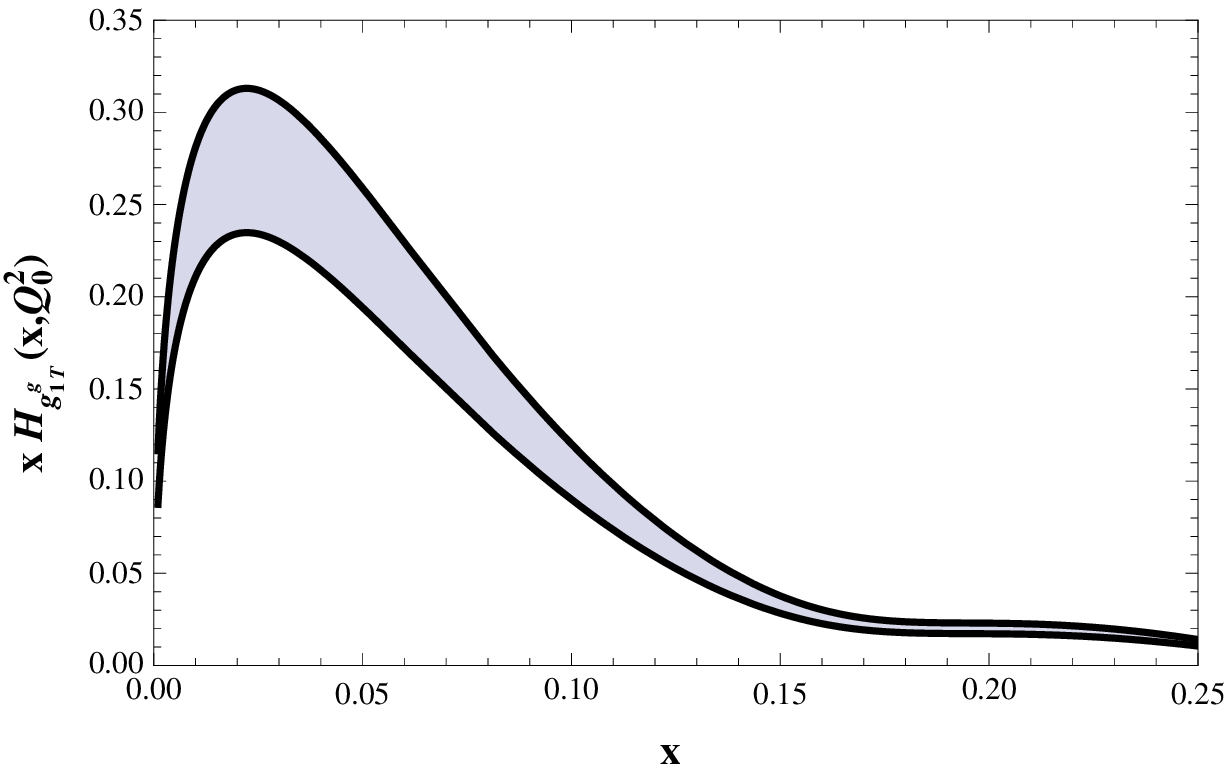,scale=.57}
\end{center}
\vspace*{-.4cm}
\noindent
\caption{$x$ dependence of the gluon GPD 
$x {\cal H}_{g_{1T}^g}(x,Q_0^2)$ 
for $Q_0^2=0.5$ GeV$^2$ (left panel) and 
for $Q_0^2=1$ GeV$^2$   (right panel).  
\label{fig19}}

\begin{center}
\epsfig{figure=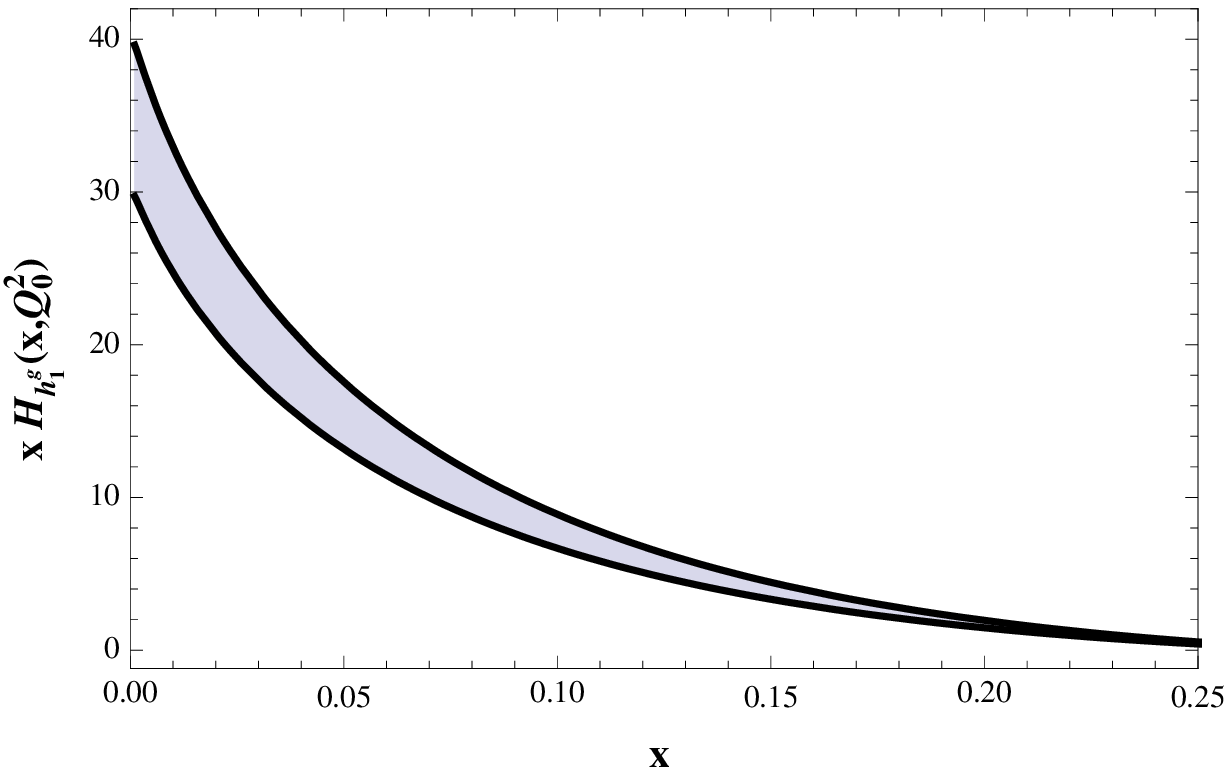,scale=.57}
\hspace*{1cm}
\epsfig{figure=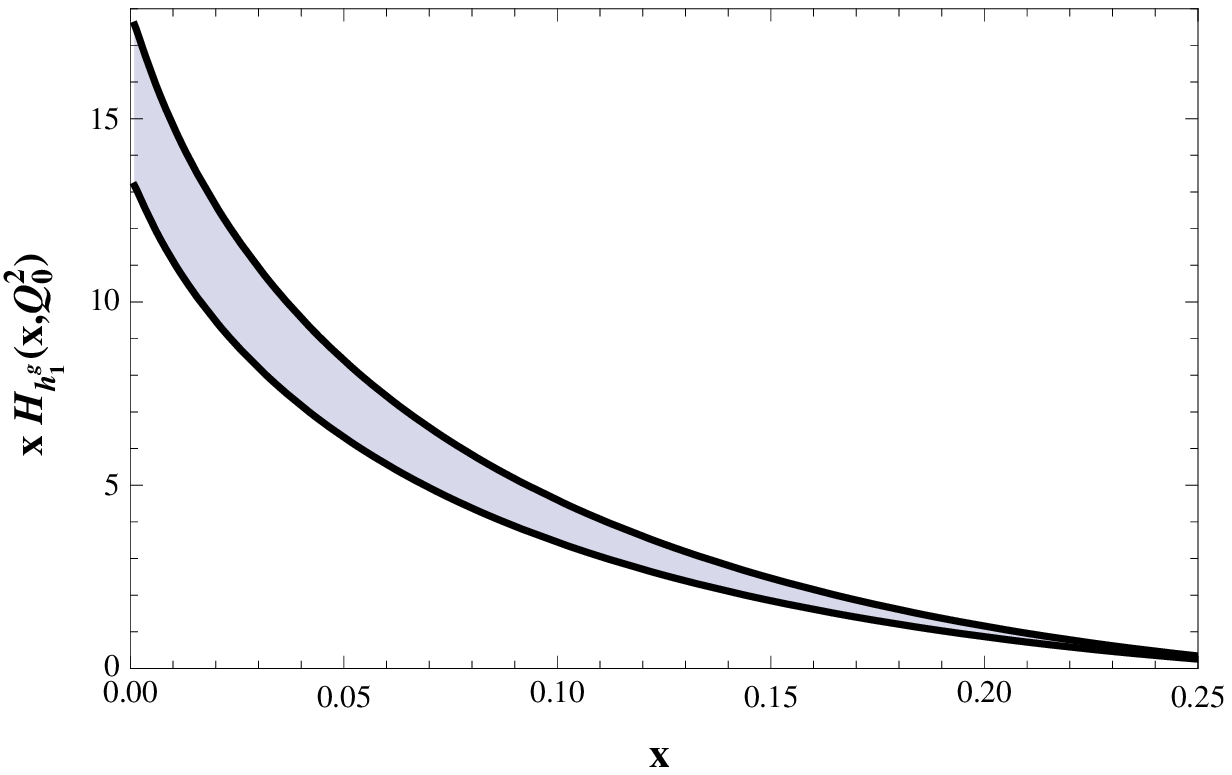,scale=.57}
\end{center}
\vspace*{-.4cm}
\noindent
\caption{$x$ dependence of the gluon GPD 
$x {\cal H}_{h_1^{\perp g}}(x,Q_0^2)$ 
for $Q_0^2=0.5$ GeV$^2$ (left panel) and 
for $Q_0^2=1$ GeV$^2$   (right panel).  
\label{fig20}}
\end{figure}

\begin{figure}
\begin{center}
\epsfig{figure=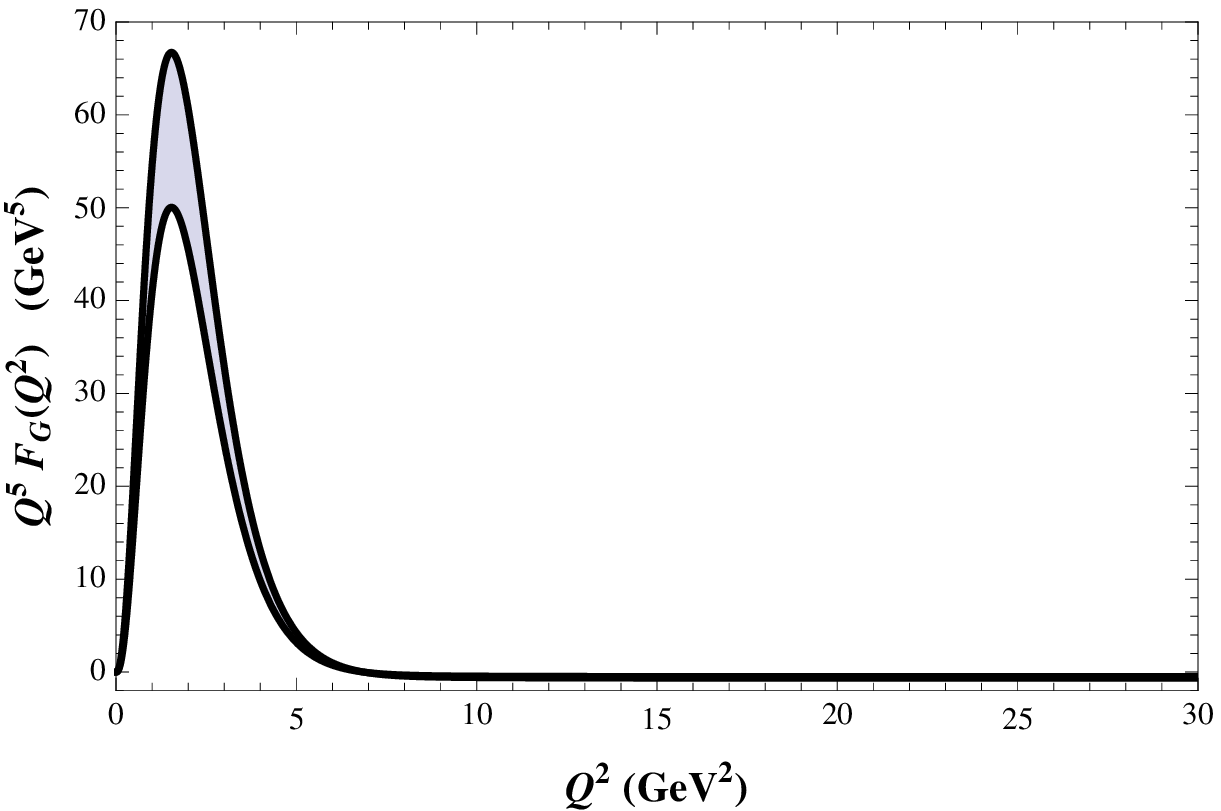,scale=.57}
\hspace*{1cm}
\epsfig{figure=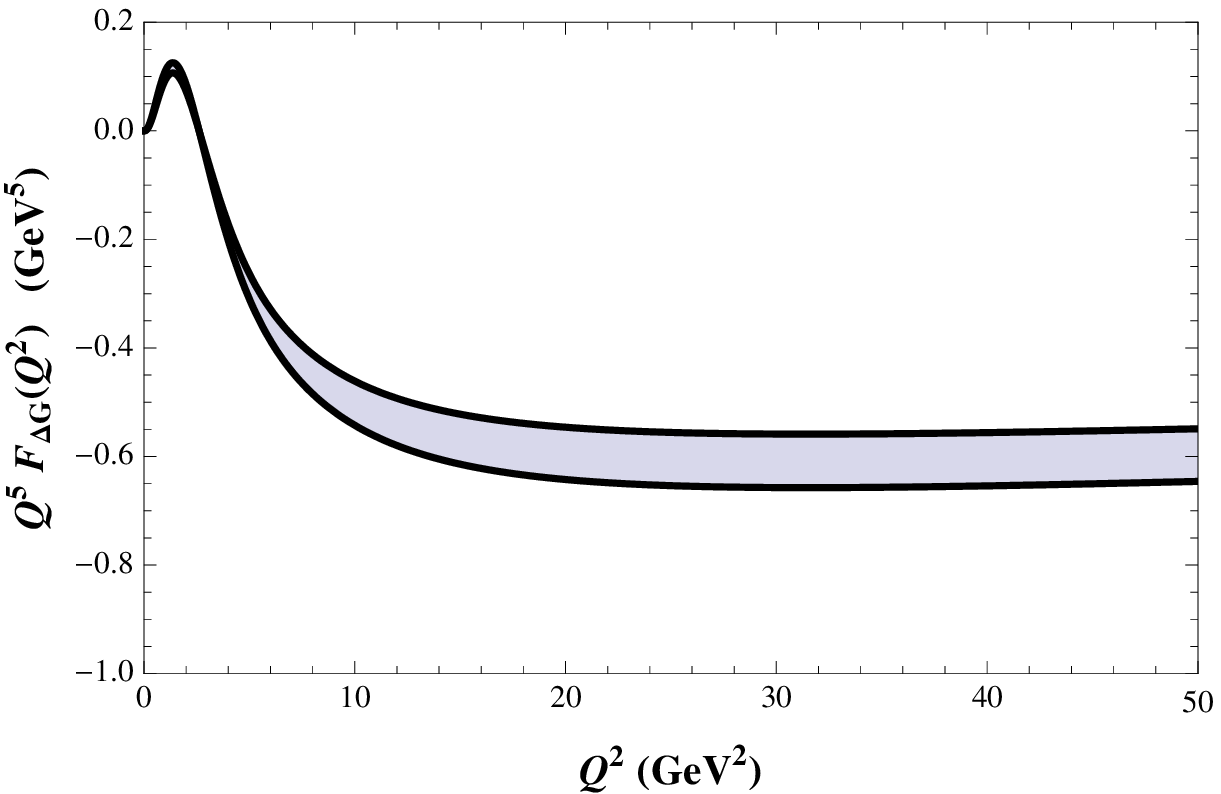,scale=.57}
\end{center}
\vspace*{-.3cm}
\noindent
\caption{$Q^2$ dependence of 
the gluon form factors 
$Q^5 F_G(Q^2)$ (left panel) and 
$Q^5 F_{\Delta G}(Q^2)$ (right panel).  
\label{fig21}}

\begin{center}
\epsfig{figure=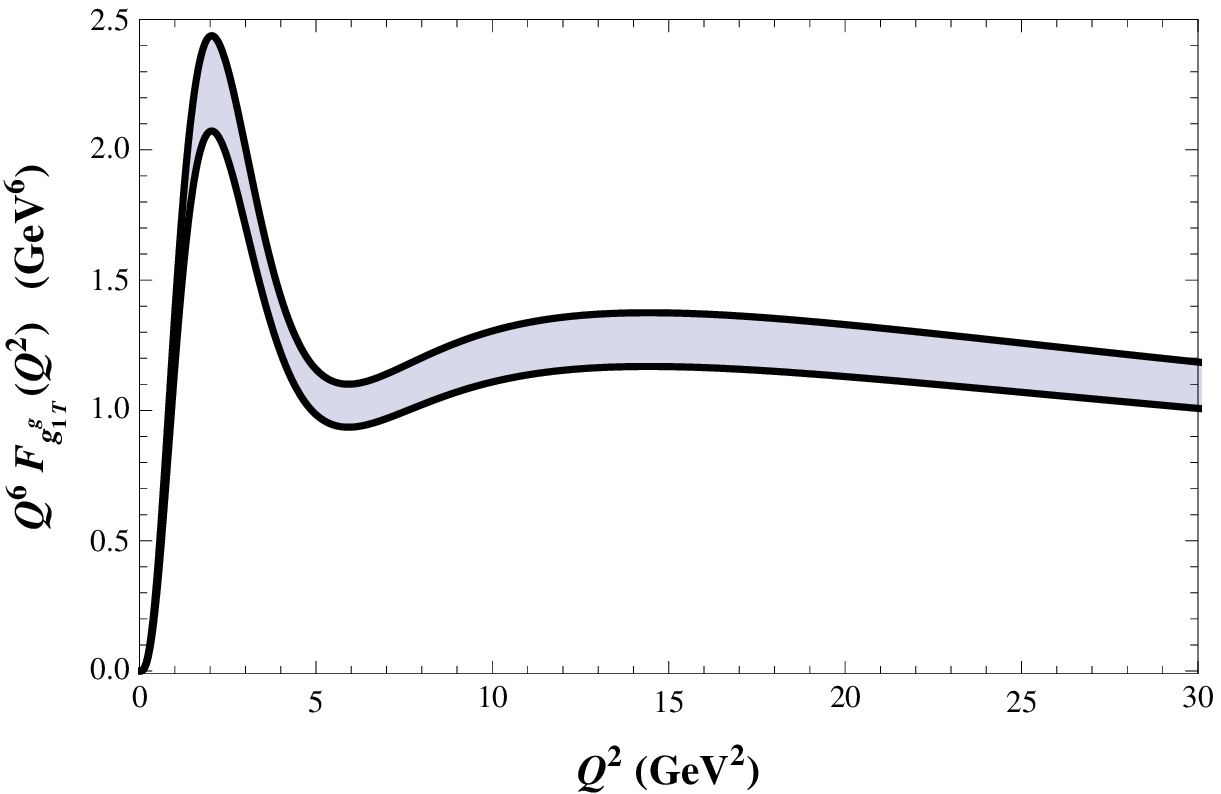,scale=.57}
\hspace*{1cm}
\epsfig{figure=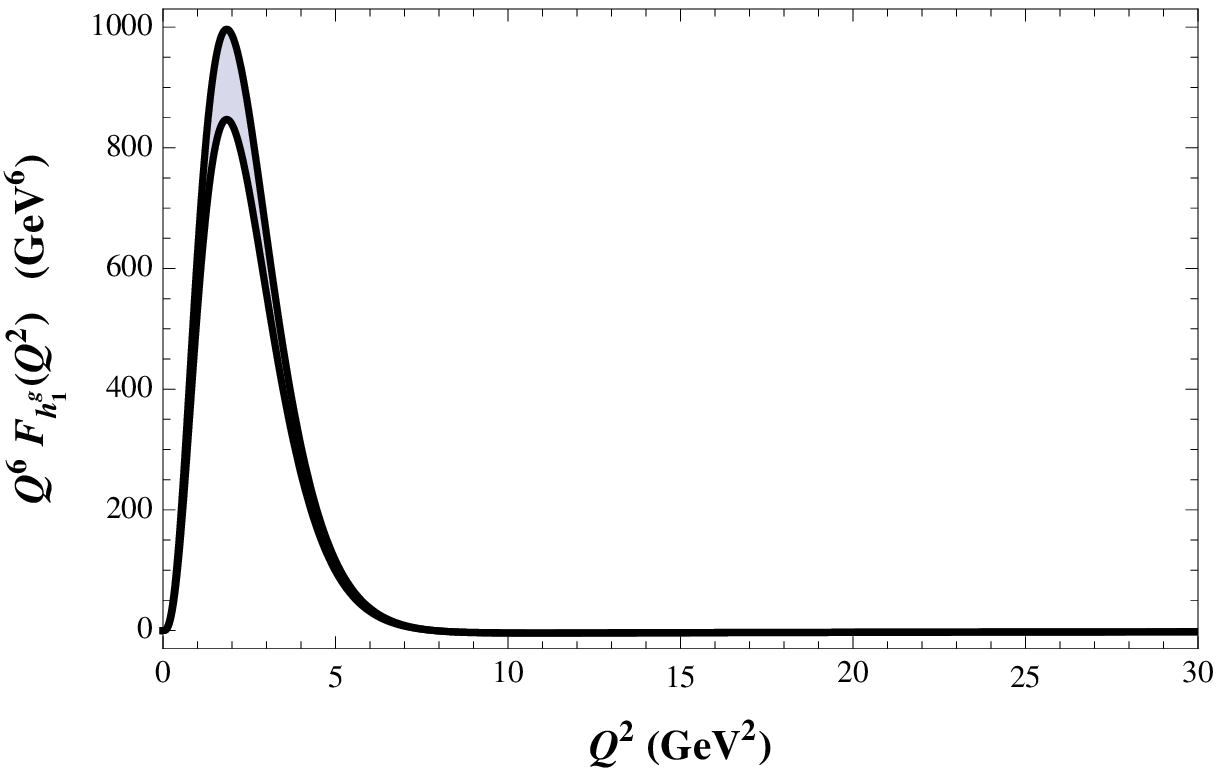,scale=.57}
\end{center}
\vspace*{-.3cm}
\noindent 
\caption{$Q^2$ dependence of 
the gluon form factors 
$Q^6 F_{g_{1T}^g}(Q^2)$ (left panel) and 
$Q^6 F_{h_1^{\perp g}}(Q^2)$ (right panel).  
\label{fig22}}
\end{figure}

Next, it is interesting to compare the $Q^2$ dependence of our GPDs 
with the ones derived in Ref.~\cite{Diehl:1998kh}:
\eq 
{\cal H}_G(x,Q^2) &=& \ \ G(x)                       
\ \exp\Big[-\frac{a_g^2}{2} \, \frac{1-x}{x} \, Q^2\Big] \,, 
\nonumber\\ 
{\cal H}_{\Delta G}(x,Q^2) &=& \Delta G(x)           
\ \exp\Big[-\frac{a_g^2}{2} \, \frac{1-x}{x} \, Q^2\Big] \,, 
\en 
where $a_g = 0.75$ GeV$^{-1}$ is the scale parameter. 
In both approaches the gluon GPDs are expressed in terms of gluon PDFs, 
therefore we just analyze the ratios of corresponding GPDs derived 
in Ref.~\cite{Diehl:1998kh} 
and by us: 
\eq\label{Ratio_GPD}
R(x,Q^2) = \frac{\exp\Big[-\frac{a_g^2}{2} \, \frac{1-x}{x} \, Q^2\Big]}
{\exp\Big[- D_g (1-x)^2 \frac{Q^2}{4 \kappa^2}\Big]} \,,
\en 
where $D_g$ is the same parameter as used in the analysis of the gluon TMDs. 
One can see that exponential of the GPDs derived in our formalism has an extra 
fall off $(1-x)$, which is consistent with large scaling of 
the gluon form factors dictated by quark counting rules. In case of GPDs 
this extra fall off is not sufficient at large $x$, because the exponentials in both approaches 
becomes 1 at $x \to 1$ and the GPDs are determined by the corresponding PDF. 
In Figs.~\ref{fig12}(a) and (b) we present a 3D plot of the ratio $R(x,Q^2)$, as function 
of $x$ and $Q^2$. It is seen that two predictions for the gluon GPDs have very good agreement 
in the region $0.2 \le x \le 1$, i.e. the ratio $R(x,Q^2) \simeq 1$ and are differed 
at small $x \le 0.2$. Note that 
the small $x$ behavior of the GPDs in our approach is consistent with small $x$ behavior 
of the PDFs and TMDs, which are in our approach are consistent with world 
data~\cite{Ball:2017nwa,Nocera:2014gqa} and similar approaches~\cite{Bacchetta:2020vty}. 
On the other hand, as we stressed before, the similar 
approaches~\cite{Meissner:2007rx,Bacchetta:2008af,Bacchetta:2020vty,Kaur:2020pvc}  
for quark and gluon TMDs and GPDs are consistent with us at small and intermediate $x$, 
while they have different large $x$ behavior inconsistent with counting rules. 
By the way, they can be easily improved by taking into account a specific 
$x$-dependence of the couplings/form factors. Such idea was 
proposed in Ref.~\cite{Jakob:1997wg}, where the spectator model was derived. 
In particular, it was suggested that the multipole form factors of
spectator diquark contain a free parameter $\alpha$, which indicates the power of 
form factors and can be clearly fixed to fulfill large $x$ counting rules. 
Therefore, our approach has advantage that it is valid in whole region of $x$ and compatible 
respectively with approaches developed for specific regions of $x$. 

In Figs.~\ref{fig13}-\ref{fig20} we present our results for the GPDs. In particular, 
in Figs.~\ref{fig13}-\ref{fig16} we display the $Q^2$ behavior of the gluon GPDs 
$x_0 {\cal H}(x_0,Q^2)$ at fixed values of $x_0=0.001$ (left panel) and $x_0=0.1$ (right panel). 
In Figs.~\ref{fig17}-\ref{fig20} we show the $x$ behavior of the gluon GPDs at fixed value of 
transversed momentum squared $Q_0^2=1$ GeV$^2$. As expected our curves for the gluon GPDs fall of 
with increasing $Q^2$ and vanish at $Q^2 \simeq 2-3$ GeV$^2$. 
Finally, in Figs.~\ref{fig21} and~\ref{fig22} we show our predictions for the gluon form factors. 
In the plots we multiply each form factor with corresponding inverse power of $Q^2$ corresponding 
to the large $Q^2$ asymptotics. In particular, in Figs.~\ref{fig21}(a), \ref{fig21}(b), \ref{fig22}(a), 
and~\ref{fig22}(b) we present the results for 
$Q^5 F_G(Q^2)$, $Q^5 F_{\Delta G}(Q^2)$, $Q^6 F_{g_{1T}^g}(Q^2)$, and 
$Q^6 F_{h_1^{\perp g}}(Q^2)$, respectively. One can see that all form factors approach the corresponding 
large $Q^2$ asymptotics at $Q^2 \sim 10$ GeV$^2$. 

\section{Summary}

In the present paper we have explicitly demonstrated how to correctly 
define gluon parton distributions (PDFs, TMDs, and GPDs) and form factors 
in the soft-wall AdS/QCD approach based on the use of quadratic
dilaton. In the description of the partonic structure of hadrons our 
approach based on ideas of Refs.~\cite{Brodsky:1989db,Brodsky:1994kg,%
Brodsky:2000ii,Brodsky:2002cx,Lyubovitskij:2020gjz} 
and consistent with constraints imposed 
by QCD~\cite{Drell:1969km,Bloom:1970xb,Brodsky:1973kr,%
Blankenbecler:1974tm,Yuan:2003fs,Aicher:2010cb,Mulders:2000sh}. 
In particular, in the case of nucleons we derive the expressions 
for the T-even gluon TMDs in terms of LFWFs describing the gluon-three quark Fock 
component in the nucleon as bound state of struck gluon and three-quark core 
(spectator). Next using expressions of the PDFs in terms of LFWfs we 
can express all gluon parton densities in terms of the set of the 
gluon PDFs $G(x)$ and $\Delta G(x)$. We demonstrated that our formalism is consistent 
with model-independent derivation of the gluon correlators 
proposed in Ref.~\cite{Mulders:2000sh} and later considered 
in Ref.~\cite{Meissner:2007rx,Bacchetta:2020vty}. 
We proved that our gluon TDMs obey the model-independent
Mulders-Rodrigues inequalities~\cite{Mulders:2000sh} without 
referring to a choice of model parameters. As a new result we derived 
the sum rule involving four T-even TMDs~(\ref{SR}), from which 
the Mulders-Rodrigues positivity bounds follow immediately. 
We checked that in similar approaches this sum rule is also 
fulfilled. E.g., in the spectator model considered 
in Ref.~\cite{Bacchetta:2020vty} the sum rule holds when the minimal 
coupling of gluon with three-quark spectator is used. 
In the quark target model discussed in Ref.~\cite{Meissner:2007rx} 
the sum rules is slightly different upon replacement of the nucleon mass 
by the constituent quark mass. For the first time, 
we derived results for the large $x$ behavior of the gluon TMDs, GPDs, 
and form factors. All gluon parton distributions are defined in terms of the unpolarized 
and polarized gluon PDFs. As numerical application we calculated the T-even 
gluon TMDs using as input the gluon PDFs extracted 
recently in Ref.~\cite{Sufian:2020wcv} based on 
ideas of QCD approaches~\cite{Brodsky:1989db} and~\cite{Brodsky:1994kg} 
and using world data analysis performed 
by the NNPDF Collaboration in Refs.~\cite{Ball:2017nwa,Nocera:2014gqa}. 
We get very good of  our results for the gluon TMDs with results of similar approach  
developed recently in Ref.~\cite{Bacchetta:2020vty}. To solidify our approach we 
calculated the electromagnetic form factors of nucleons induced by valence 
quark partonic densities and get perfect agreement with data~\cite{Diehl:2013xca,Cates:2011pz}. 
Finally, we presented our predictions for the $x$ and $Q^2$ dependence of the gluon GPDs 
and $Q^2$ dependence of the gluon form factors. 

In conclusion, we note that our approach for quark and gluon parton densities 
is very similar to the approaches developed in Refs.~\cite{Jakob:1997wg,Meissner:2007rx,%
Bacchetta:2008af,Bacchetta:2020vty,Kaur:2020pvc}. In case of quarks the consistency 
was shown in Ref.~\cite{Bacchetta:2008af}, while in case of gluons we discuss it 
in the present paper. E.g., it is supported by good agreement of description of the gluon 
TMDs in Ref.~\cite{Bacchetta:2020vty} and in our formalism. As advantage of our framework 
we mention that we are able to predict all parton densities using analytical formulas 
for the TMDs, GPDs, and form factors in terms of the set of PDFs, which are taken from 
world data analysis. Also we fulfill all constraints imposed by QCD including 
counting rules at large~$x$. We should note that in similar approaches 
consistency with counting rules was not yet implemented. However, such possibility 
exists. In particular, in Ref.~\cite{Jakob:1997wg}, where the spectator model was proposed, 
it was suggested that the multipole form factors of spectator diquark contain a free 
parameter $\alpha$. This parameter indicates the power of form factors and can be clearly 
fixed to fulfill large $x$ counting rules. 

\begin{acknowledgments}

This work was funded 
by BMBF (Germany) ``Verbundprojekt 05P2018 - Ausbau von ALICE 
am LHC: Jets und partonische Struktur von Kernen'' 
(F\"orderkennzeichen: 05P18VTCA1), by CONICYT (Chile) under
Grants No. 7912010025, No. 1180232, ANID PIA/APOYO AFB180002 (Chile),    
by FONDECYT (Chile) under Grant No. 1191103, 
and by Tomsk State and Tomsk Polytechnic University Competitiveness 
Enhancement Programs (Russia).

\end{acknowledgments} 

\appendix 
\section{TMDs in the generalized version}
\label{app_TMD} 

Here we list the gluon TMDs in soft-wall AdS/QCD using 
generalized version for the LFWFs $\varphi^{(1)}$ 
and $\varphi^{(2)}$: 
\eq\label{LFWFs_generalized} 
\varphi^{(1)}(x,\bfk^2) &=& \frac{4 \pi}{\kappa} \,
\sqrt{G^+(x)} \ \beta(x) \, \sqrt{D_{g_1}(x)}
\, \exp\biggl[- \frac{\bfk^2}{2 \kappa^2} \, D_{g_1}(x) \biggr] \,, \\
\frac{1}{M_N} \,
\varphi^{(2)}(x,\bfk^2) &=& \frac{4 \pi}{\kappa^2} \, \sqrt{G^-(x)} \, 
\frac{D_{g_2}(x)}{1-x} \, \exp\biggl[- \frac{\bfk^2}{2 \kappa^2} \, D_{g_2}(x) \biggr] 
\,,  
\en 
where $D_{g_1}(x) > 0$ and $D_{g_2}(x) > 0$ are the profile functions. 

T-even gluon TMDs in the momentum $\bfk$ space: 
\eq
f_1^g(x,\bfk^2) 
&=& \frac{1}{16 \pi^3} \, 
\biggl[ (\varphi^{(1)}(x,\bfk)\Big)^2
+ \frac{\bfk^2}{M_N^2} \, \Big(1+(1-x)^2\Big) 
\, \Big(\varphi^{(2)}(x,\bfk)\Big)^2 \biggr]
\nonumber\\
&=& 
\biggl[ G(x) + G^-(x) \, \alpha_+(x)  \, 
\biggl(\frac{\bfk^2 \, D_{g_2}^2(x)}{\kappa^2 D_{g_1}(x)} - 1 \biggr) 
\, e^{-\bfk^2 \Delta D_g(x)/\kappa^2} 
\biggr]    
\, \frac{D_{g_1}(x)}{\pi \kappa^2} 
\, \exp\biggl[- \frac{\bfk^2}{\kappa^2} D_{g_1}(x)\biggr] 
\,, \label{f1g_full}\\
g_{1L}^g(x,\bfk) 
&=& \frac{1}{16 \pi^3} \, 
\biggl[ (\varphi^{(1)}(x,\bfk)\Big)^2
+ \frac{\bfk^2}{M_N^2} \, \Big(1-(1-x)^2\Big) 
\, \Big(\varphi^{(2)}(x,\bfk)\Big)^2 \biggr]
\nonumber\\
&=& 
\biggl[ \Delta G(x) + G^-(x) \, \alpha_-(x)  \, 
\biggl(\frac{\bfk^2 \, D_{g_2}^2(x)}{\kappa^2 D_{g_1}(x)} - 1 \biggr) 
\, e^{-\bfk^2 \Delta D_g(x)/\kappa^2} 
\biggr]    
\, \frac{D_{g_1}(x)}{\pi \kappa^2} 
\, \exp\biggl[- \frac{\bfk^2}{\kappa^2} D_{g_1}(x)\biggr] 
\,, \label{g1L_full}\\
g_{1T}^g(x,\bfk^2) 
&=& \frac{1}{8 \pi^3} \, 
\varphi^{(1)}(x,\bfk) \, \varphi^{(2)}(x,\bfk) \, (1-x) 
\nonumber\\
&=& \sqrt{G^2(x)-\Delta G^2(x)} \, \beta(x) \, 
\frac{\sqrt{D_{g_1}(x)} \, D_{g_2}(x) M_N}{\pi \kappa^3} \, 
\, \exp\biggl[- \frac{\bfk^2}{2\kappa^2} \Big(D_{g_1}(x)+D_{g_2}(x)\Big) \biggr] 
\nonumber\\
&=& g_{1T}^g(x) \ \frac{D_{g_1}(x)+D_{g_2}(x)}{2 \pi \kappa^2} 
\, \exp\biggl[- \frac{\bfk^2}{2\kappa^2} \Big(D_{g_1}(x)+D_{g_2}(x)\Big) \biggr] 
\,, \label{g1T_full}\\
h_{1}^{\perp g}(x,\bfk^2) 
&=& \frac{1}{4 \pi^3}  
\, \Big[\varphi_q^{(2)}(x,\bfk)\Big]^2 \, (1-x) 
\nonumber\\
&=& \frac{G(x)-\Delta G(x)}{1-x} \,  \frac{2 D^2_{g_2}(x) M_N^2}{\pi \kappa^4} 
\,  \exp\biggl[- \frac{\bfk^2}{\kappa^2} \, D_{g_2}(x) \biggr] 
= h_1^{\perp g}(x) \ \frac{D_{g_2}(x)}{\pi \kappa^2} 
\, \exp\biggl[- \frac{\bfk^2}{\kappa^2} D_{g_2}(x)\biggr] 
\,, \label{h1g_full}
\en 
where $\Delta D_g(x) = D_{g_2}(x)-D_{g_1}(x)$. 

The gluon PDFs are written as: 
\eq\label{gluon_PDFs_general} 
f_1^g(x)    &=& \int d^2\bfk \, f_1^g(x,\bfk^2)    = G(x)        \,, 
\nonumber\\ 
g_{1L}^g(x) &=& \int d^2\bfk \, g_{1L}^g(x,\bfk^2) = \Delta G(x) \,, 
\nonumber\\ 
g_{1T}^g(x) &=& \int d^2\bfk \, g_{1T}^g(x,\bfk^2) = 
\frac{2 \sqrt{D_{g_1}(x)} D_{g_2}(x) M_N}{\Big[D_{g_1}(x) + D_{g_2}(x)\Big] \kappa} \, 
\sqrt{G^2(x)-\Delta G^2(x)} \ \beta(x) \,, 
\nonumber\\  
h_1^{\perp g}(x) &=& \int d^2\bfk \, h_1^{\perp g}(x,\bfk^2) = 
\frac{2 D_{g_2}(x) M_N^2}{\kappa^2} \, \frac{G(x)-\Delta G(x)}{1-x} 
\,.
\en  
They obey the condition 
\eq 
\frac{\Big[g_{1T}^g(x)\Big]^2}{\Big[f_1^g(x)+g_{1L}^g(x)\Big] \, h_1^{\perp g}(x)} 
= \frac{2 D_{g_1}(x) D_{g_2}(x)}{\Big[D_{g_1}(x)+D_{g_2}(x)\Big]^2} \, (1-x) \, \beta^2(x) \,. 
\en 

The results for impact TMDs are expressed in terms of 
generalized impact LFWFs: 
\eq
\tilde\varphi^{(1)}(x,\bfb^2) &=&  \frac{1}{2\pi} 
\, \sqrt{G^+(x)} \ \beta(x) \, 
\exp\Big[ - \frac{\bfb^2 \kappa^2}{8 D_{g_1}(x)}\Big]\,, \\
\tilde\varphi^{(2)}(x,\bfb^2) &=&  \frac{1}{2\pi} \, \sqrt{G^-(x)} \, 
\frac{\sqrt{D_{g_2}(x)}}{1-x} 
\, \exp\Big[ - \frac{\bfb^2 \kappa^2}{8 D_{g_2}(x)}\Big]\,. 
\en

In this case the expressions for the impact T-even gluon TMDs are given by: 
\eq 
\tilde f_1^g(x,\bfb^2) &=& \int \frac{d^2\bfk}{(2\pi)^2} 
\, e^{i\bfb\bfk} \, f_1^g(x,\bfk^2) 
\nonumber\\ 
&=& \Big[\tilde\varphi^{(1)}(x,\bfb^2)\Big]^2 
+   \Big[\tilde\varphi^{(2)}(x,\bfb^2)\Big]^2 \, 
\frac{1 + (1-x)^2}{D_{g_2}(x)} \, 
\biggl[ 1 + \frac{\bfb^2 \kappa^2}{4 D_{g_2}(x)} \biggr] 
\nonumber\\ 
&=& \frac{1}{4\pi^2} \, \biggl[ G(x) 
+ G^-(x) \, \alpha_+(x) \, \gamma(x,\bfb^2) \biggr] 
\, \exp\Big[-\frac{\bfb^2 \kappa^2}{4 D_{g_1}(x)}\Big] \,, \\
\tilde g_{1L}^g(x,\bfb^2) &=& \int \frac{d^2\bfk}{(2\pi)^2} 
\, e^{i\bfb\bfk} \, g_{1L}^g(x,\bfk^2) 
\nonumber\\
&=& 
  \Big[\tilde\varphi^{(1)}(x,\bfb^2)\Big]^2 
+ \Big[\tilde\varphi^{(2)}(x,\bfb^2)\Big]^2 \, 
\frac{1 - (1-x)^2}{D_{g_2}(x)} \, 
\biggl[ 1 + \frac{\bfb^2 \kappa^2}{4 D_{g_2}(x)} \biggr] 
\nonumber\\ 
&=& \frac{1}{4\pi^2} \, \biggl[ \Delta G(x) 
+ G^-(x) \, \alpha_-(x) \, \gamma(x,\bfb^2) \biggr] 
\, \exp\Big[-\frac{\bfb^2 \kappa^2}{4 D_{g_1}(x)}\Big] \,, \\ 
\tilde g_{1T}^g(x,\bfb^2) &=& 
\int \frac{d^2\bfk}{(2\pi)^2} \, e^{i\bfb\bfk} \, g_{1T}^g(x,\bfk^2) 
\nonumber\\
&=& \frac{2 M_N}{\kappa} \, \sigma(x,\bfb^2) \, 
\tilde\varphi^{(1)}(x,\bfb^2) \tilde\varphi^{(2)}(x,\bfb^2) (1-x)  
\nonumber\\
&=& \frac{1}{4 \pi^2} \, g_{1T}^g(x) 
\, \exp\Big[-\frac{\bfb^2 \kappa^2}{2 (D_{g_1}(x)+D_{g_2}(x))}\Big] 
\,, \\
\tilde h_1^{\perp g}(x,\bfb^2) &=& \int \frac{d^2\bfk}{(2\pi)^2}  \, e^{i\bfb\bfk} 
\, h_1^{\perp g}(x,\bfk^2) \nonumber\\
&=& \frac{4 M_N^2}{\kappa^2} \, 
\Big[\tilde\varphi^{(2)}(x,\bfb)\Big]^2 \,  (1-x)  
\nonumber\\
&=& \frac{1}{4 \pi^2} \, h_1^{\perp g}(x) 
\, \exp\Big[-\frac{\bfb^2 \kappa^2}{4 D_{g_2}(x)}\Big] \,, 
\en  
where 
\eq 
\gamma(x,\bfb^2) &=& 
\biggl[1 + \frac{\bfb^2 \kappa^2}{4 D_{g_2}(x)} \biggr] \, 
\exp\biggl[ - \frac{\bfb^2 \kappa}{4 D_{g -}(x)}\biggr] \,, 
\nonumber\\
\frac{1}{D_{g \pm}(x)} &=& 
\frac{1}{D_{g_2}(x)} \pm \frac{1}{D_{g_1}(x)} \,, 
\nonumber\\
\sigma(x,\bfb^2) &=& \frac{2 \sqrt{D_{g_1}(x) D_{g_2}(x)}}{D_{g_1}(x)+D_{g_2}(x)} \, 
\exp\biggl[ - \frac{\bfb^2 \kappa^2}{8} \, 
\frac{D_{g +}(x)}{D_{g -}^2(x)}\biggr] \,. 
\en  
One should stress again that the Mulders-Rodrigues inequalities~\cite{Mulders:2000sh} 
in the momentum $\bfk$ space hold in our approach without referring 
to a choice of the LFWF functions and in both symmetric 
$D_{g_1}(x)=D_{g_2}=D_g(x)$~(\ref{LFWFs_symmetric}) and 
generalized $D_{g_1}(x) \neq  D_{g_2}$~(\ref{LFWFs_generalized}) case. 

In impact $\bfb$ space one can also derive the inequalities between 
gluon TMDs, which involve profile functions $D_{g}$ functions and 
scale parameter $\kappa$.  
In symmetric case the inequalities were derived in 
Eqs.~(\ref{ineq1_bp})-(\ref{ineq3_bp}). 

In the generalized case the gluon TMDs satisfy the following relations: 
\eq 
& &\tilde g_{1L}^g(x,\bfb^2) \leq \tilde f_1^g(x,\bfb^2) 
\,, \label{ineq1_gbp} \\[2mm] 
& &\tilde g_{1T}^g(x,\bfb^2) \leq 
\frac{M_N}{\kappa} \, \sqrt\frac{D_{g_2}(x)}{1+\frac{\bfb^2 \kappa^2}{4 D_{g_2}(x)}} 
\, \tilde f_1^g(x,\bfb^2)
\leq 
\frac{M_N}{\kappa} \, \sqrt{D_{g_2}(x)} \, \tilde f_1^g(x,\bfb^2)
\,, \label{ineq2_gbp} \\[2mm] 
& &\tilde h_1^{\perp g}(x,\bfb^2) 
\leq \frac{2 M_N^2}{\kappa^2} 
\, \frac{D_{g_2}(x)}{1+\frac{\bfb^2 \kappa^2}{4 D_{g_2}(x)}}
\, \tilde f_1^g(x,\bfb^2) 
\leq \frac{2 M_N^2}{\kappa^2} 
\, D_{g_2}(x) \, \tilde f_1^g(x,\bfb^2) 
\,. \label{ineq3_gbp} 
\en 
One can see that the first inequality~(\ref{ineq1_gbp}) 
between $\tilde g_{1L}^g(x,\bfb^2)$ 
and $\tilde f_1^g(x,\bfb^2)$ is the same in both symmetric and 
generalized case and is similar to the one 
in momentum $\bfk$ space. The second~(\ref{ineq2_gbp})  
and third~(\ref{ineq3_gbp}) inequalities involve 
the scale parameter $\kappa$ and profile function $D_{g_2}(x)$. 
In the limit $D_{g_1}(x) = D_{g_2}(x) = D_g(x)$ 
the inequalities~(\ref{ineq2_gbp}) and~(\ref{ineq3_gbp}) 
reduce to the inequalities~(\ref{ineq2_bp}) and~(\ref{ineq3_bp}) 
in the symmetric case. 

The proof of the inequalities is straightforward. 
In particular, the inequality~(\ref{ineq1_gbp}) follows from 
decomposition of $\tilde f_1^g(x,\bfb^2)$ and 
$\tilde g_{1L}^g(x,\bfb^2)$ in terms of the LFWFs 
$\tilde\varphi^{(i)}(x,\bfb)$: 
\eq 
\tilde f_1^g(x,\bfb^2) - \tilde g_{1L}^g(x,\bfb^2) 
= 2 \Big[\tilde\varphi^{(2)}(x,\bfb^2)\Big]^2 \, \frac{(1-x)^2}{D_{g_2}(x)}
\biggl[ 1 + \frac{\bfb^2 \kappa^2}{4 D_{g_2}(x)} \biggr] \ge 0 \,. 
\en 
To derive the inequality~(\ref{ineq2_gbp}) we start with inequality 
\eq\label{ineq2_gbpA} 
\tilde f_1^g(x,\bfb^2) - \frac{\kappa}{M_N \sigma(x,\bfb^2) \sqrt{D_{g_2}(x)}} 
\, \sqrt{1+ \frac{\bfb^2 \kappa^2}{4 D_{g_2}(x)}} \, 
\tilde g_{1T}^g(x,\bfb^2) \ge 0 \,,
\en 
which holds because it can be written in the form 
\eq 
\biggl[ \tilde\varphi^{(1)}(x,\bfb) 
- \tilde\varphi^{(2)}(x,\bfb) \, \frac{1-x}{\sqrt{D_{g_2}(x)}} \, 
\sqrt{1+ \frac{\bfb^2 \kappa^2}{4 D_{g_2}(x)}}\biggr]^2 
+ \frac{1}{D_{g_2}(x)} \, \biggl[ \tilde\varphi^{(2)}(x,\bfb) \biggr]^2 \, 
\biggl[1+ \frac{\bfb^2 \kappa^2}{4 D_{g_2}(x)} \biggr] \ge 0 \,.
\en 
Using inequalities~(\ref{ineq2_gbpA}), 
\eq 
\sigma(x,\bfb^2) \le \frac{2 \sqrt{D_{g_1}(x) D_{g_2}(x)}}{D_{g_1}(x)+D_{g_2}(x)} \le 1\,, 
\qquad 
\frac{1}{\sqrt{1+ \frac{\bfb^2 \kappa^2}{4 D_{g_2}(x)}}} \le 1 
\en 
we arrive at Eq.~(\ref{ineq2_gbp}). 

To derive the inequality~(\ref{ineq3_gbp}) we start with inequality 
\eq\label{ineq3_gbpA} 
\tilde f_1^g(x,\bfb^2) - \frac{\kappa^2}{2 M_N^2 D_{g_2}(x)} 
\, \biggl[1 + \frac{\bfb^2 \kappa^2}{4 D_{g_2}(x)}\biggr] \, 
\tilde h_1^{\perp g}(x,\bfb^2) \ge 0 \,. 
\en 
The latter inequality holds because it can be rewritten as 
\eq 
\biggl[\tilde\varphi^{(1)}(x,\bfb)\biggr]^2 + 
\biggl[\frac{x \, \tilde\varphi^{(2)}(x,\bfb)}{\sqrt{D_{g_2}(x)}}\biggr]^2 \, 
\biggl[1+ \frac{\bfb^2 \kappa^2}{4 D_{g_2}(x)} \biggr] \ge 0 \,.
\en 
Using inequalities~(\ref{ineq3_gbpA}) and 
\eq 
\frac{1}{1+ \frac{\bfb^2 \kappa^2}{4 D_{g_2}(x)}} \le 1 
\en 
we arrive at Eq.~(\ref{ineq3_gbp}).

\end{document}